\documentclass[a4paper,11pt]{article}

\usepackage{jcappub} 

\usepackage[T1]{fontenc} 

\usepackage{bm}
\usepackage{latexsym}
\usepackage{dcolumn}
\usepackage{amsfonts,amssymb}
\usepackage{graphicx}
\usepackage{epsfig}
\usepackage{psfrag}
\usepackage{amsmath}
\usepackage{enumerate}
\usepackage[hang,flushmargin]{footmisc}
\usepackage[titletoc,toc]{appendix}
\usepackage{lipsum}
\usepackage{subcaption}
\usepackage{xfrac}
\usepackage{hyperref}
\usepackage{rotating}
\usepackage{multirow}
\usepackage{mathtools}

\newcommand{\be}{\begin{equation}}
\newcommand{\ee}{\end{equation}}
\newcommand{\bea}{\begin{eqnarray}}
\newcommand{\eea}{\end{eqnarray}}
\newcommand{\bse}{\begin{subequations}}
\newcommand{\ese}{\end{subequations}}
\newcommand{\bce}{\begin{center}}
\newcommand{\ece}{\end{center}}
\newcommand{\bfg}{\begin{figure}}
\newcommand{\efg}{\end{figure}}
\newcommand{\bi}{\begin{itemize}}
\newcommand{\ei}{\end{itemize}}
\newcommand{\bed}{\begin{description}}
\newcommand{\eed}{\end{description}}
\newcommand{\ben}{\begin{enumerate}}
\newcommand{\een}{\end{enumerate}}
\newcommand{\nn}{\nonumber}

\newcommand{\la}{\label}
\newcommand{\pa}{\partial}
\newcommand{\fr}{\frac}
\newcommand{\sq}{\sqrt}
\newcommand{\no}{\noindent}

\newcommand{\rarr}{\rightarrow}

\newcommand{\lra}{\longrightarrow}

\def\a  {\alpha}
\def\b  {\beta}
\def\c  {\gamma}
\def\C  {\Gamma}
\def\d  {\delta}
\def\D  {\Delta}
\def\e  {\epsilon}

\def\f  {\phi}
\def\F  {\Phi}
\def\k  {\kappa}
\def\l  {\lambda}
\def\L  {\Lambda}
\def\m  {\mu}
\def\n  {\nu}

\def\O  {\Omega}

\def\r  {\rho}

\def\s  {\sigma}

\def\vph {\varphi}

\def\le {\left}
\def\ri {\right}
\newcommand{\cA}{\mathcal A}

\newcommand{\cK}{\mathcal K}
\newcommand{\cL}{\mathcal L}

\newcommand{\cO}{\mathcal O}

\newcommand{\cQ}{\mathcal Q}
\newcommand{\cR}{\mathcal R}
\newcommand{\cS}{\mathcal S}
\newcommand{\cT}{\mathcal T}
\newcommand{\cU}{\mathcal U}
\newcommand{\cV}{\mathcal V}

\newcommand{\fH}{\mathfrak H}
\newcommand{\fL}{\mathfrak L}

\newcommand{\fv}{\mathfrak v}
\newcommand{\fw}{\mathfrak w}
\newcommand{\nab}{\nabla\!}
\newcommand{\nt}{\widetilde{\nabla}\!}

\newcommand{\Rt}{\widetilde{\cR}}
\newcommand{\Ct}{\widetilde{\C}}
\newcommand{\ha}{\widehat{a}}
\newcommand{\hg}{\widehat{g}}

\newcommand{\wht}{\widehat{t}}

\newcommand{\hu}{\widehat{u}}

\newcommand{\hT}{\widehat{T}}

\newcommand{\hR}{\widehat{\cR}}
\newcommand{\hS}{\widehat{\cS}}
\newcommand{\hLm}{\widehat{\cL}^{(m)}}

\newcommand{\hr}{\widehat{\rho}}

\newcommand{\hnab}{\widehat{\nabla}}

\newcommand{\vx}{\vec{\pmb x}}

\newcommand{\Geff}{G_{\text{\scriptsize eff}}}
\newcommand{\Veff}{V_{\text{\scriptsize eff}}}

\newcommand{\mA}{m_{\!\!_\cA}}
\newcommand{\lmean}{\langle \l \rangle}

\newcommand{\rmt}{\r^{(m)}}

\newcommand{\rx}{\r_{\!_X}}
\newcommand{\px}{p_{\!_X}}

\newcommand{\rJ}{\r_{\!_J}}
\newcommand{\Om}{\O^{(m)}}

\newcommand{\OX}{\O^{\text{\scriptsize ({\it X})}}}

\newcommand{\rmp}{\r^{(m)}_{_0}}
\newcommand{\Omp}{\O^{(m)}_{_0}}
\newcommand{\rmf}{\r^{(m)}_{\text{\footnotesize eff}}}

\newcommand{\Omf}{\O^{(m)}_{\text{\footnotesize eff}}}
\newcommand{\Omff}{\O^{(m)}_{\text{\scriptsize eff}}}
\newcommand{\OXf}{\O^{(X)}_{\text{\footnotesize eff}}}
\newcommand{\OXff}{\O^{(X)}_{\text{\scriptsize eff}}}
\newcommand{\sw}{\mathsf w}

\newcommand{\wx}{\sw_{\!_X}}

\newcommand{\Fp}{\F_{\!_0}}

\newcommand{\Hp}{H_{_0}}
\newcommand{\fp}{\f_{_0}}
\newcommand{\tp}{t_{_0}}

\newcommand{\bh}{\overline{h}}

\newcommand{\bw}{\overline{\sw}}

\newcommand{\bOm}{\overline{\O}}
\newcommand{\bH}{\overline{H}}

\newcommand{\bY}{\overline{Y}}
\newcommand{\Omb}{\bOm^{(m)}}
\newcommand{\OXb}{\bOm^{(X)}}
\newcommand{\Ombp}{\bOm^{(b)}_{_0}}
\newcommand{\Omcp}{\bOm^{(c)}_{_0}}
\newcommand{\bHp}{\bH_{\!_0}}

\newcommand{\dvph}{\dot{\varphi}}

\newcommand{\nbig}{\textit{\footnotesize n}}
\newcommand{\sbig}{\textit{\footnotesize s}}
\newcommand{\zbig}{\textit{\footnotesize z}}
\newcommand{\nmax}{n_{\text{max}}}
\newcommand{\smax}{s_{\text{max}}}
\newcommand{\nmaxb}{\textit{\normalsize n}_{\text{\scriptsize max}}}
\newcommand{\smaxb}{\textit{\normalsize s}_{\text{\scriptsize max}}}
\newcommand{\wxmin}{\wx^{\mbox{\scriptsize min}}}
\newcommand{\fwmin}{\fw_{\mbox{\scriptsize min}}}

\newcommand*\rfra[2]{{}^{\scriptstyle{#1}}\!\!\diagup_{\!\!\scriptstyle{#2}}}
\newcommand*\rfraa[2]{{}^{\displaystyle #1}\!\!\!\diagup_{\!\!\displaystyle #2}}

\newcommand*\rfr[2]{{}^{\displaystyle #1}\!/{\displaystyle #2}}
\newcommand{\bdm}{\begin{displaymath}}
\newcommand{\edm}{\end{displaymath}}

\long\def\symbolfootnote[#1]#2{\begingroup%
\def\thefootnote{\fnsymbol{footnote}}\footnote[#1]{#2}\endgroup}
\numberwithin{equation}{section}

\title{\boldmath Weakly dynamic dark energy via metric-scalar couplings with torsion}

\author{Sourav Sur}
\author{Arshdeep Singh Bhatia}

\affiliation{Department of Physics \& Astrophysics, University of Delhi,\\ New Delhi - 110 007, India}

\emailAdd{sourav.sur@gmail.com, arshdeepsb@gmail.com}

\abstract{We study the dynamical aspects of dark energy in the context of a non-minimally 
coupled scalar field with curvature and torsion. Whereas the scalar field acts as the 
source of the trace mode of torsion, a suitable constraint on the torsion pseudo-trace 
provides a mass term for the scalar field in the effective action. In the equivalent 
scalar-tensor framework, we find explicit cosmological solutions representing dark energy 
in both Einstein and Jordan frames. We demand the dynamical evolution of the dark energy 
to be weak enough, so that the present-day values of the cosmological parameters could be 
estimated keeping them within the confidence limits set for the standard $\L$CDM model 
from recent observations. For such estimates, we examine the variations of the effective 
matter density and the dark energy equation of state parameters over different redshift 
ranges. In spite of being weakly dynamic, the dark energy component differs significantly 
from the cosmological constant, both in characteristics and features, for e.g. it interacts 
with the cosmological (dust) fluid in the Einstein frame, and crosses the phantom barrier 
in the Jordan frame. We also obtain the upper bounds on the torsion mode parameters and 
the lower bound on the effective Brans-Dicke parameter. The latter turns out to be fairly 
large, and in agreement with the local gravity constraints, which therefore come in support 
of our analysis.}

\begin{document}
\maketitle
\flushbottom



\section{Introduction       \la{sec:intro}}

Despite a host of recent developments, cosmology is faced with a vexed question:
{\em to what extent a dynamical evolution of dark energy} (DE) {\em is tenable, as 
opposed to a cosmological constant $\L$, in driving the late-time cosmic acceleration?} 
\cite{ries,perl,cope,amtsuj,wols,bamba,li-etal}. 
While the well-known ``fine tuning" problem with $\L$ 
\cite{carr-cc,paddy-cc,cline,polch,kraus} 
prompts one to look into the theoretical aspects of a dynamical DE, observations 
generally favour the $\L$CDM model (with $\L$ and the cold dark matter (CDM) as the 
dominant energy contributors in the universe)
\cite{dodel,peack,pebrat,sola}. 
Nevertheless, the type Ia supernovae data
\cite{kow-union,suz-union,jla}, 
as well as the WMAP 
\cite{hin-wmap9,ben-wmap9} 
and PLANCK results 
\cite{ade-pln15-13,ade-pln15-15}, 
do provide some room for a DE with at least a slowly time-varying equation of state 
(EoS) parameter $\wx$, on an average close to the $\L$CDM value (equal to $-1$).

Extensive studies of the dynamical aspects of DE have explored a wide variety of its 
candidates  
\cite{cope,amtsuj,wols,bamba,li-etal}. 
Most common are of course the scalar fields, e.g. quintessence, k-essence, dilaton, 
tachyon, chameleon, etc. 
\cite{{cald-quin,carr-quin,cope-quin,zlat-quin,tsuj-quin,khur-quin,
AMS-kess-2000,AMS-kess-2001,mal-kess,scher-kess,sssd-kess,shar-kess,
piaz-dil,kar-dil, bag-tach,cal-tach,mart-tach, khour-cham,brax-cham}}, 
and the aerodynamic fluids, such as the Chaplygin gas and its variants 
\cite{bent-chap,xu-chap-1,xu-chap-2}. 
Some of these have their origin traced from underlying string/M theories 
\cite{wit,bous,shgreen}
or brane-world scenarios involving Dirac-Born-Infeld (DBI) actions
\cite{gar-dbi,mart-dbi,ahn-dbi,ss-dbi}. 
Besides, a considerable attention have been drawn in recent years by the DE models 
in modified/alternative theories of gravity 
\cite{chib-mg,noj-mg-2005,noj-mg-2006,noj-mg-2007,sant-mg,soti-mg,clif-mg,jain-mg} 
(e.g. the $f(\cR)$ theories\footnote{Characterized by arbitrary functional dependence 
of the gravitational Lagrangian on the Ricci scalar $\cR$.} 
\cite{fay-fR,fauk-fR,feltsuj-fR,cat-fR,noj-fR}). 
In many occasions, such purely geometric theories have one-to-one mapping with 
scalar-tensor theories 
\cite{BD,frni,fujii,eliz-st,camp-st,bois-st,sar-st}, 
and as such can give rise to interacting (or unified) dark energy-matter scenarios 
\cite{amen-int,com-int,farpeb-int,zha-int,cai-int,bert-int,bertbrun-int,
li-int,guen-int} 
under conformal transformations. A class of theories that comes under the alternative 
gravity category, is that of scalar field(s) coupled to gravity with torsion, viz. 
the {\em metric-scalar-torsion} (MST) theories 
\cite{shap,neto}. 
Our objective here is to look for cosmological models that emerge out of these 
theories, particularly from the DE perspective.

Torsion, which geometrically manifests the classical spin density, is a third rank 
(partially) antisymmetric tensor field introduced via the covariant derivative in 
theoretical formulations of gravity in the Riemann-Cartan space-time 
\cite{traut,hehlrev1976,akr,hehl1995,hehlobu2001,sabgasp,sabsiv,popl,west}. 
Such theories, unlike General Relativity (GR), are characterized by an asymmetric 
(but metric-compatible) affine connection, a two-index antisymmetrization of which is 
identified as the torsion. The study of torsion has a very long history that traces 
back to the attempts of Cartan and Einstein, in early 1920s, to unify the fundamental 
interactions in the spirit of general covariance 
\cite{traut,hehlrev1976,akr,hehl1995,hehlobu2001,sabgasp,sabsiv,popl,west}. 
Modern conception of torsion is often apprehended as a classical background being provided 
for quantized matter with spin, and as such torsion could be a low energy manifestation 
of a fundamental (quantum gravitational) theory 
\cite{shap,ham1991,ham2000,ham2002}. 
For instance, the massless antisymmetric Kalb-Ramond (KR) field in closed string theory is 
argued to be the source of a completely antisymmetric torsion in the low energy limit 
\cite{pmssg,saa}. 
Many implications of such a KR-induced torsion have been studied in detail, both in the 
four dimensional framework as well as in the extra dimensional scenarios
\cite{bmssg,ssgss2001,ssgas,das-etal,skpmssgas,skpmssgss,bmssgss,bmssgsen,skssgss,ssgss2003,
ssgss2004,rubhehl,dmssg,dmpmssg,bmssssgss,dmssgss-EPJ,dmssgss-PRD,ssgsdss,hehl-etal-2005,
acpm,alex,tgssg,bmssssg,balani,sbac,adbmssg,scssg2015,cgssg,scssg2016}. 
Cosmology in particular being a major testing area for torsion, a few pertinent works worth 
special mention, viz. the non-singular cosmological models with torsion induced from vacuum 
quantum effects 
\cite{buchodshap1985,zarodshap,buchodshap1992}, 
the torsion quintessence models 
\cite{capo2002,capovig}, 
the $f(T)$ models in {\em teleparallel} theories 
\cite{capo-fT-2007,boem-fT,cai-fT,genglee-fT,li-fT,card-fT,ior-fT,gengleesar-fT,xusarleon-fT,
capo-fT-2013,skusartop-fT,kofpapsar-fT,sarrev-fT}, 
and the models based on the extended gravity theories 
\cite{alle-ext,capo-ext,car-ext}
and the Poincar\`e gauge theory of gravity (see 
\cite{blaghehl} 
for a classic compilation of contemporary works, see also 
\cite{mink1980,mink2007,yonest,nest,shie,mink2009,li-2009-PRD,li-2009-JCAP,baekhehlnest,tseng}). 

We in this paper study the extent to which torsion, via its coupling with curvature and 
a scalar field (viz. the MST coupling), can provide a self-consistent explanation of the 
DE evolution. In particular, restricting ourselves to homogeneous and isotropic cosmologies, 
we resort to a specific MST formalism that can accommodate a viable model of DE with  
presumably slow dynamics, not differing much from $\L$CDM. Conversely, from a supposition 
that such a DE model is indeed tenable in a MST cosmological setup, and the predicted 
cosmological parameters are within the limits given by the $\L$CDM error estimates from 
recent observations, we look to determine the bounds on the torsion mode parameters, and 
hence constrain the MST theory. Our main assumption is of a specific non-minimal coupling 
of the scalar field with the four-dimensional Riemann-Cartan (or $U_4$) Lagrangian. One 
motivation for this is to avoid a well-known uniqueness problem encountered while defining 
equivalent actions under the (usual) minimal coupling scheme, upon integrating out the 
boundary terms in four dimensions \cite{shap}. The non-minimal MST action inevitably implies
that the (vector) trace mode of torsion is sourced by the (supposedly primordial) scalar 
field, and hence gets eliminated via a constraint relation. Moreover, the assumption of a 
homogeneous and isotropic universe, described by the well-known Friedmann-Roberstson-Walker 
(FRW) metric, puts certain restrictions on the torsion degrees of freedom. Specifically, 
since the four-dimensional FRW space-time is foliated into three-dimensional maximally 
symmetric hypersurfaces of constant time, only those torsion tensor components that uphold 
the maximal symmetry can exist, and that too with a constrained coordinate dependence 
\cite{ssasb,tsim,dgssg1994,dgssg1997-1,dgssg1997-2,dgssg1999,bloom}. 
Suitable implementation of such constraints (or a few alternative ones) in the action may 
result in the scalar field picking up an effective mass, in addition to its inherent mass 
(if any), at the expense of one of the torsion modes, viz. the completely antisymmetric or 
the pseudo-trace mode. One is thus left with a scalar-tensor theory equivalent action that 
exhibits a non-minimal coupling of a massive scalar field with the Riemannian curvature 
scalar $\cR$ in four dimensions. We obtain exact particular solutions of the field equations 
derived from such an action, and make their parametric estimations, in course of building up 
a viable DE model. 

While dealing with the effective scalar-tensor theory, we resort to the formulations in both 
the {\em Einstein} frame and the {\em Jordan} frame (in which the corresponding MST action 
gets reduced to the Brans-Dicke (BD) form 
\cite{BD,frni,fujii}). 
Whereas the Jordan frame exhibits an explicit non-minimal coupling of the scalar field with 
curvature, in the Einstein frame the scalar has an inherent coupling with the cosmological 
matter (which we take as the non-relativistic {\em dust}). Being related by a conformal 
transformation, these two frames are mathematically equivalent. However, their physical 
equivalence is not ensured and can be reinstated only by a redefinition of the units of mass, 
length and time under the transformation. This makes the gravitational coupling constant in 
one frame, evolving with time in the other. As the formulation of a physical theory in a 
frame with a running coupling parameter is generally not feasible, in the literature one 
often finds a rigid set of units chosen for either of the frames, which of course destroys 
their physical equivalence 
\cite{frni,fujii}. 
So it becomes difficult to decide which frame is suitable for interpreting observational 
results. Without getting into this much debated issue (which has no clear resolution till 
date), we carry out the MST cosmological analysis in both Einstein and Jordan frames, 
taking one or the other to be physical. Of course, for the Jordan frame cosmology we 
actually resort to certain definitions and parametrizations that in a way enable us to 
avoid a direct confrontation with the running gravitational coupling parameter. 

As mentioned earlier, the assumption of weak DE dynamics allows for a parametric estimation 
in comparison with $\L$CDM. We take into account the $\L$CDM parametric marginalizations due 
to the WMAP nine year data 
\cite{hin-wmap9,ben-wmap9} 
and the PLANCK 2015 data 
\cite{ade-pln15-13,ade-pln15-15} 
(with combinations of high redshift supernovae of type Ia, lensing, and Baryon Acoustic 
Oscillations (BAO) data). We follow certain procedures in which we impose, for example, the 
condition that the effective (MST) Hubble constant $\Hp$ to be within the $68\%$ confidence 
limits of the corresponding $\L$CDM Hubble constant $\bHp$, for a given dataset. This in 
principle enables us to work out the bounds on the quantities of interest, such as the 
non-minimal coupling parameter $\b$ (or equivalently, the effective BD parameter $\fw$), 
and the torsion parameters (that are taken to be lengths of the trace and pseudo-trace mode 
vectors). Although approximate, for not being derived via a full fledged likelihood analysis, 
these bounds are of importance qualitatively (and to some extent quantitatively) in 
demonstrating the following:
\bi 
\item A slowly dynamical DE may result from even a small amount of torsion, via the coupling of the 
latter with a scalar field. As such, one may have an indirect observational signature of a fundamental 
theory (e.g. string theory), if torsion or(and) the scalar field is(are) considered as low energy 
manifestation(s) of the same.
\item It suffices to consider only a mass term for the scalar field. No other form of its potential, 
nor a cosmological constant, is necessary for a slight deviation from the $\L$CDM model.
\item The scalar field mass (or part thereof) may effectively be due to torsion's pseudo-trace 
vector $\cA^\m$, via for e.g. the norm fixation or higher order effects of the latter (which would 
then play a key role in the DE evolution, as opposed to torsion's trace mode).
\ei
Besides these, the very assumption of the weak dynamics of DE gets independent support in both Einstein 
and Jordan frames, as the upper limit of the non-minimal coupling parameter turns out to quite low, 
implying a large lower limit of the effective BD parameter $\fw$, consistent with the local gravity 
constraints on the latter 
\cite{acq-BD,avi-BD,chen-BD,als-BD}.

The organization of this paper is as follows: we start with the general discussion on the MST 
formalism in $\S$\ref{sec:mst-gen}, and in particular motivate the non-minimally coupled 
gravity-scalar field action in the Riemann-Cartan space-time. The standard cosmological setup 
for the MST coupling is described in $\S$\ref{sec:MST-dust}, in which the MST action is reduced 
to an equivalent scalar-tensor form by using the constraints on the torsion modes in a spatially 
flat FRW universe. We also make propositions for a scalar field mass term out of the torsion 
pseudo-trace mode, and define the effective torsion mode parameters in Jordan and Einstein 
frames. In $\S$\ref{sec:E-MST}, we make a comprehensive study of the DE evolution in the 
Einstein frame. We work out first the appropriate cosmological solution for the same. Then the 
parametric estimations are done using a comparative analysis, taking the six parameter base 
$\L$CDM model as the reference. We follow two procedures that depend on the choice of the 
cosmological parameter(s) relevant for determining the statistical bound on our model parameter. 
For different datasets, this bound is utilized to obtain the plots for various quantities of 
interest, and to study their characteristic difference with those for $\L$CDM. Finally, we work 
out the bounds on the torsion mode parameters and the effective BD parameter (or, rather the 
one for the non-minimal coupling, from which the Einstein frame cosmology stems out). A 
similar analysis is done in $\S$\ref{sec:J-MST}, for the Jordan frame MST cosmology. The most 
striking feature we notice in the Jordan frame is the effective crossing of the so-called 
`{\em phantom barrier}' which leaves the universe in a `{\em superaccelerating}' regime at the 
present epoch. For correlating the results of the analysis in Einstein and Jordan frames, we 
discuss in $\S$\ref{sec:transf}, the difference between our approach and that of an alternative 
one which takes into consideration the transformation of the model parameters (and other derived
quantities) under the redefinition of length, mass and time in terms of the conformal factor. 
We conclude with a brief summary and some open questions in $\S$\ref{sec:concl}.

{\em Notations and conventions} : We choose to work in units with the speed of light $c = 1$. The 
Newtonian gravitational coupling factor is denoted by $\, \k^2 \equiv 8 \pi G$. A subscript `$0$' 
affixed to the cosmological quantities denotes their values at the present epoch. The metric 
signature throughout is $\, (-,+,+,+)$, and the determinant of the metric tensor $g_{\m\n}$ is 
denoted by $g$.

\section{Metric-torsion coupling with a scalar field --- the general MST formalism} 
\la{sec:mst-gen}

Let us briefly review some important aspects of the four dimensional Riemann-Cartan ($U_4$) 
space-time, which is characterized by an asymmetric affine connection $\Ct^\l_{\m\n} 
\le(\neq \Ct^\l_{\n\m}\ri)$. 

The torsion tensor, defined by
\be \la{tor-def}
T^\l_{~\m\n} \,:=\, 2 \, \Ct^\l_{[\m\n]} \,\equiv\, \Ct^\l_{\m\n} \,-\, \Ct^\l_{\n\m} \,\,,
\ee
can in general be decomposed in its irreducible modes as \cite{shap,ssasb,capo}
\be \la{tor-decom}
T^\l_{~\m\n} \,=\, \fr 1 3 \le(\d^\l_\n \, \cT_\m \,-\, \d^\l_\m \, \cT_\n\ri) 
+\, \fr 1 6 \, \e^\l_{~\m\n\s} \, \cA^\s \,+\, \cQ^\l_{~\m\n} \,\,,
\ee
where $\, \cT_\m \,:=\, T^\n_{~\m\n} \,$ is torsion {\em trace} vector mode, $\, \cA^\s 
\,:=\, \e^{\a\b\c\s} \,T_{\a\b\c} \,$ is the torsion {\em pseudo-trace} vector mode, and 
$\, \cQ_{\m\n\s}\,$ is the {\em (pseudo-)tracefree} tensorial mode of torsion. 

The $U_4$ curvature scalar is expressed as\footnote{We correct a couple of misprints in 
Eq. (A.22) of ref. \cite{ssasb}.}
\be \la{U4-curv}
\Rt \,:=\, \cR \,-\, 2 \, \nab_\m \, \cT^\m \,-\, \fr 2 3 \, \cT_\m \cT^\m 
\,+\, \fr 1 {24} \, \cA_\m \cA^\m \,+\, \fr 1 2 \cQ_{\m\n\s} \cQ^{\m\n\s} \,\, 
\ee
where $\cR$ is the four dimensional Riemannian ($R_4$) curvature scalar, and $\nab_\m$ 
denotes the usual $R_4$ covariant derivative defined in terms of the Riemannian Levi 
Civita connection $\C^\l_{\m\n} \le(= \C^\l_{\n\m}\ri)$. The $U_4$ covariant derivative 
$\nt_\m$, on the other hand, is defined in terms of the asymmetric connection
$\Ct^\l_{\m\n}$, so that the metricity condition $\, \nt_\m g_{\a\b} = 0$ holds.

Our starting point is the  $U_4$ action non-minimally coupled to a scalar field $\f$:
\be \la{U4-f-ac}
\cS \,=\, \int d^4 x \, \sq{-g} \le[\fr {\b \f^2} 2 \, \Rt \,-\, \fr 1 2 \, g^{\m\n} \, 
\pa_\m \f \, \pa_\n \f \,-\, V (\f)\ri] \,,
\ee
where $\b$ is a dimensionless coupling parameter, $V (\f)$ is a self-interacting potential
for $\f$. The non-minimal coupling may be motivated as follows:

In the four-dimensional Minkowski ($M_4$) space-time, with metric $\eta_{\m\n} = \mbox{diag}
(-1,1,1,1)$, the scalar field Lagrangian is given in the standard form and an equivalent form
(up to a total divergence term) respectively as 
\be \la{M4-f-ac}
\cL_{(\f)} \,=\, - \, \fr 1 2 \, \eta^{\m\n} \pa_\m \f \, \pa_\n \f \,-\, V(\f) \,, \qquad
\mbox{and} \qquad
\cL'_{(\f)} \,=\, \fr 1 2 \, \eta^{\m\n} \f \, \pa_\m \pa_\n \f \,-\, V(\f) \,.
\ee
When coupled to curvature in the Riemannian ($R_4$) space-time, via the minimal coupling (MC) 
scheme ($\pa_\m \rightarrow \nab_\m$), the scalar field has Lagrangian given by either of the 
equivalent forms, viz. the generalization of $\cL_{(\f)}$ and the generalization of $\cL'_{(\f)}$, 
which differ by a $R_4$ divergence term. However, in the Riemann-Cartan ($U_4$) space-time 
having torsion, the MC generalizations ($\pa_\m \rightarrow \nt_\m$) of $\cL_{(\f)}$ and 
$\cL'_{(\f)}$ no longer remain equivalent, as their difference is not just a total divergence 
in $U_4$, but in addition a term proportional to $\, \sq{-g} \, \f^2 \, \nab_\m \cT^\m$ 
\cite{shap}. The choice of the Lagrangian for a minimally coupled metric-scalar-torsion (MST) 
theory therefore becomes ambiguous. The remedy to this is to modify the overall MST action by 
including a term that would cancel the above additional term (plus a total divergence at most). 
Such a modification may require one to resort to the whole exercise of suitably generalizing 
the standard Gibbons-Hawking-York (GHY) boundary term in presence of the torsion and scalar 
degrees of freedom. As an alternative however, one may simply look beyond the MC prescription, 
and propose a MST action which explicitly has a counter-term given as a contact interaction of 
the scalar field $\f$ with the torsion trace mode\footnote{Note that a scalar field could be 
minimally coupled to the $U_4$ action without any ambiguity when torsion is traceless (or 
completely antisymmetric, such as that induced by the string theoretic Kalb-Ramond field).}, 
or at least with its (Riemannian) covariant derivative $\nab_\m \cT^\m$. Now, it would seem 
quite a contrived setup if only the $\nab_\m \cT^\m$ term is picked up among the host of terms 
in $\Rt$ (given by Eq. (\ref{U4-curv})) and made to couple with $\f$ via a contact interaction. 
So to maintain regularity in the terms, we choose to couple the field $\f$ non-minimally with 
the entire $U_4$ Lagrangian $\, \sq{-g} \, \Rt$ in  Eq. (\ref{U4-f-ac}) above\footnote{More 
general MST actions are also there in the literature \cite{shap}. However we restrain from using 
those as they would involve more than one coupling parameter, which may affect the predictability 
of the theory.}. Using now Eq. (\ref{U4-curv}) we rewrite the MST action (\ref{U4-f-ac}), up to a
divergence term, as 
\bea \la{MST-ac}
\cS \,=\, \int d^4 x \, \sq{-g} \le[\fr{\b \f^2} 2 \le(\cR \,+\, 4 \, \cT^\m \, 
\fr{\pa_\m \f}{\f} \,-\, \fr 2 3 \, \cT_\m \cT^\m \,+\, \fr 1 {24} \, \cA_\m \cA^\m 
\,+\, \fr 1 2 \cQ_{\m\n\s} \cQ^{\m\n\s}\ri) \ri. \nn\\
\le. -\, \fr 1 2 \, g^{\m\n} \, \pa_\m \f \, \pa_\n \f \,-\, 
V (\f) \,+\, \cL^{(m)} \ri] \,,
\eea
where $\cL^{(m)}$ stands for the Lagrangian density of the cosmological fluid matter,
fermions, etc., that are to be coupled as well.

\section{MST theory in the standard cosmological setup} \la{sec:MST-dust}

To study the cosmological consequences of the MST coupling, let us resort to the 
following widely accepted scenario:
\ben 
\item The universe is homogeneous and isotropic at large scales, with the spatially flat
FRW metric structure
\be \la{FRW}
ds^2 \,=\, - \, dt^2 \,+\, a^2 (t) \le(dr^2 \,+\, r^2 \, d\O_2^2\ri) \,,
\ee
where $t$ is the comoving time coordinate, and $a(t)$ is the cosmological scale factor.
\item The cosmological fluid is of the form of a pressureless {\em dust}, with energy density 
$\r^{(m)} (t)$.
\een
As is well-known, the FRW metric describes a four-dimensional space-time foliated into 
three-dimensional maximally symmetric subspaces identified as hypersurfaces of constant 
time. For the maximal symmetry to be preserved in presence of torsion, the latter gets 
constrained from the requirement of its form-invariance under such symmetry. An in-depth 
study in ref. \cite{ssasb} (see also 
\cite{tsim,dgssg1994,dgssg1997-1,dgssg1997-2,dgssg1999,bloom}) 
reveals that for determining such constraints on torsion, one may in principle resort to 
two schemes depending on whether torsion plays a passive or an active role in implementing 
the very concept of maximal symmetry. In one scheme, in which maximal symmetry is attributed 
solely to the metric properties of space-time, the FRW metric structure restricts torsion 
to have at most four non-vanishing independent components, viz. $\,T_{110}, T_{220}, 
T_{330}$ and $\, T_{123}$, each of which can depend on time only. In the other scheme, 
in which torsion directly affects maximal symmetry, following strictly the principle of 
general covariance, some more independent torsion components with only time-dependence, 
viz. $\, T_{001}, T_{002}$ and $T_{003}$, are allowed in the FRW context \cite{ssasb}. 
Hence, for the irreducible torsion modes, one finds that both the schemes are in 
agreement yielding $\, \cQ_{\m\n\s} = 0$ and $\, \cA^\m = \d^\m_0 \l (t)$ where 
$\, \l(t)$ is a pseudoscalar function of time. The disagreement comes when one scheme 
allows only the $\cT_0$ component of the torsion trace vector $\cT_\m$, whereas the 
other scheme allows all the components (with some restrictions though). Nevertheless, 
this is not a concern for us since we are considering the cosmological matter to be the 
non-relativistic dust, having no explicit dependence on torsion. As such, $\cT_\m$ being 
an auxiliary field, satisfies the constraint 
\be \la{tr-eqn}
\cT_\m \,=\, 3 \, \fr{\pa_\m \f} \f \,\,,
\ee 
which (in conjunction with $\, \cQ_{\m\n\s} = 0$) implies the MST action (\ref{MST-ac}) 
reducing to
\be \la{MST-ac1}
\cS \,=\, \int d^4 x \, \sq{-g} \le[\fr{\b \f^2} 2 \le(\cR \,+\, \fr 1 {24} 
\, \cA_\m \cA^\m\ri) -\, \fr{\le(1 - 6\b\ri)} 2 \, g^{\m\n} \, \pa_\m \f \, \pa_\n \f 
\,-\, V (\f) \,+\, \cL^{(m)} \ri] \,.
\ee
This has an arbitrariness left in it though, viz. the choice of the potential $V (\f)$. Now, 
in analogy with the gravitational field, torsion is conventionally taken to be massless 
\cite{shap,hehlrev1976,hehl1995,hehlobu2001,ham1991,ham2000,ham2002}. In the MST context therefore, since the scalar field $\f$ acts as the source 
of the trace mode of torsion (by virtue of the relation (\ref{tr-eqn})), one expects $\f$ to 
be massless as well. We however make the allowance for $\f$ to have a mass $m_\f$, i.e.  
\be \la{scal-mass}
V (\f) \,=\, \fr 1 2 \, m_\f^2 \, \f^2 \,\,. 
\ee
In fact, a scalar field mass term may be induced effectively in the FRW space-time, at the 
expense of the torsion pseudo-trace mode $\cA^\m$, as demonstrated below.

\subsection{Effective scalar field mass via torsion pseudo-trace} \la{sec:A-mass}

The above constraint $\, \cA^\m = \d^\m_0 \l (t)$ can in principle be implemented in the 
theory by using a Lagrange multiplier field. Now, in the literature one often finds that 
phenomenological studies are done with the approximation of a constant (or very weakly 
evolving) torsion background 
\cite{shap,kost,obu,sandsg,sark,ces,ivan,derel}. 
In the same spirit we may therefore make an additional assumption that the pseudoscalar 
$\l (t)$ is rolling so slowly that it may effectively be replaced with its average 
$\lmean$ over a very large cosmological time-scale (possibly stretching well past the 
redshift of last scattering $z_{_{LS}} \simeq 1089$, so that dust cosmologies would not 
get affected anyway). In other words, it is reasonable to invoke a modified constraint 
$\, \cA^\m = \d^\m_0 \lmean$, by augmenting the action (\ref{MST-ac}) [or (\ref{MST-ac1})] 
with a Lagrange multiplier term, so that the $\cA^\m$-dependent part of (\ref{MST-ac}) is 
\be \la{A-ac}
\cS_{\!\!_\cA} = \!\int \!\! d^4 x \sq{-g} \, \fr{\b \f^2} 2 \le[\fr 1 {24} \cA_\m \cA^\m 
+ \le(\cA^\m - \d^\m _0 \lmean \ri) \fL_\m \ri] \,, \qquad \fL_\m := \, 
\mbox{Lagrange multiplier} \,.
\ee
It then follows that in the FRW space-time $\fL_\m = \rfraa{- \cA_\m}{24} = 
\le(\rfraa{\lmean}{24}, \, 0, \, 0, \, 0\ri) \,$, whence
\be \la{A-ac-on}
\cS_{\!\!_\cA} \,=\, - \int dt \int d^3 x \, a^3(t) \le[\fr 1 2 \, \mA^2 \, \f^2(t) \ri]\,,
\qquad \mbox{with} \qquad \mA \,=\, \lmean \, \sq{\fr{\b}{24}} \,\,.
\ee
This corresponds to the part of the scalar field action, due to a mass-like potential, which 
however has conformal properties different from that of the usual scalar field mass term 
(\ref{scal-mass}). In fact, note that Eq. (\ref{A-ac-on}) holds only for the FRW space-time 
metric in comoving coordinates (with constant lapse function). In other space-times (or 
coordinate systems), the Lagrangian would carry a factor proportional to the metric component 
$g_{_{00}}$, and would therefore get affected in a different way than a scalar field mass term
under conformal transformations\footnote{This is reminiscent of the rather simple metric-torsion
cosmological setup in 
\cite{shap,buchodshap1985,buchshap}, 
involving only the completely antisymmetric (or the pseudo-trace) mode $\cA^\m$ of torsion. 
For a contact interaction of $\cA^\m$ with a conformally constant spinor current $J^\m$, 
assumed therein, the ensuing constraint equation implicates the norm of $J^\m$ as a 
cosmological constant, albeit with different conformal properties.}.

As possible alternative ways of inducing, via the $\cA^\m$ mode, a mass term for the scalar 
field $\f$ with its usual conformal properties preserved, we may resort to one of the following:

\vskip 0.1in
\noindent
{\bf Proposition 1}:
Refer for instance to some variants of the vector-tensor gravity theories, viz. the 
Einstein-{\ae}ther (EA) theories 
\cite{jacob2001,jacob2004,jacob2006,carr2009-1,carr2009-2,gao,hag-etal,jacob2015,won}, 
in which a {\it fixed} expectation value for a time-like vector field (dubbed ``{\ae}ther") is 
enforced via a norm-fixing constraint, using a Lagrange multiplier\footnote{In fact, Lagrange 
multipliers have been commonly used to enforce suitable constraints in quite a few cosmological 
studies or in the formulation of gravitational theories, for e.g. the `dusty dark energy' models 
\cite{limsawvik-dde,cidlab-dde,myrzvik-dde}, 
some higher derivative gravity theories 
\cite{capo-etal-lm,klu-lm,klunojod-lm,noj-lm,capomakod-lm}, 
and the mimetic gravity scenarios 
\cite{chamuk-mim,chamukvik-mim,nojod-mim,myrzsebvag-mim,myrz-etal-mim,ijj-mim,noj-mim-CQG,
noj-mim-PRD,sebvagmyrz-mim}.}. 
The objective is to see the consequences of breaking the local Lorentz invariance, 
albeit with the general covariance preserved. In fact, a preferred rest frame at each 
point in the space-time can be apprehended in view of the untested nature of the local 
flatness condition in certain domains. The boost parameter for example, unlike the 
rotation group, remains unbounded from above, thus making it physically impossible to 
perform a uniform check. Actually, the problem with the exact Lorentz invariance is 
that it leads to divergences in quantum field theories relating to states with very 
high energies and momenta. Lorentz violation is hence desirable (for the ultraviolet 
completion of a gravitational theory), provided the preferred frames are determined 
by a suitably constrained {\ae}ther field, in a way that the general covariance holds 
\cite{jacob2001,jacob2004,jacob2006,hag-etal,jacob2015,won}. 
The space-time metric itself may serve the purpose, however that would lead to an 
indefinitely non-local nature of the field equations resulting in a highly unstable 
theory. So it is convenient to consider the {\ae}ther to be an independent local field, 
usually a time-like vector $\fv^\m$, satisfying $\, \fv_\m \fv^\m = -1$ 
\cite{jacob2001,jacob2004,jacob2006,jacob2015}). 
The integral curves of $\fv^\m$ supposedly represent the flow of an ``{\ae}ther fluid", 
and $\fv^\m$ itself measures the four-velocity tangent to the flow \cite{jacob2015}. 
In principle, $\fv^\m$ need not necessarily be dynamical. It could be a background field, 
non-minimally coupled to a scalar field in the theory, which would nevertheless be 
dynamically over-constrained then, and hence inconsistent \cite{jacob2015}.    

Considering torsion or(and) non-metricity to play an active or supportive role in the EA 
theories has been an alluring proposition, right from the pioneering works of Gasperini 
\cite{gasp1987,gasp1998} 
to the rather recent work of for e.g. Heinicke, Baekler and Hehl \cite{heinhehl}. By the
same instinct, we may ponder over the possibility that the torsion pseudo-trace mode 
$\cA^\m$ may effectively lead to the realization of the {\ae}ther $\fv^\m$, and hence of the 
preferred frame at every space-time location. The fact that $\cA^\m$ is inherently time-like 
in a cosmological setup, fits well with such aspiration. Also, inducing the {\ae}ther via a 
pseudovector (or the gradient of a pseudoscalar, for e.g. an {\it axion}) is nothing new in 
the literature 
\cite{balalem,alpbala,bala}. 
We may therefore consider making an identification (via a Lagrange multiplier constraint) 
of $\fv^\a$ with the unit vector $\rfraa{\cA^\m}{\le|\cA\ri|}$ (where $\le|\cA\ri| := 
\sq{- g^{\m\n} \cA_\m \cA_\n}$), whence it is automatically ensured that $\, \fv_\m \fv^\m 
= -1$. However, the consistency requirement of $\fv^\m$ to be dynamical then necessitates 
$\cA^\m$ to be a {\it propagating} mode of torsion. Propagating torsion theories have been 
studied extensively in the literature 
\cite{sabgasp,shap,saa,hojrosryan-prop,carrfld-prop,dobmar-prop,belshap-prop,popl-prop,
blagcvet-prop}
and have had several interesting consequences. Nevertheless, in the context of our MST 
formalism developed so far, the rather enticing scenario is that in which the apriori 
non-propagating torsion mode $\cA^\m$ acts as the source of a `dynamical' {\ae}ther field 
$\fv^\m$. This could be realized by simply taking into account certain interaction term(s) 
of $\cA^\m$ and $\fv^\m$ in the effective MST action, alongwith a Lagrange multiplier term 
that imposes the constraint $\, \fv_\m \fv^\m = -1$. The ($\cA^\m, \fv^\m$)-dependent part 
of the MST action would then read:
\be \la{Av-ac}
\cS_{\!\!_{\cA, \fv}} = \!\int \! d^4 x \sq{-g} \, \fr{\b \f^2} 2 \le[\fr 1 {24} 
\cA_\m \cA^\m \,+\, \cK \le(\fv^\m, \nab_\m \fv_\n, \, g_{\m\n}\ri) +\, \cV 
\le(\cA^\m, \fv^\m\ri) + \le(\fv_\m \fv^\m + 1\ri) \fL\ri] \,, 
\ee
where $\fL$ is a scalar Lagrange multiplier field, $\cK \le(\fv^\m, \nab_\m \fv_\n, 
\, g_{\m\n}\ri)$ denotes the kinetic term for the {\ae}ther $\fv^\m$, having the 
contributions of the expansion, shear, vorticity and acceleration of the flow 
\cite{jacob2015}, and $\cV \le(\cA^\m, \fv^\m \ri)$ is the interaction potential. 
Note that we have kept for brevity the factor $\b \f^2$ as representing an overall 
coupling of the scalar field $\f$ with both $\cA^\m$ and $\fv^\m$. Therefore, when 
$\f$ is trivial (i.e. {\it frozen} in to a particular vacuum expectation value), 
the torsion trace $\cT_\m$ is trivial as well, and the MST action effectively 
reduces to that of the EA theory in presence of a completely antisymmetric torsion 
(having only the mode $\cA^\m$). Note also that the {\ae}ther kinetic term would 
in general depend on the $U_4$ covariant derivative $\nt_\m \fv_\n$. Decomposing 
this in to the Riemannian covariant derivative $\nab_\m \fv_\n$ and the corrections 
due to torsion, one would get the term $\cK \le(\fv^\m, \nab_\m \fv_\n, \, g_{\m\n}
\ri)$ plus a plethora of interaction terms of the individual torsion modes with the 
{\ae}ther $\fv^\m$ (and possibly with $\nab_\m \fv_\n$ as well).

Let us for instance consider 
\be \la{EA-int}
\cV \le(\cA^\m, \fv^\m \ri) =\, \a \, \fv_\m \, \cA^\m \,+\, \s \le(\fv_\m \, 
\cA^\m\ri)^2 \,,
\ee
where $\a$ and $\s$ are coupling constants, $\a$ having the dimension of $\le|\cA\ri|$
whereas $\s$ is dimensionless (since $\fv^\m$ is a dimensionless (unit) vector). The 
variation of the MST action (or the part (\ref{Av-ac}) thereof) with respect to the 
auxiliary field $\cA^\m$ gives 
\be \la{A-eqn1}
\cA_\m \,+\, 12 \le(\a \,+\, 2 \s \, \fv_\n \cA^\n\ri) \fv_\m = 0 \,\,. 
\ee
whence with the constraint $\fv_\m \fv^\m = - 1$, and provided $\s \neq \rfraa 1 {24}$ 
and $\a \neq 0$, we have 
\be \la{A-eqn2}
\cA_\m \, \fv^\m \,=\, - \, \fr{12 \, \a}{\le(24 \, \s \,-\, 1\ri)} \,\,, \qquad 
\mbox{so that} \qquad \cA^\m \,=\, \fr{12 \, \a \, \fv^\m}{24 \, \s \,-\, 1} \,\,. 
\ee
Now, the host of terms constituting the {\ae}ther kinetic term $\cK \le(\fv^\m, \nab_\m 
\fv_\n, \, g_{\m\n}\ri)$ supposedly have little significance in the infrared limit. 
Moreover, going with our prior assumption that torsion evolves very weakly (almost as a 
background field) over the cosmological time-scales, it is reasonable to neglect the 
effect of the term $\cK \le(\fv^\m, \nab_\m \fv_\n, \, g_{\m\n}\ri)$ in the action. All 
that counts then is the norm-fixing constraint for the {\ae}ther field, which preserves 
the general covariance of the theory. With the relationship (\ref{A-eqn2}), the ($\cA^\m, 
\fv^\m$)-dependent part (\ref{Av-ac}) becomes
\be \la{A-ac-on1}
\cS_{\!\!_{\cA,\fv}} \,=\, - \int d^4 x \sq{-g} \le[\fr 1 2 \, \mA^2 \, \f^2(t,\vx) \ri]\,,
\qquad \mbox{with} \qquad \mA \,=\, \fr{\a \, \sq{6 \b}}{\sq{24 \, \s \,-\, 1}} \,\,.
\ee
This is similar to Eq. (\ref{A-ac-on}), but without any particular reference to the FRW 
metric in co-moving coordinates. So the scalar field $\f$ effectively picks up a mass
$\mA$, which now has its `generic' interpretation, i.e. preserves its usual properties 
under a conformal transformation. Note however that the term $\fv_\m \cA^\m$ in the above 
exemplary form (\ref{EA-int}) of the interaction potential $\cV \le(\cA^\m, \fv^\m \ri)$ 
breaks the spatial parity symmetry explicitly\footnote{Parity violation has always been 
an intriguing aspect of gravitational theories with torsion 
\cite{bmssg,bmssgss,bmssssgss,baekhehlnest,hojmuksay-pv,holst-pv,frei-pv,merc-pv,kaul-pv,
lucper-pv,baekhehl-pv,shapteix-pv,fab-pv}. 
However, it is not a necessity in the cosmological studies in general.}. Also, one has
to exclude $\s < \rfraa 1 {24}$, since $\mA^2 < 0$ then. On the other hand, for $\a = 0$ 
(whence parity is preserved) the condition $\fv_\m \fv^\m = -1$ follows automatically 
from Eq. (\ref{A-eqn1}) if we set $\s = \rfraa 1 {24}$. That is, there is no need 
of the ad-hoc term $\le(\fv_\m \fv^\m + 1\ri) \fL \,$ in the effective action. Anyway, 
it is obvious that Eq. (\ref{A-eqn2}) would not hold then, and the effective mass $\mA = 0$
identically. More general forms of the interaction potential can be suggested from the 
point of view of having $\fv_\m \fv^\m = -1$ naturally (i.e. not via any $\fL$-term), and 
plausibly with the parity symmetry kept intact. However, such an exercise also may not lead 
to a mass term for the field $\f$.

Observe further that the entire construction above (for this proposition) may conform to
the general perception that the pseudo-vector part of torsion has its significance only in 
presence of spinning matter fields. In accord with the fermionic {\ae}ther theories, or the
fermion-{\ae}ther interactions studied in the literature
\cite{carr-aeth,gom-aeth,puj-aeth,alex-aeth}, 
one may treat the above expression (\ref{EA-int}) for $\cV \le(\cA^\m, \fv^\m\ri)$ as that 
of a spin-torsion coupling, once the {\ae}ther $\fv^\m$ is identified as a fermion (spinor) 
current. Of course, the back-reaction of the fermion dynamics on the cosmologies is presumed
to be feeble enough to neglect the effect of the {\ae}ther kinetic term $\cK \le(\fv^\m, 
\nab_\m \fv_\n, \, g_{\m\n}\ri)$.

\vskip 0.1in
\noindent
{\bf Proposition 2}:
Consider incorporating higher order torsion term(s) in the MST action (\ref{MST-ac}). 
For simplicity, if we take only the square of $\cA_\m \cA^\m$, then the $\cA^\m$-dependent 
part of (\ref{MST-ac}) is  
\be \la{A-ac-quad}
\cS_{\!\!_\cA} \,=\, \int d^4 x \sq{-g} \, \fr{\b \f^2} 2 \le[\fr 1 {24} \, \cA_\m \cA^\m 
\,+\, \c \le(\cA_\m \cA^\m\ri)^2 \ri] \,,
\ee
where $\c$ is a coupling constant having the dimension of $\f^{-2}$. Note that we have
once again kept $\b \f^2$ as an overall coupling factor for all torsion and curvature
terms. While $\b$ is dimensionless, $\f^{-1}$ most importantly determines the length
scale of the theory (viz. it leads to the running gravitational coupling factor, see
the discussion in $\S$\ref{sec:scaltens} below). Variation of $\cS_{\!\!_\cA}$, Eq. 
(\ref{A-ac-quad}), with respect to the auxiliary field $\cA^\m$ leads to the only 
non-trivial constraint $\, \cA_\m \cA^\m = \rfraa {- 1} {48 \c}$. Consequently the 
action (\ref{A-ac-quad}) becomes 
\be \la{A-ac-on2}
\cS_{\!\!_\cA} \,=\, - \int d^4 x \sq{-g} \le[\fr 1 2 \, \mA^2 \, \f^2(t,\vx) \ri]\,,
\qquad \mbox{with} \qquad \mA \,=\, \fr 1 {48} \sq{\fr{\b}{\c}} \,\,.
\ee
This is again of the form (\ref{A-ac-on}), but without any specific allusion to the 
FRW metric structure in co-moving coordinates. 

More higher order torsion terms may be incorporated in the MST action, however the effects 
of those would mostly be insignificant in the context of late-time cosmologies (or in the 
low energy limit of the theory). For e.g. consider augmenting the action (\ref{MST-ac}) with
\be \la{A-ac-quad1}
\cS_{\!_{T^2}} \,=\, \int d^4 x \sq{-g} \, \fr{\b \f^2} 2 \le[\a \le(\cT_\m \cT^\m\ri)^2 
\,+\, \c \le(\cA_\m \cA^\m\ri)^2 \ri] \,\,,
\ee
where $\a$ is a coupling constant having the same dimension of $\c$, i.e. of the 
fundamental squared-length \footnote{Note however that we restrain ourselves from 
taking account of the mixing of the (hitherto independent) $\cT_\m$ and $\cA^\m$ 
modes, i.e. a term such as $\le(\cT_\m \cA^\m\ri)^2$. Neither we are looking for 
an explicit parity violation via say, the Holst term $\e^{\m\n\a\b} \Rt_{\m\n\a\b}$, 
which yields the terms like $\nab_\m \cA^\m$ and $\cT_\m \cA^\m$ (not higher order 
though) \cite{shapteix-pv}.}. While the variation of the full action with respect 
to $\cA^\m$ would lead to the part (\ref{A-ac-on2}) above, varying with $\cT_\m$ 
would give
\be \la{tr-eqn1}
\le(1 \,-\, 3 \, \a \, \cT_\n \cT^\n\ri) \cT_\m \,=\, 3 \, \fr{\pa_\m \f} \f \,\,,
\ee 
whence the kinetic part of the action would be
\be \la{MST-ac-kin}
\cS_{_{\mbox{\scriptsize kin}}} \!= \!\int \! d^4 x \sq{-g} \le[-\, \fr{\le(1 - 6\b\ri)} 2 \, 
\pa_\m \f \, \pa^\m \f \,+\, \fr{\a \b} 2 \le(\fr 9 \f\ri)^2 \!\! \le(\pa_\m \f \, 
\pa^\m \f\ri)^2 +\, \cO (\a^2) \ri] \,.
\ee
So the leading correction due to the $\le(\cT_\m \cT^\m\ri)^2$ term is, as expected, of
the order $\a$, whereas the result of the $\le(\cA_\m \cA^\m\ri)^2$ term, viz. the 
effective mass squared is of the order of $\c^{-1}$. The fact that both $\a$ and $\c$
have the same dimensional suppression, and presumably of the same order of magnitude,
therefore implies the insignificance of the term $\le(\cT_\m \cT^\m\ri)^2$ (and similar 
others) in the cosmological scale. So it would suffice one to take account of only the 
$\le(\cA_\m \cA^\m\ri)^2$ term to see the higher order torsion effects, such as the
effective mass $\mA$ for the scalar field $\f$.   
 
Overall thus we have, following either of these propositions, an effective potential
\be \la{eff-pot}
\Veff (\f) \,=\, V (\f) \,+\, \fr{\b \f^2}{48} \, \cA_\m \cA^\m \,+\, \mbox{suitable
augmentations} \,=\, \fr 1 2 \, m^2 \, \f^2 \,\,, 
\ee
for the scalar field, implicating the latter's effective mass $\, m = \sq{m_\f^2 + \mA^2} \,$, 
in general. As it seems, the proposition 2 is much more robust compared to the proposition 1,
although some tuning of the numerical value of the coupling constant $\c$ is required, so as 
to avoid unnaturalness in the theory. Moreover, much alike the proposition 1, the torsion
pseudo-trace $\cA^\m$ may be perceived as a direct manifestation of matter fields with spin
for the proposition 2 as well. For instance, $\cA^\m$ can be identified (via a contact 
interaction) with a spinor current $J^\m$
\cite{shap,buchodshap1985,buchshap}, 
or can have its source in for e.g. the massless Kalb-Ramond antisymmetric tensor field 
$B_{\m\n}$ in string theory
\cite{pmssg,saa}.   
However, note that the form of the action (\ref{A-ac-quad}) considered above (for the 
proposition 2) does not make this a compulsion. In other words, Eq. (\ref{A-ac-quad})
implies that $\cA^\m$, although non-propagating, can play a significant role in governing
the cosmological evolution, regardless of its coupling with spinning matter fields. 

In principle, more alternative ways to assign a mass term for the scalar field may be looked upon, 
since no claim is there that the torsion pseudo-trace $\cA^\m$ would always induce such a mass. 
The above propositions are solely for leading up to the scenarios in which both the trace and 
pseudo-trace modes of torsion can have active roles in viable cosmologies. Our interest in this 
work conform to such scenarios, and in particular the ones in which the scalar field mass is 
entirely due to the mode $\cA^\m$, i.e. $\, m = \mA$. In other words, we look for the optimization 
of the effect of torsion in a cosmological setup, rather than persisting with a superfluous 
parameter $m_\f$. Henceforth, we shall neither make any allusion to $m_\f$, nor attempt to assert 
any other form of the effective potential $\Veff$. Note also that with $\, m = \mA$, we have $\, 
\cA_\m \cA^\m = \rfraa{- 24 m^2}{(24 \s - 1) \b} \,$ for proposition 1 and $\, \cA_\m \cA^\m = 
\rfraa{- 48 m^2} \b \,$ for proposition 2. That is, for a given $m$, the quantity $\cA_\m \cA^\m$ 
differ for the two propositions by a numerical factor (dependent on the parameter $\s$ though). 
It would therefore suffice us to allude only to, say, the proposition 2 in our subsequent 
analysis. The results corresponding to the proposition 1 would follow simply by taking account 
the numerical factor.

\subsection{Effective scalar-tensor theoretical framework} \la{sec:scaltens}

Considering the mass term for the scalar field $\f$, and the elimination of the mode $\cA^\m$ 
via a constraint (as discussed above), we have the effective action
\be \la{MST-ac2}
\cS \,=\, \int d^4 x \, \sq{-g} \le[\fr{\b \f^2} 2 \, \cR \,-\, \fr{\le(1 - 6\b\ri)} 2 
\, g^{\m\n} \, \pa_\m \f \, \pa_\n \f \,-\,\fr 1 2 \, m^2 \f^2 \,+\, \cL^{(m)} \ri] \,.
\ee
This is of course the action of a scalar-tensor theory in the original {\em Jordan} 
frame \cite{frni,fujii}, which is characterized by a running gravitational coupling parameter
\be \la{G-eff}
\Geff (t) \,=\, G \le[\fr{\fp}{\f (t)}\ri]^2 \,, \qquad \le[G := \, \mbox{Newton's constant}\ri] \,,
\ee
defined so that at the present epoch $\, t = \tp$, we have $\, \Geff (\tp) = G$ under the 
stipulation
\be \la{phi0}
\f (\tp) \, \equiv \, \fp \,=\, \fr 1 {\k \sq{\b}} \,\,, \qquad \mbox{with} \qquad 
\k = \sq{8 \pi G} \,\,.
\ee
If one redefines the scalar field as
\be \la{J-phi}
\F (t) \,:=\, \Fp \le[\fr{\f (t)} \fp\ri]^2 \,\,, \qquad \mbox{with} \qquad
\Fp \,\equiv\, \F (\tp) \,=\, \b \, \fp^2 \,=\, \k^{-2} \,\,, 
\ee
then Eq. (\ref{MST-ac2}) reduces to an equivalent {\em Brans-Dicke} (BD) action \cite{BD,frni,fujii}:
\be \la{J-ac}
\cS \,=\, \int d^4 x \, \sq{-g} \le[\fr{\F \, \cR} 2 \,-\, \fr \fw {2 \F} \, g^{\m\n} \, 
\pa_\m \F \, \pa_\n \F \,-\, \cV(\F) \,+\, \cL^{(m)} \ri] \,,
\ee
albeit with a potential for $\F$:
\be \la{J-pot}
\cV (\F) \,=\, \fr {\L \, \F} \Fp \,\,, \qquad \mbox{where} \qquad
\L \,\equiv\, \cV (\F) \big|_{t=\tp} =\, \fr 1 2 m^2 \fp^2 \,=\, \fr{m^2}{2 \k^2 \b} \,\,.
\ee
The effective BD parameter $\fw$ is related to the non-minimal parameter $\b$ as
\be \la{BD-param}
\fw \,=\, \fr 1 {4 \b} \,-\, \fr 3 2 \,\,.  
\ee
The corresponding equations of motion are given by
\bea 
\cR_{\m\n} &=& \fr 1 \F \le[T_{\m\n}^{(m)} \,- \le(\fw + 1\ri) g_{\m\n} \, \Box \F \,+\, 
\nab_\m \pa_\n \F \,+\, \fr \fw \F \, \pa_\m \F \, \pa_\n \F\ri]\,, \la{J-eom1} \\
\Box \F &=& \fr 1 {2 \fw + 3} \le[T^{(m)} \,-\, 2 \, \cV (\F)\ri] \,, \la{J-eom2}
\eea
where $\, T^{(m)} = g^{\m\n} T_{\m\n}^{(m)}$ is the trace of the matter stress-energy tensor
\be \la{J-emtens}
T_{\m\n}^{(m)} \,=\, - \, \fr 2 {\sq{- g}} \, \fr \d {\d g^{\m\n}} \le[\sq{- g} \,
\cL^{(m)}\ri] \,,
\ee
which is conserved:
\be \la{J-consv}
\nab_\a \le(g^{\a\n} \, T_{\m\n}^{(m)}\ri) =\, 0 \,\,.
\ee
As the cosmological fluid is considered here to be the pressureless dust, we have
\be \la{J-dust}
T_{\m\n}^{(m)} \,=\, \r^{(m)} \, u_\m \, u_\n \,\,, \qquad \mbox{and} \qquad
T^{(m)} \,=\, - \r^{(m)} \,\,,
\ee
where $\, \r^{(m)}$ is the dust energy density, and $\, u^\m$ is the four velocity
vector $\le(g_{\m\n} u^\m u^\n = - 1\ri)$.
 
Refer back again to Eq. (\ref{MST-ac2}), which under a conformal transformation
\be \la{conf}
g_{\m\n} \,\lra\, \hg_{\m\n} \,= \le(\fr \f \fp\ri)^2 g_{\m\n} \,\,,
\ee 
reduces to the scalar-tensor action in the so-called {\em Einstein} frame:
\be \la{E-ac}
\hS = \int d^4 x \sq{-\hg} \le[\fr{\hR}{2 \k^2} - \fr 1 2 \le(\fr \fp \f\ri)^2
\hg^{\m\n} \pa_\m \f \, \pa_\n \f \,- \L \le(\fr \fp \f\ri)^2 + 
\le(\fr \fp \f\ri)^4 \cL^{(m)} \!\le(\hg,\f\ri) \ri] \,,
\ee
where $\, \hg \equiv \text{det} \le(\hg_{\m\n}\ri)$, and $\, \hR = \hg^{\m\n} \hR_{\m\n}$ 
with $\, \hR_{\m\n}$ the Ricci tensor constructed using $\, \hg_{\m\n}$. Note that 
$\, \cL^{(m)}$ is in general dependent on both $\hg_{\m\n}$ and $\f$. Redefining the scalar 
field as
\be \la{E-phi}
\vph (t) \,:=\, \fp \, \ln \le[\fr{\f (t)} \fp\ri] =\, \fr 1 {\k \sq{\b}} \, 
\ln \le[\k \sq{\b} \, \f (t)\ri] \,, \qquad \mbox{such that} \qquad \vph (\tp) = 0 \,\,,
\ee
the Einstein frame action (\ref{E-ac}) can be expressed in a more convenient form:
\be \la{E-ac1}
\hS \,=\, \int d^4 x \, \sq{-\hg} \le[\fr{\hR}{2 \k^2} \,-\, \fr 1 2 \, \hg^{\m\n} \, 
\pa_\m \vph \, \pa_\n \vph \,-\, \cU (\vph) \,+\, \hLm \!\le(\hg,\vph\ri) \ri] \,,
\ee
with the corresponding matter Lagrangian density 
\be \la{E-mat}
\hLm \!\le(\hg,\vph\ri) \,=\, e^{- 4 \k \sq{\b} \vph} \, \cL^{(m)} \!\le(\hg,\f (\vph)\ri) \,,
\ee
and the scalar field potential given as
\be \la{E-pot}
\cU (\vph) \,=\, \L \, e^{- 2 \k \sq{\b} \vph} \,\,, \qquad \mbox{with} \qquad 
\L \,\equiv\, \cU (\vph) \big|_{t=\tp} =\, \fr{m^2}{2 \k^2 \b} \,=\, \fr 1 2 m^2 \fp^2 \,\,.
\ee
One derives the following equations of motion in the Einstein frame:
\bea 
&& \hR_{\m\n} \,=\, \k^2 \le[\hT_{\m\n}^{(m)} \,-\, \fr 1 2 \, \hg_{\m\n} \, \hT^{(m)} 
\,+\, \pa_\m \vph \, \pa_\n \vph \,+\, g_{\m\n} \, \cU (\vph)\ri]\,, \la{E-eom} \\
&& \hnab_\a \le(\hg^{\a\n} \, \hT_{\m\n}^{(m)}\ri) =\, - \, \k \sq{\b} ~ \hT^{(m)} \, 
\pa_\m \vph \,\,. \la{E-consv}
\eea
The corresponding matter stress-energy tensor, viz. 
\be \la{E-emtens}
\hT_{\m\n}^{(m)} \,=\, - \, \fr 2 {\sq{- \hg}} \, \fr \d {\d \hg^{\m\n}} \le[\sq{- \hg}
\, \hLm\ri] \,, \qquad \mbox{with trace} ~~ \hT^{(m)} = \hg^{\m\n} \hT_{\m\n}^{(m)} \,,
\ee
is not conserved though, as is evident from Eq. (\ref{E-consv}). One can express 
\be \la{E-dust}
\hT_{\m\n}^{(m)} \,=\, \hr^{(m)} \, \hu_\m \, \hu_\n \,\,, \qquad \mbox{with} ~~~
\hg_{\m\n} \hu^\m \hu^\n = - 1 \,, \qquad \mbox{so that} ~~
\hT^{(m)} \,=\, - \hr^{(m)} \,\,,
\ee
where $\, \hr^{(m)}$ and $\, \hu^\m$ are respectively the Einstein frame dust energy 
density and four velocity. 

It is easy to verify that $\hT_{\m\n}^{(m)}, \, \hr^{(m)}$ and $\hu^\m$ have the following 
relationships with the corresponding quantities in the Jordan frame \cite{frni,fujii}:
\be \la{relations}
T_{\m\n}^{(m)} \,=\, e^{2 \k \sq{\b} \vph} \, \hT_{\m\n}^{(m)} \,\,, \qquad 
\r^{(m)} \,=\, e^{4 \k \sq{\b} \vph} \, \hr^{(m)} \,\,, \qquad \mbox{and} \qquad
u^\m \,=\, e^{\k \sq{\b} \vph} \, \hu^\m \,\,.
\ee
The Einstein frame is of course more convenient to work on, than the Jordan frame in 
which the running gravitational coupling parameter causes some abstruseness in the 
corresponding Einstein's equations. Consequently, there are characteristic differences 
between the cosmologies in the two frames. However, amidst its bizarreness, the Jordan 
frame cosmology shows one regularity, viz. the conserved stress-energy tensor of the 
fluid matter. In the Einstein frame, the scalar field is made to couple with gravity in 
the usual minimal way, but at the expense of its interaction with the fluid. Moreover, 
the original non-minimal scalar-gravity coupling parameter $\b$ is still traced in the 
Einstein frame, since it appears in both the scalar field potential expression (\ref{E-pot})
and the matter conservation relation (\ref{E-consv}).

In what follows, we shall analyse in $\S$\ref{sec:E-MST} the solutions of the Einstein 
frame MST cosmological equations, in course of building up a viable DE model. A similar 
analysis in the Jordan frame would subsequently be carried out in $\S$\ref{sec:J-MST}, 
not just for completeness but also to restrain ourselves from making a preference whilst a 
longstanding debate on the physical relevance of the two frames prevails. Our focus would 
be on the estimation of the cosmological model parameters, and hence on obtaining specific 
bounds on the effective Brans-Dicke parameter and the torsion mode parameters (see the next 
subsection for their precise expressions).

\subsection{Effective torsion mode parameters} \la{sec:tor-param}

The effective scalar-tensor formulation of the MST theory somewhat obscures the torsion 
modes $\cT_\m$  and $\cA^\m$, which are the key MST constituents. A quantitative study of 
the individual effects of these modes requires their appropriate parametrization in the 
MST-cosmological setup. One may look to define the density parameters corresponding to 
$\cT_\m$ and $\cA^\m$, however note that these modes are interacting with the scalar 
field (and also with the cosmological fluid in the Einstein frame). A rather safe 
alternative is to consider (as the torsion parameters) the norms of the vectors $\cT_\m$ 
and $\cA^\m$. In general a difficulty is still there though, due to the lack of uniqueness 
in defining the lengths of vector fields in the conformally transformed frames. Nevertheless, 
for torsion we may choose to take the norms of $\cT_\m$ and $\cA^\m$ as defined in the 
Jordan frame. We then simply have in the Einstein frame, the expressions for these norms 
recast in terms of the corresponding metric tensor components. This is justified, since our 
starting point had been an explicit non-minimal coupling of the scalar field $\f$ with 
curvature and torsion, which is perceptible in the equivalent scalar-tensor formulation 
of the MST theory only in the Jordan frame. Moreover, in such a formulation the mode $T_\m$ 
has its norm contributing to the kinetic part of the scalar field Lagrangian density, by 
virtue of Eq. (\ref{tr-eqn}). On the other hand, the norm of $A^\m$ contributes to 
the scalar field potential (or the mass). 
\bed 
\item
{\bf Norm of torsion trace vector}:
Using Eq. (\ref{tr-eqn}), we have in the Jordan frame   
\be \la{J-T-norm}
\le|\cT\ri| :=\, \sq{- \, g^{\m\n} \, \cT_\m \, \cT_\n} \,=\, 
\fr 3 {\f} \, \sq{- \, g^{\m\n} \, \pa_\m \f \, \pa_\n \f} \,\,,
\ee
which, in terms of the Brans-Dicke scalar field $\F$ ~[{\it cf.} Eq. (\ref{J-phi})], is
expressed as
\be \la{J-T-norm1}
\le|\cT\ri| =\, \fr 3 {2 \F} \, \sq{- \, g^{\m\n} \, \pa_\m \F \, \pa_\n \F} \,\,.
\ee
In the Einstein frame, the norm of $\cT_\m$ is expressed in accord with Eq. (\ref{conf}): 
\be \la{E-T-norm}
\le|\cT\ri| :=\, \fr{\f}{\fp} \, \sq{- \, \hg^{\m\n} \, \cT_\m \, \cT_\n} \,=\, 
\fr 3 {\fp} \, \sq{- \, \hg^{\m\n} \, \pa_\m \f \, \pa_\n \f} \,\,.
\ee
Since $\fp = (\k \sq{\b})^{-1}$, we have in terms of the Einstein frame scalar field 
$\vph$ [Eq. (\ref{E-phi})]:
\be \la{E-T-norm1}
\le|\cT\ri| = \, 3 \, \k \sq{\b} \, e^{\k \sq{\b} \vph} \, \sq{- \, \hg^{\m\n} 
\, \pa_\m \vph \, \pa_\n \vph} \,\,.
\ee
\item
{\bf Norm of torsion pseudo-trace vector}: Both the propositions above imply that the norm 
of $A^\m$ is constant in the FRW space-time. For our analysis, following the exemplary 
scenario of the proposition 2, say, we have in both Jordan and Einstein frames:
\be \la{A-norm}
\le|\cA\ri| :=\, \sq{- \, g^{\m\n} \, \cA_\m \, \cA_\n} \,=\, \fr{\f}{\fp} \, 
\sq{- \, \hg^{\m\n} \, \cA_\m \, \cA_\n} \,=\, 4 \k \, \sq{6 \L} \,\,, 
\ee
since $\, \L = \rfraa{m^2 \fp^2 \,} 2 = \rfraa{m^2}{2 \k^2 \b} \,$, and $\, g^{\m\n} \cA_\m 
\cA_\n = \rfraa{- 48 m^2} \b \,$ by our prior assumption that the scalar field mass $m$ is 
solely due to the mode $\cA^\m$, i.e. $m = \mA$. 
\eed
It is worth mentioning here that the DE interpretation of the scalar degree of freedom, within 
the FRW framework, implies the mass of the scalar field playing the key role in the evolution
of the universe, compared to the kinetic part of the scalar field Lagrangian. Therefore, while
we take $m$ to be due only to $\cA^\m$, it becomes evident that $\le|\cA\ri|$ would be much 
more dominant than $\le|\cT\ri|$ in an ensuing DE model. Else, one would require some other 
parameter in the effective scalar field potential, such as the inherent mass $m_\f$, to prevail 
over both $\le|\cT\ri|$ and $\le|\cA\ri|$, if they are to be of comparable order of magnitude. 
Now the scalar field DE models with arbitrary mass parameter $m_\f$ are nothing new in the 
literature. What is intriguing here in the context of MST cosmology is the plausible aspect of 
torsion inducing the scalar field mass, and hence becoming a dominant energy contributor of the 
universe. This is the very reason that has prompted the above propositions and subsequently the 
assumption $m = \mA$.

\section{MST Cosmology in the Einstein frame}  \la{sec:E-MST}

The general setup discussed above enables us to look for exact solution(s) of the cosmological 
equations in presence of the MST coupling. In this section, we choose to work in the Einstein 
frame, supposing it to be suitable for interpreting the results of physical observations. Our 
objective is to construct a MST-DE model consistent with the observations, or conversely to 
determine the bounds on the torsion parameters from the supposition that such a DE model emerges 
out in the MST cosmological scenario, under the reappraisal of the following:
\ben[(i)]
\item The time-evolution of the DE, resulting from the torsion trace mode\footnote{As is evident
from the constraint (\ref{tr-eqn}), which is due to the interlinking of the kinetic part of the 
scalar field Lagrangian and the torsion trace $\cT_\m$.}, would expectedly be weak enough, given 
the miniscule evidence of the dynamical effects of torsion in various other searches 
\cite{shap,kost,flanros-tor,stef-tor,babfrol-tor,hehlobupuet-tor,camcorrad-tor,cast-etal-tor,
jaf-etal-tor,lucprok-tor}. 
\item The DE evolution would expectedly not differ much from $\L$CDM, in view of the fairly
concordant cosmological parameteric limits set for the latter from various observations.
\een
We therefore take the observed best-fit values of the $\L$CDM cosmological parameters as the 
basic measures for parametric estimations of the MST-DE model, which however has one unusual 
feature in the Einstein frame, viz. the interaction
of the DE and the cosmological matter. So we need to define the effective uncoupled 
DE and matter densities for the comparative study of the corresponding quantities 
in the $\L$CDM case. This is a mathematical exercise though, and for notational 
simplicity we also drop the hats over all the Einstein frame quantities and 
operators throughout this section.

\subsection{Cosmological equations and solution} \la{sec:E-sol}

The Friedmann and Raychaudhuri equations follow from Eqs. (\ref{E-eom}), on using Eqs. (\ref{E-dust}):
\bea 
&& H^2 (t) \,=\, \fr{\k^2} 3 \le[\rmt (t) \,+\, \fr{\dvph^2 (t)} 2 \,+\, \cU 
\le(\vph (t) \ri)\ri] \,,
\la{E-eq1} \\
&& \dot{H} (t) \,=\, - \, \fr{\k^2} 2 \le[\rmt (t) \,+\, \dvph^2 (t)\ri] \,,
\la{E-eq2}
\eea
where $\, H (t) := \rfraa{\dot{a} (t)}{\!\! a (t)}$ is the Einstein frame Hubble parameter  
(the overhead dot $\{\cdot\} \equiv \rfraa d {\!\! dt}$).

The conservation equation (\ref{E-consv}) integrates to give
\be \la{E-matdens}
\rmt (a) \,=\, \fr \rmp {a^3} \, e^{- \k \sq{\b} \, \vph (a)} \,\,,
\ee
where $\, \rmp = \rmt \rvert_{t=\tp} = \rmt \rvert_{a=1}$ is the present-day value of 
the matter density. 

Assuming now a simple power-law ansatz:
\be \la{E-ans}
e^{\k \sq{\b} \, \vph (a)} \,=\, a^\sbig \,\,, \qquad \mbox{with} \qquad
s = ~ \text{a constant} \,\,,
\ee  
a few rearrangement of terms in Eqs. (\ref{E-eq1}) and (\ref{E-eq2}), and algebraic 
steps, would lead to
\bea 
&& H^2 (a) \,=\, \fr{\k^2} 3 \le(1 - \fr{s^2}{6\b}\ri)^{-1} \le[\fr \rmp {a^{3 + \sbig}} \,+\, 
\fr{\L}{a^{2 \sbig}}\ri] \,, \la{E-eq1a} \\
&& \le[a^6 \, H^2 (a)\ri]' \,=\, \k^2 \le[\rmp \, a^{2 - \sbig} \,+\, 2 \, \L \, a^{5 - 2 \sbig}\ri] 
\,, \la{E-eq2a}
\eea
where the prime $\{'\}$ denotes $\rfraa d {\!\! da}$, and we have recalled Eq. (\ref{E-pot}) 
for the potential $\cU (\vph)$.

It is easy to verify that a non-trivial solution can be obtained only for 
\be \la{E-solcond}
s \,=\, 2 \, \b \,\,,
\ee
whence the Friedmann equation (\ref{E-eq1}) [or (\ref{E-eq1a})] becomes
\be \la{E-Hub}
H^2 (a) \,=\, \fr{\k^2}{3 - s} \le[\fr \rmp {a^{3 + \sbig}} \,+\, \fr{\L}{a^{2 \sbig}}\ri] \,.
\ee
This is however of a rather unusual form, because of the interaction of the scalar 
field with the dust. Also, the validity of Eq. (\ref{E-Hub}) requires $\, s < 3$,
i.e. $\, \b < \rfraa 3 2$, which puts a (somewhat less stringent) theoretical restriction 
$\, \fw > \rfraa {- 4} 3$, on the effective Brans-Dicke parameter $\fw$ [given by Eq. 
(\ref{BD-param})]. Rigorous bounds on $\fw$ can in principle be obtained while
matching the theoretical predictions with the observational results, as we shall 
see shortly. A comparison of such bounds with the local gravity constraints on 
$\fw$ would be an independent consistency check of the MST-cosmological DE model
we are studying here. 

Let us, for convenience, define an effective matter density (in the usual form) as
\be \la{E-m-dens}
\rmf (a) \,:=\, \rmp \, a^{-3} \,\,, 
\ee
and a surplus density (considered to be the effective DE density) as
\be \la{E-de-dens}
\rx (a) \,:= \le(1 - \rfraa s 3\ri)^{-1} \le[\fr \L {a^{2 \sbig}} \,+\, \fr \rmp {a^3} 
\le(a^{- \sbig} - 1 + \rfraa s 3\ri)\ri] \,, 
\ee
so that the critical density of the universe is given by
\be \la{E-crit}
\r (a) \,:=\, \fr{3 H^2 (a)}{\k^2} \,=\, \rmf (a) \,+\, \rx (a) \,\,.
\ee
Defining further an effective DE pressure $\px$ as
\be \la{E-effpres}
\px (a) \,:=\, - \le(1 - \rfraa s 3\ri)^{-1} \le[\le(1 - \rfraa{2s} 3\ri) \fr \L {a^{2 \sbig}} 
\,-\, \fr s 3  \fr \rmp {a^{3 + \sbig}}\ri] \,,
\ee
we have the standard conservation equation
\be \la{E-critconsv}
\r' (a) \,+\, \fr 3 a \le[\r (a) \,+\, \px (a) \ri] =\, 0 \,\,.
\ee

\subsection{Effective cosmological parameters} \la{sec:E-param}

Using Eqs. (\ref{E-Hub}) -- (\ref{E-effpres}) one obtains the effective DE equation of state (EoS) 
parameter $\wx$, as a function of the redshift $\, z = \le(a^{-1} - 1\ri)$:
\be \la{E-de-eos}
\wx (z) \,:=\, \fr{\px (z)}{\rx (z)} \,=\, - \, 1 \,+\, \rfraa{2s} 3 \,+\, \fr{\Omf (z)}
{1 \,-\, \Omf (z)} \le[\le(1 + z\ri)^\sbig -\, 1 \,+\, \rfraa{2s} 3\ri] \,.
\ee
Here, $\,\Omf (z)$ is the effective matter density parameter, defined by
\be \la{E-m-eff}
\Omf (z) \,:=\, \fr{\rmf (z)}{\r (z)} \,=\, \Omp \, \fr{\le(1 + z\ri)^3}{\fH^2 (z)} \,,
\ee
with $\, \Omp$ the value of $\, \Omf$ at the present epoch $\, t = \tp$ (i.e. $a = 1$, 
or $z = 0$), and 
\be \la{E-RHub} 
\fH (z) \,\equiv\, \fr{H (z)} \Hp \,= \le(1 - \rfraa s 3\ri)^{- \rfra 1 2} \le[\Omp 
\le(1 + z\ri)^{3 + \sbig} + \le(1 - \rfraa s 3 - \Omp\ri) 
\le(1 + z\ri)^{2 \sbig}\ri]^{\rfra 1 2} \,\,,
\ee
is the rationalized Hubble parameter, $~ \Hp \equiv H (z=0) \,$ being the Hubble constant.

Substituting Eq. (\ref{E-RHub}) back in  Eq. (\ref{E-m-eff}) and simplifying, we have
the following expression for the effective matter density parameter
\be \la{E-m-eff1}
\Omf (z) \,=\, \fr{\Omp \le(1 + z\ri)^{3 - 2 \sbig}}{1 \,+\, \Omp \le(1 - \rfraa s 3\ri)^{-1}
\le[\le(1 + z\ri)^{3 - \sbig} -\, 1\ri]} \,,
\ee
and the actual (interacting) matter density parameter is related to this as
\be \la{E-m-dp}
\Om (z) \,:=\, \fr{\rmt (z)}{\r (z)} \,= \le(1 + z\ri)^\sbig \Omf (z) \,,
\ee
so that at the present epoch ($z = 0$) we have $\Omf (0) = \Om (0) = \Omp$.

The EoS parameter for the system is worked out as 
\be \la{E-eos}
\sw (z) \,:=\, \fr{\px (z)}{\r (z)} \,= \le[1 \,-\, \Omf (z)\ri] \wx (z) 
\,=\, - \, 1 \,+\, \rfraa {2s} 3 \,+\, \Om (z) \,\,,
\ee
so that the condition for cosmic acceleration at the present epoch ($z=0$) 
\be \la{E-accl} 
\sw (0) < - \fr 1 3  \qquad \mbox{imples} \qquad s \,+\, \fr 3 2 \, \Omp \,<\, 1 \,\,,
\ee
or more specifically, $\, 0 < s < 1$, because $\Omp$ is positive definite and the 
parameter $s$ is presumably such\footnote{As $s = 2\b$, a negative value of $s$ would 
imply $\b < 0$, i.e. the kinetic term of the scalar coupled gravity action having a 
{\it wrong} sign, rendering the underlying quantum theory unbounded from below.}. So 
we have a constraint on $s$ tighter than the previous one ($s < 3$). As such, the bound 
on the BD parameter gets tighter, viz. $\, \fw > -1$, than that obtained previously 
($\fw > \rfraa {-4} 3$). Tighter still it could be, considering the observational 
prediction that the universe is dominated by both dark energy and (baryonic plus cold 
dark) matter at present. That is, $\, \Omp \sim 1$, meaning $s$ cannot have a large 
fractional value (close to $1$). If, let us say, we take $\, \Omp \simeq 0.3$ as the 
fiducial value of the matter density parameter at the present epoch, then Eq. 
(\ref{E-accl}) gives $\, s \lesssim \rfraa 1 2$, whence $\, \fw \gtrsim \rfraa {-1} 
2$. Moreover, by Eq. (\ref{E-de-eos}) the present-day value of the DE EoS parameter 
is constrained for $\, \Omp \simeq 0.3$ as
\be \la{E-deos-cons}
\wx (0) \,=\, - \, 1 \,+\, \fr{2 \, s}{3 \le(1 \,-\, \Omp\ri)} \,\lesssim\, - \,
\fr 1 2 \,\,.
\ee
This is of course nothing new --- the same bound applies to all barotropic DE models 
simply by virtue of the relation $\, \sw (z) = \le[1 \,-\, \Omf (z)\ri]\! \wx (z) \,$
and by the requirement $\sw (0) < \rfraa {-1} 3$. Therefore, like in any other DE model, 
we have here a fairly large scope of deviation from $\L$CDM (in which the DE EoS 
parameter is fixed at $-1$). We however restrict ourselves to the MST induced small 
parametric fluctuations over $\L$CDM, relying on the general consensus that the latter 
presently stands out as the observationally favoured model of DE. In other words, we 
presume the parameter $s$ to be quite small, which is also corroborated by a very large 
lower bound on the BD parameter ($\fw \gtrsim 40$ to even $\fw \gtrsim 40000$) obtained 
independently\footnote{Mostly for the massless Brans-Dicke theory however.} from the solar 
system tests and other probes \cite{acq-BD,avi-BD,chen-BD,als-BD}.

\subsection{Parametric estimation using the $\L$CDM observational limits} \la{sec:E-est}

As mentioned earlier, we in this paper are interested primarily on a order of magnitude 
estimate of $s$ (and hence on the bounds on the torsion parameters). Therefore, instead
of a full fledged likelihood analysis with the observational data, it would suffice us 
to compare the parameters of our model with those of a reference model. Our chosen
reference is the six parameter $\L$CDM model, which is generally taken as the base 
theoretical model for the observational probes. Assuming the cosmological parameters of 
our MST-DE model to be within the $68\%$ confidence limits of the corresponding ones for 
$\L$CDM, obtained from various datasets (or combinations thereof), we can then determine 
the respective upper bound on the parameter $s$. Of course, the legitimacy of such a bound 
would be implicated by the smallness of its numerical value, since the $\L$CDM equations 
are recovered in the limit $s \rarr 0$ of the MST-DE model. Refer for instance to the 
Friedmann equation (\ref{E-Hub}), whose limiting version (expressed in terms of the 
redshift $z$):
\be \la{E-Hub-lim}
\bH^2 (z) \,\equiv\, \lim_{s \rarr 0} H^2 (z) \,=\, \fr{\k^2} 3 \le[\rmp \le(1 + z\ri)^3 
+\, \L\ri] \,,
\ee
is nothing but the Friedmann equation of the $\L$CDM model, with $\L$ as the cosmological 
constant, and $\bH$ denoting the $\L$CDM Hubble parameter. In fact, from now on we shall 
denote every $\L$CDM cosmological parameter by an overbar $\{\bar{~}\}$, for e.g. the 
$\L$CDM matter density parameter as $\Omb$, the $\L$CDM EoS parameter as $\bw$, the 
$\L$CDM Hubble constant as $\bHp$, and so on. The relationship between $\bHp$ and the
Hubble constant $\Hp = H (z=0)$ of our model can be traced out from Eqs. (\ref{E-Hub})
and (\ref{E-Hub-lim}):
\be \la{E-Hubrel}
\bHp \,=\, \Hp \sq{1 - \rfraa s 3} \,\,.
\ee
Similarly, Eq. (\ref{E-m-eff1}) can be expressed as a relation between $\Omf (z)$ and 
$\Omb (z)$:
\be \la{E-m-dprel}
\Omf (z) \,=\, \fr{\le(3 - s\ri) \le(1 + z\ri)^{3 - 2 \sbig} \Omb (z)}
{\le[3 \le(1 + z\ri)^{3 - \sbig} -\, s\ri] \Omb (z) \,+ \le(3 - s\ri) 
\le(1 + z\ri)^3 \le[1 - \Omb (z)\ri]} \,,
\ee
with
\be \la{E-m-dplim}
\Omb (z) \,\equiv\, \lim_{s \rarr 0} \Omf (z) \,=\, \fr{\Omp \le(1 + z\ri)^3}
{1 \,+\, \Omp \le[\le(1 + z\ri)^3 -\, 1\ri]} \,.
\ee
Now, for the comparative error analysis, we shall denote the statistical error on a given 
quantity $Y$ here by $\d Y$, and that on the corresponding $\L$CDM quantity $\bY$ by 
$\d \bY$. To determine the maximum value $\smax$ of the parameter $s$, using such an 
analysis, we are required to pick a suitable cosmological parameter that depends 
explicitly on $s$, and use the best fit value and the $68\%$ confidence limit of the 
corresponding $\L$CDM parameter for a given dataset. Of course, the parameter $\Omp$ would 
not serve the purpose, since by definition $\, \Omp = \Omf (0) = \Omb (0)$ is independent 
of $s$. Let us therefore resort to one of the following: 
\bed
\item {\bf Procedure 1}: The error propagation equation corresponding to Eq. 
(\ref{E-Hubrel}) is given by
\be \la{E-H-err}
\le(\fr{\d \bHp}{\bHp}\ri)^2 \,= \le(\fr{\d \Hp}{\Hp}\ri)^2 \,+ 
\le(\fr{\Hp}{\bHp}\ri)^4 \le(\fr{\d s} 6\ri)^2 \,\,.
\ee
Let us stipulate $\, \Hp = \bHp + \d \Hp$, and the minimum and maximum fractional 
error
\be \la{E-H-stip}
\le|\fr{\d \Hp}{\Hp}\ri|_{\mbox{\scriptsize min}} =\, 0 \,\,, \qquad \mbox{and}
\qquad \le|\fr{\d \Hp}{\Hp}\ri|_{\mbox{\scriptsize max}} =\, \D_h \,\equiv\,
\fr{\le|\d \bHp\ri|}{\bHp} \,\,.
\ee
For a given dataset, taking $\, \bHp$ as the best fit $\L$CDM Hubble constant, and 
$\le|\d \bHp\ri|$ as its $68\%$ error limit, we have a {\it fixed} value assigned 
to $\D_h$. It then follows from Eq. (\ref{E-H-err}): 
\be \la{E-H-s-err}
\d s \,=\, 6 \le(1 \,- \le|\fr{\d \Hp}{\Hp}\ri|\ri)^2 \le(\D_h^2 - 
\le|\fr{\d \Hp}{\Hp}\ri|^2\ri)^{\!\!\sfrac 1 2} \,.
\ee
Under the above stipulation, the second factor has the least value equal to zero, 
which ensures $\d s \geq 0$. Moreover, the parameter $s$ being positive, its upper 
bound $\smax$ is given by the maximum value of $\d s$, i.e.
\be \la{E-smax1}
\smax \,=\, 6 \, \D_h \le(1 \,+\, \D_h^2\ri)  \,\,.
\ee
\item {\bf Procedure 2}: Let us re-express Eq. (\ref{E-Hubrel}) as
\be \la{E-mrel}
\Omp \, \bh^2 \,=\, \Omp \, h^2 \le(1 - \rfraa s 3\ri)^2 \,\,,
\ee
where $\, \bh = \rfraa{\bHp \,}{[100 \, \mbox{Km s$^{-1}$ Mpc$^{-1}$}]} \,$ and $\, h = 
\rfraa{\Hp \,}{[100 \, \mbox{Km s$^{-1}$ Mpc$^{-1}$}]} \,$ are the reduced Hubble constants.
We have the corresponding error propagation equation 
\be \la{E-M-err}
\le[\fr{\d\! \le(\!\Omp \bh^2\ri)}{\Omp \bh^2}\ri]^2 \,= \le[\fr{\d\! \le(\Omp h^2\ri)}
{\Omp h^2}\ri]^2 \,+\, \fr{\le(\Omp h^2\ri)^2}{\le(\Omp \bh^2\ri)^2} \le(\fr{\d s} 
3\ri)^2 \,\,.
\ee
Stipulate now $\, \Omp h^2 = \Omp \bh^2 + \d \le(\Omp h^2\ri)$, and 
\be \la{E-M-stip}
\le|\fr{\d\! \le(\Omp h^2\ri)}{\Omp h^2}\ri|_{\mbox{\scriptsize min}} =\, 0 \,\,, 
\qquad \mbox{and} \qquad 
\le|\fr{\d\! \le(\Omp h^2\ri)}{\Omp h^2}\ri|_{\mbox{\scriptsize max}} =\, \D_m 
\,\equiv\, \le|\fr{\d\! \le(\!\Omp \bh^2\ri)}{\Omp \bh^2}\ri| \,\,,
\ee
as the minimum and maximum fractional errors. Keeping $\D_m$ fixed, by taking $\Omp \bh^2$ 
as the best fit value and $\, \d\! \le(\!\Omp \bh^2\ri)$ the corresponding $68\%$ margin 
for a given dataset, it follows that $\d s \geq 0$. Working out now the maximum value of 
$\d s$ one finds
\be \la{E-smax2}
\smax \,=\, 3 \, \D_m \,\,.
\ee
\eed
A caveat is there though in the above ways of indirectly marginalizing the parameter $s$. That is, 
neither $\bHp$ nor $\Omp \bh^2$ belong to the set of six independent parameters of the base $\L$CDM 
model. We may in principle choose, for the error analysis, some parameter other than $\bHp$ and 
$\Omp \bh^2$. However, that would also not be among the basic six. This is evident from our 
compelling need to the use Eq. (\ref{E-Hubrel}) in determining $\smax$ from the comparative study 
of cosmological parameters. Nevertheless, among the plausible parametric choices, preference may be 
given to $\Omp \bh^2$ (used in the procedure 2 above). The reason is the simple (additive) 
relationship $\, \Omp = \Ombp + \Omcp$, where $\Ombp$ and $\Omcp$ are the present-day values of the 
$\L$CDM baryon density and cold dark matter (CDM) density parameters respectively. Since 
$\Ombp \bh^2$ and $\Omcp \bh^2$ are two of the six basic $\L$CDM parameters, such a simple 
relationship may lead to a low error limit on $\Omp \bh^2$ (compared to that on $\bHp$ used in the 
procedure 1 for instance):
\be \la{E-errprop}
\le|\d\! \le(\!\Omp \bh^2\ri)\ri| =\, \sq{\le|\d\! \le(\!\Ombp \bh^2\ri)\ri|^2 
\,+ \le|\d\! \le(\!\Omcp \bh^2\ri)\ri|^2 \,+ 
\le\langle \Ombp \bh^2 , \Omcp \bh^2 \ri\rangle} \,\,,
\ee
where $\le\langle \Ombp \bh^2 , \Omcp \bh^2 \ri\rangle$ is the $1\s$ error 
covariance of the base $\L$CDM parameters $\Ombp \bh^2$ and $\Omcp \bh^2$. 
The smallness of $\le|\d\! \le(\!\Omp \bh^2\ri)\ri|$ does not however ensure 
that the corresponding percentage error $\D_m$ would be the least among those 
on all the other $\L$CDM parameters suitable for the determination of $\smax$ 
using the comparative error analysis. In fact, for the two procedures above, 
we mostly have $\D_m > \D_h$, as shown in Table \ref{E-tab1}. The $\smax$ values
also shown therein, are obtained from $\D_h$ and $\D_m$ using Eqs. (\ref{E-smax1})
and (\ref{E-smax2}) respectively. All the values of $\D_h$ and $\D_m$ are computed 
using the $\L$CDM parametric estimates from the following:
\bed  
\item {\it Dataset 1}: WMAP 9 year data with the Baryon Acoustic Oscillation (BAO) 
and the $\L$CDM Hubble constant $\bHp$ priors 
\cite{hin-wmap9,ben-wmap9}.
\item {\it Dataset 2}: WMAP 9 year data with BAO and $\bHp$ priors, combined with 
results of the Atacama Cosmology Telescope (ACT) \cite{act}, the South Pole Telescope 
(SPT) \cite{spt}, and the Supernova Legacy Survey (SNLS) 3 year sample of high 
redshift supernovae 1a \cite{snls}.
\item {\it Dataset 3}: PLANCK 2015 Cosmic Microwave Background (CMB) power 
spectra (TT, TE, EE) with polarization information from low multipole range 
temperature + polarization pixel-based likelihood (LowP) 
\cite{ade-pln15-13,ade-pln15-15}.
\item {\it Dataset 4}: PLANCK 2015 CMB power spectra (TT, TE, EE) and LowP, 
combined with weak lensing data (Lensing) \cite{lensing} and external data (Ext) 
that includes the BAO and $\bHp$ priors and the Joint Lightcurve Analysis (JLA) 
sample \cite{jla} constructed from the SNLS \cite{snls} and the Sloan Digital Sky 
Survey (SDSS) \cite{ries-sdss,macd-sdss} data, and many low redshift type-1a supernovae samples. 
\eed
%
\begin{table*}[!htb]
\centering
\renewcommand{\arraystretch}{1.2}
{\small
\begin{tabular}{|cc||c||c||c|c|}
\hline
\multicolumn{2}{|c||}{Observational} & $\L$CDM parameters 
& Fractional error & \multicolumn{2}{c|}{Value of $\smax$ via} \\
\cline{5-6}
\multicolumn{2}{|c||}{datasets} & (best fit \& $68\%$ limits) & estimates 
& {\footnotesize procedure 1} & {\footnotesize procedure 2} \\
\hline\hline 
1. & \hspace{-25pt} {\footnotesize WMAP\_9y} & $\Omp = 0.2880 \pm 0.0100$ & & &  \\
& \hspace{-25pt} {\footnotesize +BAO+$\bHp$} & $\bHp = 69.33 \pm 0.88$ 
& $\D_h = 0.01269$ & $0.07615$ & \\
& & $\Omp \bh^2 = 0.1383 \pm 0.0025$ & $\D_m = 0.01808$ & & $0.05424$ \\
\hline\hline
2. & \hspace{-15pt} {\footnotesize WMAP\_9y+SPT} & $\Omp = 0.2835 \pm 0.0094$ & & &  \\
& \hspace{-15pt} {\footnotesize +ACT+SNLS\_3y} & $\bHp = 69.55 \pm 0.78$ 
& $\D_h = 0.01121$ & $0.06727$ & \\
& \hspace{-15pt} {\footnotesize +BAO+$\bHp$} & $\Omp \bh^2 = 0.1371 \pm 0.0019$ 
& $\D_m = 0.01386$ & & $0.04158$ \\
\hline\hline
3. & \hspace{-15pt} {\footnotesize PLANCK\_2015} & $\Omp = 0.3156 \pm 0.0091$ & & &  \\
& \hspace{-15pt} {\scriptsize TT,TE,EE}{\footnotesize +LowP} & $\bHp = 67.27 \pm 0.66$ 
& $\D_h = 0.00981$ & $0.05887$ & \\
& & $\Omp \bh^2 = 0.1427 \pm 0.0014$ & $\D_m = 0.00981$ & & $0.02943$ \\
\hline\hline
4. & \hspace{-15pt} {\footnotesize PLANCK\_2015} & $\Omp = 0.3089 \pm 0.0062$ & & &  \\
& \hspace{-15pt} {\scriptsize TT,TE,EE}{\footnotesize +LowP} & $\bHp = 67.74 \pm 0.46$ 
& $\D_h = 0.00679$ & $0.04074$ & \\
& \hspace{-15pt} {\footnotesize +Lensing+Ext} & $\Omp \bh^2 = 0.1417 \pm 0.00097$ 
& $\D_m = 0.00684$ & & $0.02052$ \\
\hline
\end{tabular}
}
\caption{\footnotesize Best fit values and $68\%$ confidence limts of $\L$CDM cosmological parameters 
$\Omp, \bHp$ and $\Omp \bh^2$ for different observational datasets, alongwith the corresponding 
parametric upper bound $\smax$ obtained by the procedures 1 and 2 in the Einstein frame. The 
estimated fractional errors $\D_h$ and $\D_m$, from the $\bHp$ and $\Omp \bh^2$ limits 
respectively, used in the procedures 1 and 2, are also shown for each dataset.}
\la{E-tab1}
\end{table*} 
%
Using the estimated $\smax$ values, one can work out the maximal effective deviations
from the $\L$CDM matter density parameter $\Omb (z)$ and the DE EoS value ($= -1$):
\be \la{E-changes}
\d \Omf (z) \,:=\, \Omf (z, \smax) \,-\, \Omb (z) \,\,, \qquad \mbox{and} \qquad 
\d \wx (z) \,:=\, 1 \,+\, \wx (z, \smax) \,\,.
\ee
Following are the features we observe here for the Einstein frame DE evolution:

\vskip 0.1in
\no 
{\bf In the near past:}
The matter density deviation from $\L$CDM, $\d \Omf$, is negative but increases in magnitude 
with gradually diminishing rate as we go back in the past, up to say a redshift $z = 2.5$. 
The variations of $\d \Omf$ with $z$ for all the datasets are shown in Fig. \ref{E-fig1} (a) 
and (b), for $\smax$ obtained via the procedures $1$ and $2$ respectively. We see that the 
greater the value of the estimated $\smax$, the greater is the magnitude of $\d \Omf$ at a 
given $z \,(> 0)$.
%
\begin{figure}[!htb]
\centering
   \begin{subfigure}{0.495\linewidth} \centering
     \includegraphics[scale=0.475]{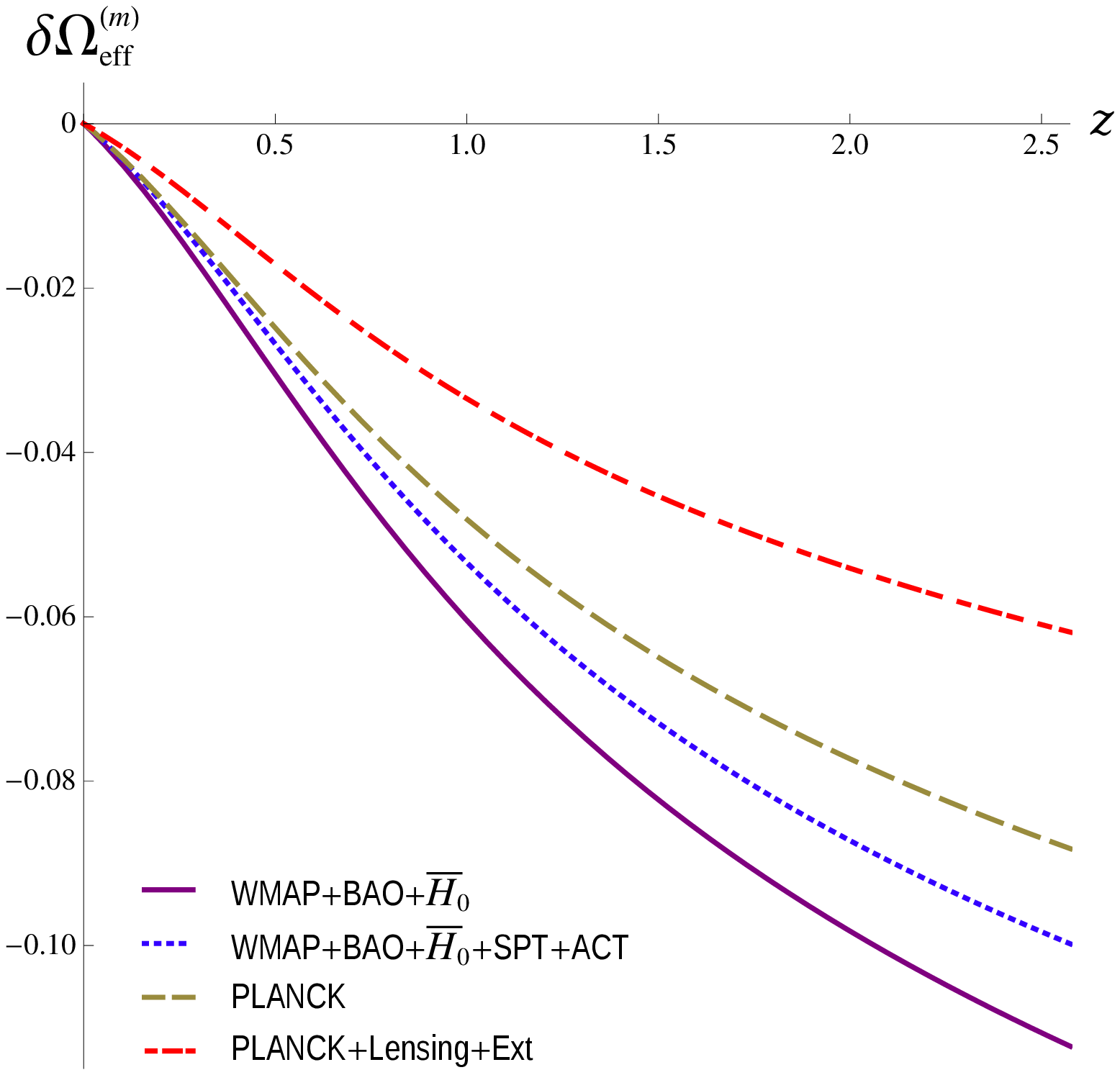}
     \caption{{\footnotesize $\d \Omf (z)$ for $\smax$ determined by procedure 1}}\label{E-fig1a}
   \end{subfigure}
   \begin{subfigure}{0.495\linewidth} \centering
     \includegraphics[scale=0.475]{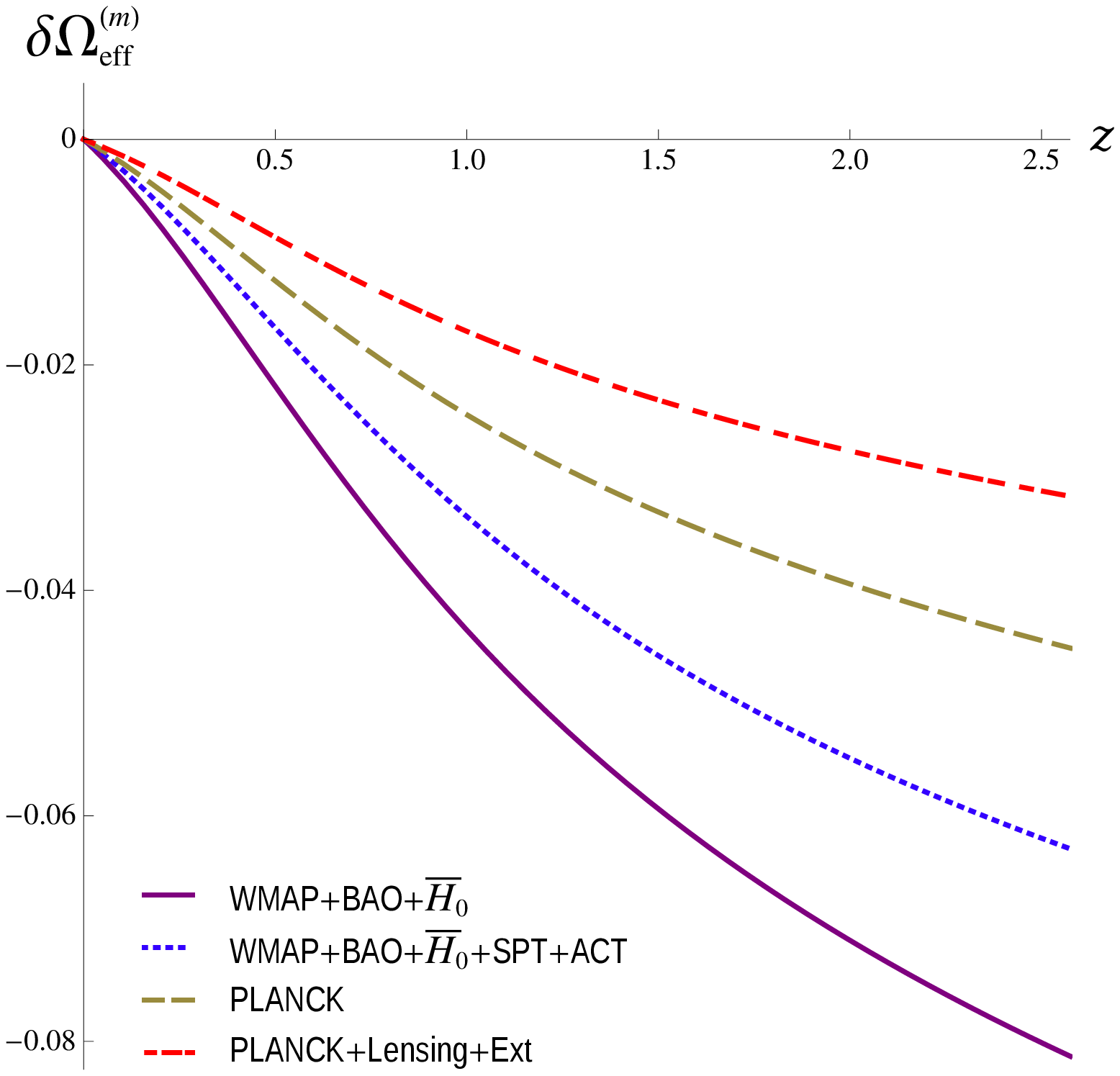}
     \caption{{\footnotesize $\d \Omf (z)$ for $\smax$ determined by procedure 2}}\label{E-fig1b}
   \end{subfigure}
\caption{\footnotesize Variation (with redshift $z$) of $\d \Omff$, the Einstein frame maximal 
effective deviation from the $\L$CDM matter density, corresponding to the sets of values of 
$\smax$ obtained using the procedures $1$ and $2$ respectively for the different datasets.}
\vskip -0.15in
\la{E-fig1}
\end{figure}
%
\begin{figure}[!htb]
\centering
   \begin{subfigure}{0.495\linewidth} \centering
     \includegraphics[scale=0.475]{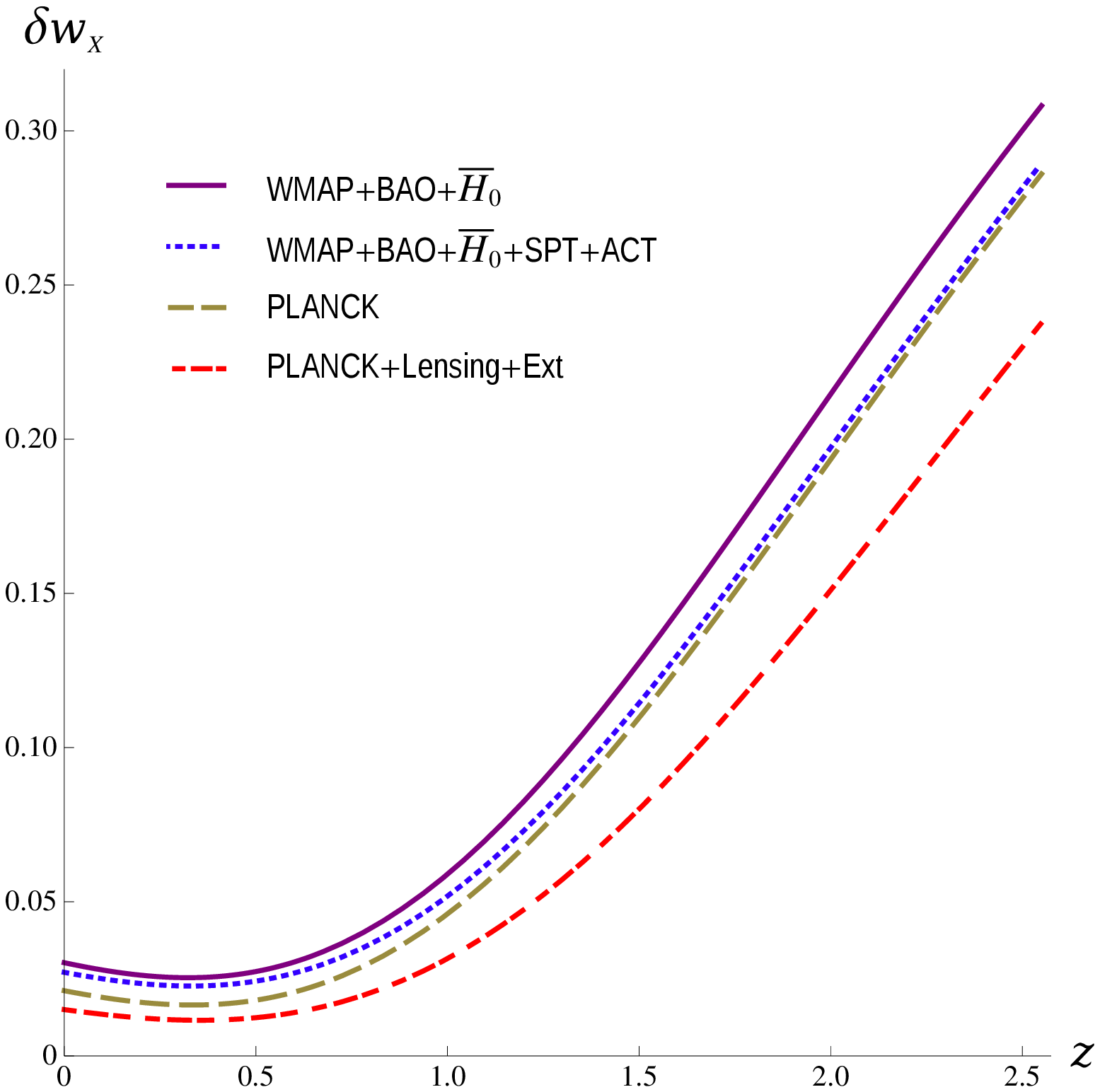}
     \caption{{\footnotesize $\d \wx (z)$ for $\smax$ determined by procedure 1}}\label{E-fig2a}
   \end{subfigure}
   \begin{subfigure}{0.495\linewidth} \centering
     \includegraphics[scale=0.475]{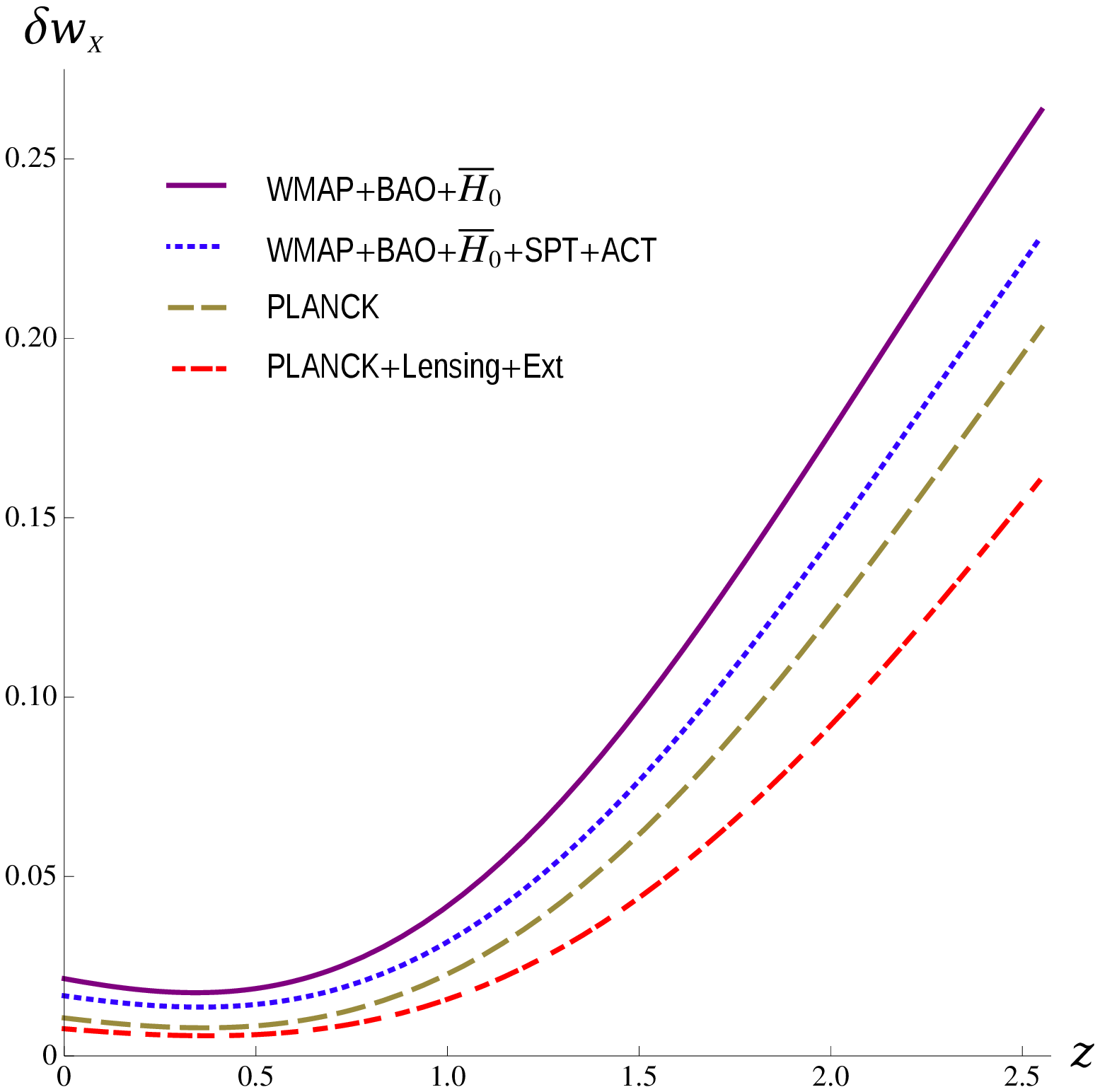}
     \caption{{\footnotesize $\d \wx (z)$ for $\smax$ determined by procedure 2}}\label{E-fig2b}
   \end{subfigure}
\caption{\footnotesize Variation (with redshift $z$) of $\d \wx$, the Einstein frame maximal 
change over the $\L$CDM dark energy EoS ($= -1$), corresponding to the sets of values of 
$\smax$ obtained using the procedures $1$ and $2$ respectively for the different datasets.}
\vskip -0.1in
\la{E-fig2}
\end{figure}
%
The DE EoS deviation from $\L$CDM, $\d \wx$, on the other hand, remains positive at all 
redshifts $z > 0$, i.e. there is no revelation of an effective {\em phantom} regime 
($\wx < - 1$) in the near past. For all the datasets, Fig. \ref{E-fig2} (a) and (b) show 
the variation of $\d \wx$ with $z$, respectively for $\smax$ obtained via the procedures 
$1$ and $2$. As we go back in the past (from the present epoch $z = 0$), $\d \wx$ 
decreases initially and reaches a minimum value $\in \le[0.0056, 0.0254\ri]$ at a redshift 
$z \in \le[0.3736, 0.3225\ri]$, corresponding to $\, \smax \in \le[0.02052,0.07615\ri]$ 
(obtained via the two procedures). Thereafter $\d \wx$ increases rapidly with $z$, till 
beginning to slow down around $z \simeq 2$. The greater the estimated $\smax$, the greater 
is the value of $\d \wx$ at a given $z$. However, one can see an unevenness in the nature 
of the $\d \wx$ curves for the different $\smax$ values. This is due to the fact that 
$\d \wx$ depends not only on $\smax$ but also on the best fit value of $\Omp$.
  
\vskip 0.1in
\no 
{\bf Up to a moderately distant past:} Let us now examine the DE evolution from the present 
epoch ($z = 0$) up to a fairly distant past ($z = 20$ say). As a reference $\smax$ value, we 
choose to take the largest estimate, viz. $0.07615$ (obtained by the proceduce $1$, for the 
dataset $1$). 
%
\begin{figure}[!htb]
\centering
   \begin{subfigure}{0.495\linewidth} \centering
     \includegraphics[scale=0.465]{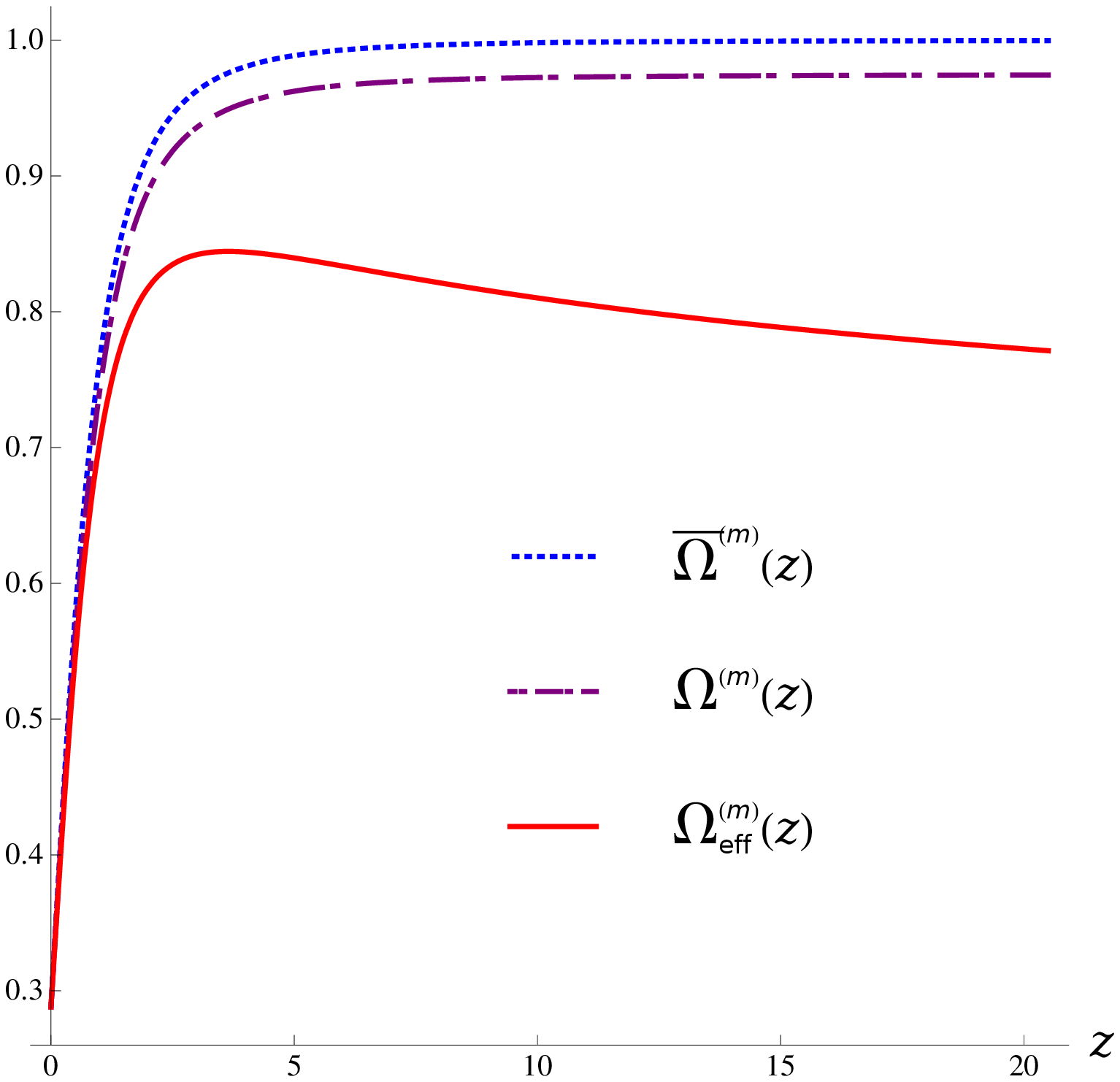}
     \caption{{\footnotesize Matter density parameters for $z \in [0, 20]$}}\label{E-fig3a}
   \end{subfigure}
   \begin{subfigure}{0.495\linewidth} \centering
     \includegraphics[scale=0.485]{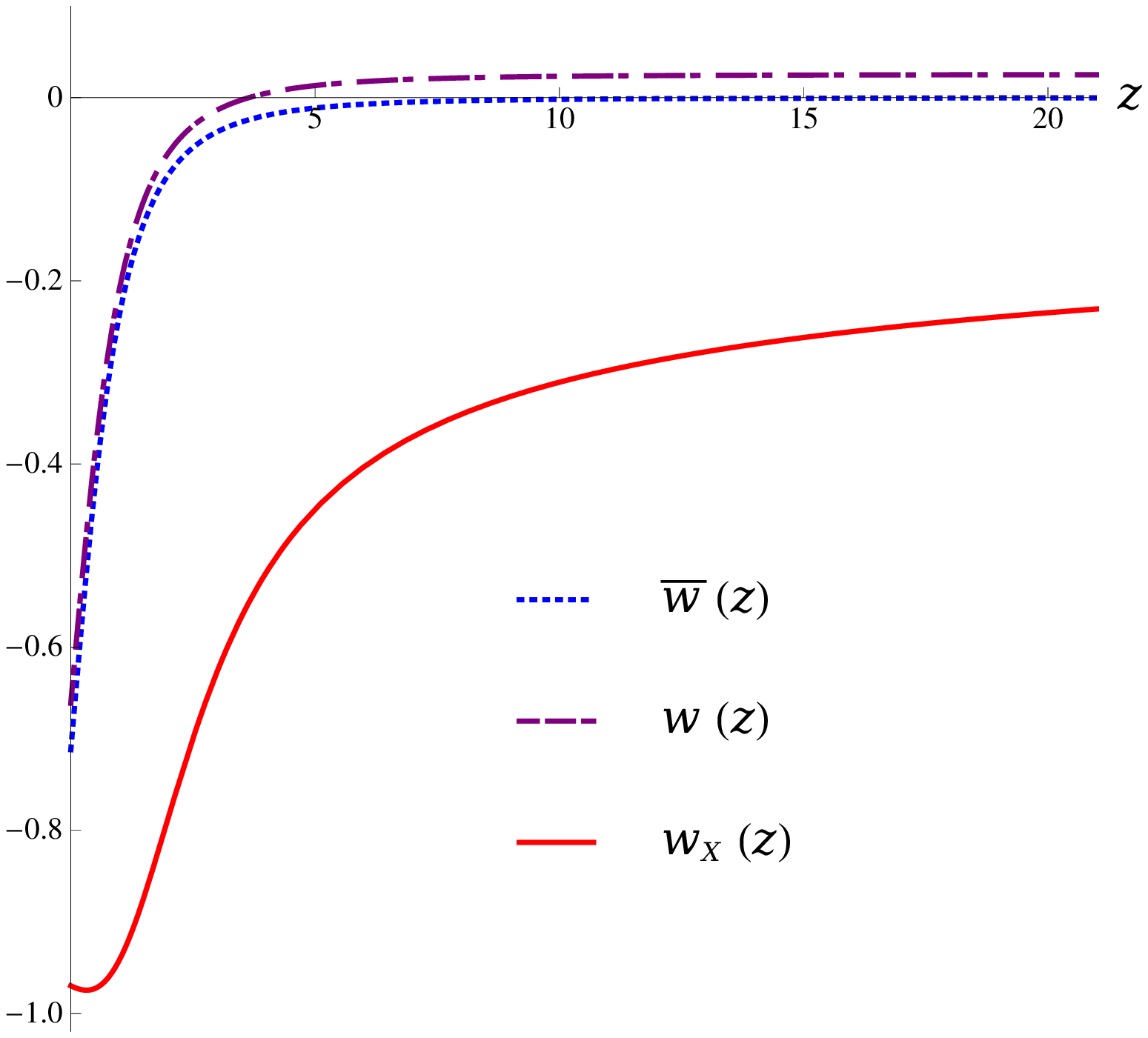}
     \caption{{\footnotesize EoS parameters for $z \in [0, 20]$}}\label{E-fig3b}
   \end{subfigure}
\caption{\footnotesize Einstein frame evolution, from the present ($z = 0$) upto a redshift 
$z = 20$ in the past, of (a) the matter density parameters $\Omff$, $\Om$ and $\Omb$, and 
(b) the EoS parameters $\wx$, $\sw$ and $\bw$. The value of $\smax$ used is the largest 
estimated one ($= 0.07615$ in Table \ref{E-tab1}).}
\vskip -0.1in
\la{E-fig3}
\end{figure}
%
For this value of $\smax$, the $z$-variations of the effective and the actual matter density 
parameters, $\Omf$ and $\Om$ respectively, as well as that of the $\L$CDM matter density 
parameter $\Omb$, are shown in Fig. \ref{E-fig3} (a). We see that the differences in the 
evolution pattern of all these paramters are noticiable only for redshifts $z \gtrsim 1$. 
The evolution of $\Om$ is similar to that of $\Omb$ --- increasing rapidly with $z$ until 
saturating to a value close to $1$ for $z \gtrsim 5$. The evolution pattern of $\Omf$ is 
different however --- it increases rapidly from its value $\Omp = 0.288$ at $z = 0$, attains 
a maximum value $0.843$ at $z = 3.6253$, and then decreases slowly with increasing $z$. Fig. 
\ref{E-fig3} (b) shows the Einstein frame evolution of the effective DE EoS parameter $\wx$ 
and the total EoS parameter $\sw$ of the system (for $\smax = 0.07615$), as well as the 
evolution of the $\L$CDM total EoS parameter $\bw$. Although the evolution of $\sw$ is 
similar to that of $\bw$, it saturates to a value slightly greater than zero for $z \gtrsim 
5$, implying no dust-like behaviour at high redshifts. This is expected, since in the Einstein 
frame the barotropic fluid is not actually the dust, but the one whose energy density varies 
as $(1 + z)^{3 + \sbig}$ as a consequence of the interaction with the scalar field. The 
evolution of $\wx$ is altogether different however. At the present epoch its value is 
$- 0.9698$, then it decreases with increasing $z$ and attains a minimum value $- 0.9746$ at 
$z = 0.3225$, turns around and increases rapidly to about $- 0.6$ at $z \simeq 3$, and 
thereafter increases with $z$ at a gradually diminishing rate. Most notably, $\wx$ does not 
get saturated to a particular value till $z = 20$, and in fact keeps on increasing (albeit 
very slowly) even when $z$ approaches its value at the {\em last scattering} (viz. $\simeq 
1100$)\footnote{Although not evident from Fig. \ref{E-fig3} (b), one may verify this by 
explicitly working out $\rfraa{d\wx}{dz}$ from Eq. (\ref{E-de-eos}).}.  

\vskip 0.1in
\no 
{\bf Near past to the extreme future:} Finally, let us study the DE evolution from a redshift 
of near past (say $z = 1.5$) to the extreme future ($z = - 1)$. 
%
\begin{figure}[!htb]
\centering
   \begin{subfigure}{0.495\linewidth} \centering
     \includegraphics[scale=0.525]{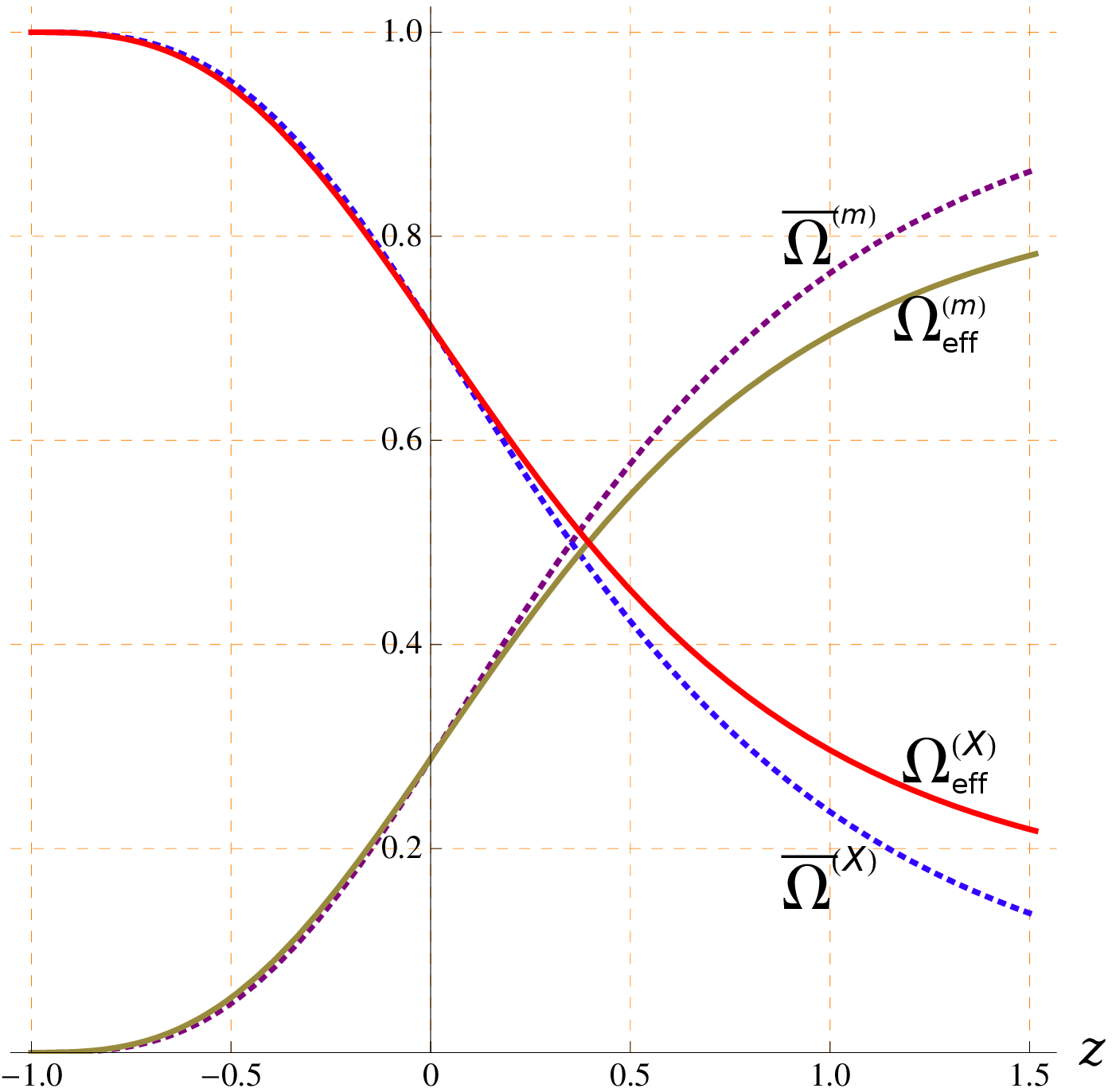}
     \caption{{\footnotesize The density parameters for $z \in (-1, 1.5]$}}\label{E-fig4a}
   \end{subfigure}
   \begin{subfigure}{0.495\linewidth} \centering
     \includegraphics[scale=0.5]{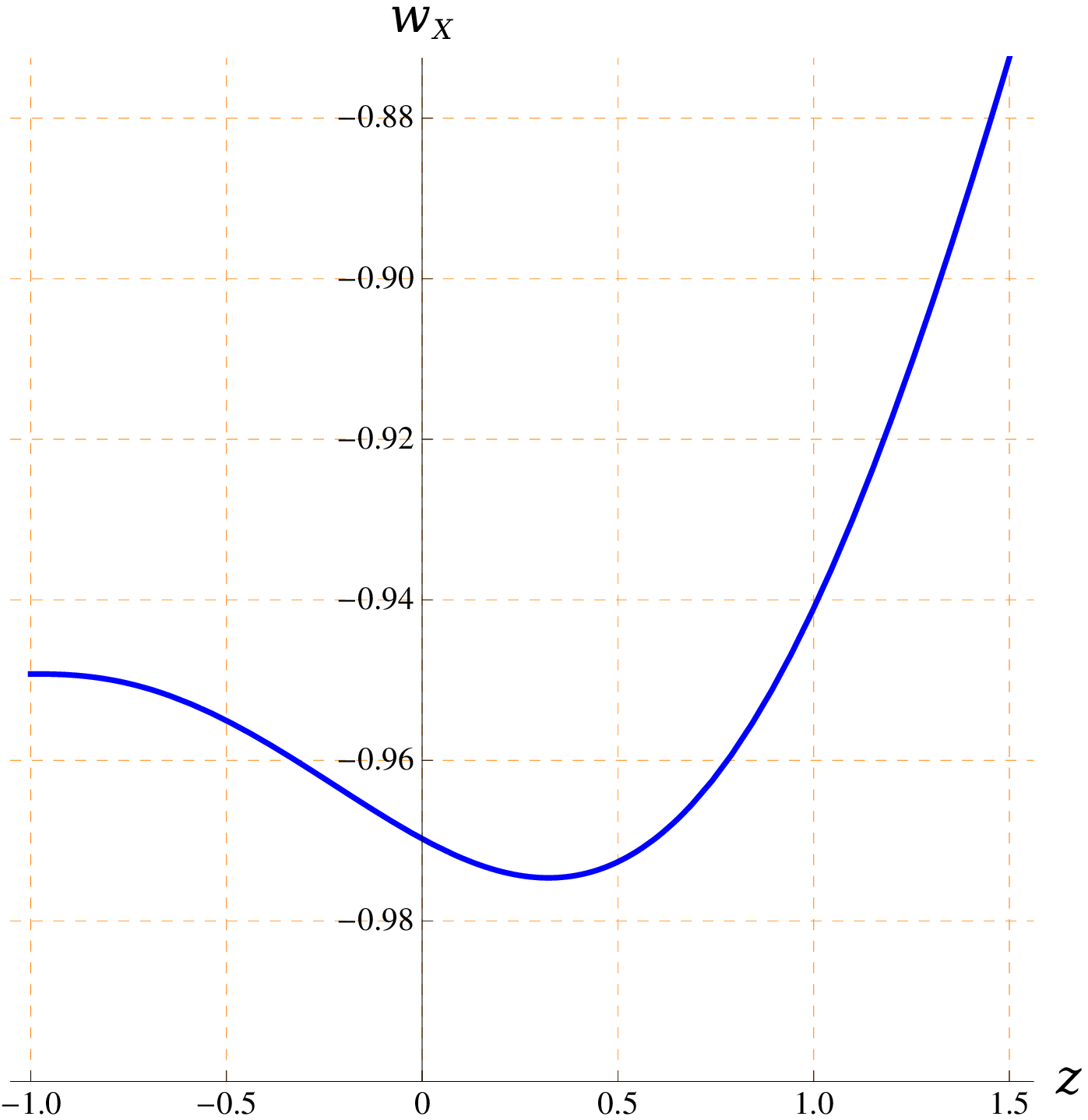}
     \caption{{\footnotesize Dark energy EoS parameter for $z \in (-1, 1.5]$}}\label{E-fig4b}
   \end{subfigure}
\caption{\footnotesize  Einstein frame evolution, from a redshift $z = 1.5$ in the near past to 
the extreme future $z = -1$, of (a) the density parameters $\Omff$ and $\OXff$, vis-a-vis the
corresponding $\L$CDM parameters $\Omb$ and $\OXb$, and (b) the DE EoS parameter $\wx$. The 
$\smax$ value used is the largest estimated one ($= 0.07615$ in Table \ref{E-tab1}).}
\vskip -0.1in
\la{E-fig4}
\end{figure}
%
Once again we consider, for brevity, the largest estimated value of $\smax \,(= 0.07615)$. The 
Fig. \ref{E-fig4} (a) shows the evolutions of $\Omf$ and $\, \OXf = 1 - \Omf$, the effective 
density parameters for matter and the DE respectively, and those of the corresponding $\L$CDM 
parameters $\Omb$ and $\OXb = 1 - \Omb$ respectively. While in the entire future regime, $\Omf$ 
is seen to be almost coincident with $\Omb$, its deviation from the latter become more and more 
significant as we go back in the past. Obviously, the same is true for $\OXf$ and $\OXb$. Hence, 
the future evolutions are $\L$CDM-like, with the DE dominating over the barotropic matter, and 
asymptotically $\, \Omf \rightarrow 0$ and $\, \OXf \rightarrow 1$ as $z \rightarrow -1$. The 
redshift of the matter-DE equality ($\Omf = \OXf = 0.5$) for the dataset $1$  considered here 
(corresponding to which $\smax = 0.07615$ by the procedure $1$) turns out to be $z_{eq} = 0.3952$. 
This is not much different from that for $\L$CDM, viz. $z_{eq} = 0.3521$, for the same dataset. 
As to the {\em coincidence} problem encountered in the $\L$CDM model, we therefore have only a 
slightly improved situation here. In fact, one may verify that this is true not only for $\smax = 
0.07615$ (corresponding to the dataset $1$), but also for the other $\smax$ values (corresponding 
to the other datasets) in Table \ref{E-tab1}. Nevertheless, for our reference $\smax (= 0.07615)$,
the evolution of the effective DE EoS parameter $\wx$ over the range $z \in \le[-1, 1.5\ri]$ is 
shown in Fig. \ref{E-fig4} (b). As $z$ decreases from the value $1.5$, we have $\wx$ decreasing 
rapidly to a minimum value $\, \wxmin = - 0.9746$ at $z = 0.3225$. Thereafter it increases (albeit 
rather slowly) as $z$ decreases further. At the present epoch, $\wx (0) = - 0.9698$, and in the 
future (i.e. for $z < 0$, or positive blueshifts), $\wx$ continues to increase, untill saturating 
to a value $\simeq - 0.95$ at $z \simeq - 0.8$. The overall evolution of $\wx$ is smooth all the 
way to $z = -1$, with no discontinuities whatsoever.

\subsection{Bounds on torsion parameters and the effective Brans-Dicke 
parameter} \la{sec:E-tbounds}

Refer back to Eqs. (\ref{E-T-norm1}) and (\ref{A-norm}) for the norms of the torsion mode 
vectors $\cT_\m$ and $\cA^\m$ respectively. Remember that we have the assumption of the 
scalar field mass being solely due to $\cA^\m$, and are alluding to the proposition 1 (in 
$\S$\ref{sec:A-mass}) as an exemplary scenario. Using the Einstein frame solution 
ansatz (\ref{E-ans}), and the subsequent equation (\ref{E-Hub}), we express these norms 
(referred to as the {\em torsion parameters}) in terms of the Hubble rate $H$ as
\bea
&& \le|\cT\ri| (z) \,=\, 3 \, \k \, \sq{\b} \, e^{\k \, \sq{\b} \, \vph (z)} \, 
\dot{\vph} (z) \,=\, 3 \, s \le(1 + z\ri)^{-\sbig} \, H (z) \,\,, \la{E-Tnormz} \\
&& \le|\cA\ri| =\, 4 \, \k \, \sq{6 \, \L} \,=\, 12 \sq{2} \le(1 + z\ri)^{-\sbig}
\, H (z) \le[1 \,-\, \rfraa s 3 \,- \le(1 + z\ri)^\sbig \Omf (z)\ri]^{\rfra 1 2} \,\,.
\la{E-Anormz}
\eea 
%
\begin{table*}[!htb]
\centering
\renewcommand{\arraystretch}{1.5}
\begin{tabular}{|cc||c c c c|}
\hline
\multicolumn{2}{|c||}{Observational dataset} & Estimations:
& {$\rfraa{\le|\cT\ri|_{_0} \,}{\bHp} \Big|_{\smaxb}$} 
& {$\rfraa{\le|\cT\ri|_{_0} \,}{\le|\cA\ri|} \Big|_{\smaxb}$}
& {$\fwmin$} \\ [4pt]
\hline\hline 
1. & \hspace{-15pt} {\small WMAP\_9y+BAO+$\bHp$} & {\footnotesize (by procedure 1)} & $0.2314$ 
& $0.0163$ & $5.0660$ \\
\cline{3-6}
& & {\footnotesize (by procedure 2)} & $0.1642$ & $0.0115$ & $7.7183$ \\
\hline
2. & \hspace{-15pt} {\small WMAP\_9y+SPT+ACT} & {\footnotesize (by procedure 1)} & $0.2041$ 
& $0.0143$ & $5.9327$ \\
\cline{3-6}
& \hspace{-15pt} {\small SNLS\_3y+BAO+$\bHp$} & {\footnotesize (by procedure 2)} & $0.1256$ 
& $0.0088$ & $10.5250$ \\
\hline
3. & \hspace{-25pt} {\small PLANCK} {\scriptsize TT,TE,EE} & {\footnotesize (by procedure 1)} 
& $0.1784$ & $0.0128$ & $6.9933$ \\
\cline{3-6}
& \hspace{-25pt} {\small +LowP} & {\footnotesize (by procedure 2)} & $0.0887$ & $0.0063$ 
& $15.4895$ \\
\hline
4. & \hspace{-25pt} {\small PLANCK} {\scriptsize TT,TE,EE} & {\footnotesize (by procedure 1)} 
& $0.1231$ & $0.0088$ & $10.7729$ \\
\cline{3-6}
& \hspace{-25pt} {\small +LowP+Lensing+Ext} & {\footnotesize (by procedure 2)} & $0.0618$  
& $0.0044$ & $22.8665$ \\
\hline
\end{tabular}
\caption{\small Torsion trace parameter at the present epoch, $\le|\cT\ri|_{_0}$ (in units 
of $\bHp$), and its ratio with the torsion pseudo-trace parameter $\le|\cA\ri|$, in the 
Einstein frame. The quantities are evaluated at $s = \smax$ and the corresponding minimum 
values of the parameter $\fw$ are shown for the various datasets.}
\vskip -0.05in
\la{E-tab2}
\end{table*} 
%

\vskip -0.1in
\noindent
Let us denote the norm of $\cT_\m$ at the present epoch as $\le|\cT\ri|_{_0} \! \equiv 
\le|\cT\ri|_{\zbig \,= 0}$, the norm of $\cA^\m$ is of course fixed at all $z$. Our 
interest is in working out $\le|\cT\ri|_{_0}$ and $\le|\cA\ri|$ in terms of the 
independent parameters $\, \smax, \, \Omp$ and(or) the $\L$CDM Hubble constant $\bHp$, Eq. 
(\ref{E-Hubrel}). Table \ref{E-tab2} shows
\bea 
&& \le[\fr{\le|\cT\ri|_{_0}}{\bHp}\ri]_{\smaxb} =\, 3 \, \smax
\le(1 \,-\, \fr{\smax} 3\ri)^{- \rfra 1 2} \,, \qquad \mbox{and} \la{E-Tmax} \\
&& \le[\fr{\le|\cT\ri|_{_0}}{\le|\cA\ri|}\ri]_{\smaxb} =\, \fr{\smax}{4 \sq{2}} 
\le(1 \,-\, \fr{\smax} 3 \,-\, \Omp\ri)^{- \rfra 1 2} \,, \la{E-TAmax}
\eea
computed using the estimates of $\smax$ and the best fit values of $\bHp$ and $\Omp$ for 
the various datasets (in Table \ref{E-tab1}). It is easy to verify that the expressions in 
(\ref{E-TAmax}) are in fact the upper limits of $\le|\cT\ri|_{_0}$ (in units of $\bHp$) and  
the ratio $\rfraa{\le|\cT\ri|_{_0} \,}{\le|\cA\ri|}$. Note also that while $\smax$ (obtained by 
the two procudures) has values $\, \in \le[0.02052, 0.07615\ri]$ for the different datasets 
considered, the effective BD parameter $\, \fw = \rfraa{\le(1 - 3 s\ri)}{2 s} \,$ has its 
minimum value $\fwmin \in \le[5, 23\ri]$ approximately (as shown in Table \ref{E-tab2}). 
This matches fairly (by order of magnitude) with $\, \fwmin \simeq 40$ obtained in other
independent studies \cite{acq-BD,avi-BD,chen-BD}. Further high values on $\fwmin$ \cite{als-BD} 
would imply $s \lesssim 10^{-2}$, which is perceptable as well, within the domain of validity 
of our parametric estimation.

\section{MST Cosmology in the Jordan frame}  \la{sec:J-MST}

Let us now consider, in this section, the Jordan frame to be suitable for interpreting the 
results of physical observations. Our approach would be similar to that in the previous 
section, presuming once again that the MST formalism, under certain circumstances, gives 
rise to a DE model which is a slight modification of $\L$CDM. Of course, while analyzing 
the Jordan frame action (\ref{MST-ac2}) (or more conveniently (\ref{J-ac})) and the ensuing 
field equations, we encounter the running gravitational coupling constant $\Geff$, which is 
given explicitly by a function of the scalar field $\F$ [{\it cf.} Eq. (\ref{G-eff})]. As 
such, the total energy-momentum tensor due to the cosmological (dust) matter and the scalar 
field combined, is not conserved. The matter Lagrangian though, being minimally coupled to 
gravity in the Jordan frame, renders the corresponding (matter) energy-momentum tensor 
$T_{\m\n}^{(m)}$ conserved [{\it cf.} Eq. (\ref{J-consv})]. All these are quite unlike what 
we had in the Einstein frame. So the cosmology in the Jordan frame is expected to differ
characteristically from that in the Einstein frame.

\subsection{Cosmological equations and solution} \la{sec:J-sol}

Eqs. \!\!(\ref{J-eom1}), (\ref{J-eom2}) and (\ref{J-dust}) lead to the Jordan frame Friedmann 
and Raychaudhuri equations
\bea 
&& H^2 (t) \,=\, \fr 1 {3 \F (t)} \le[\rmt (t) \,-\, 3 \, H (t) \, \dot{\F} (t)
\,+\, \fr{\fw \, \dot{\F}^2 (t)}{2 \, \F (t)} \,+\, \cV \le(\F (t) \ri)\ri] \,,
\la{J-eq1} \\
&&\dot{H} (t) \,=\, - \, \fr 1 {2 \F (t)} \le[\rmt (t) \,-\, H (t) \, \dot{\F} (t)
\,+\, \fr{\fw \, \dot{\F}^2 (t)}{\F (t)} \,+\, \ddot{\F} (t)\ri] \,,
\la{J-eq2}
\eea
where $\, H (t) := \rfraa{\dot{a} (t)}{\!\! a (t)}$ is now the Jordan frame Hubble parameter.

The conservation equation (\ref{J-consv}) yields the usual dust matter density
\be \la{J-matdens}
\rmt (a) \,=\, \fr \rmp {a^3} \,\,,
\ee
with its present-day value $\, \rmp = \rmt (t=\tp) = \rmt (a=1)$. 

Assuming once again a simple power-law ansatz:
\be \la{J-ans}
\F (a) \,=\, \Fp \, a^n \,\,, \qquad \mbox{with} \qquad
n = ~ \text{a constant} \,\,,
\ee  
we obtain 
\bea 
&& H^2 (a) \,=\, \fr{\k^2} 3 \le(1 + n - \fr{\fw n^2}{6}\ri)^{-1} 
\le[\fr \rmp {a^{n + 3}} \,+\, \L\ri] \,, \la{J-eq1a} \\
&& \le[a^{2n + 6} \, H^2 (a)\ri]' \,=\, \fr{2 \, \k^2}{\le(2 \fw + 3\ri) n} 
\le[\rmp \, a^{n + 2} \,+\, 2 \, \L \, a^{2n + 5}\ri] 
\,,\la{J-eq2a}
\eea
by recalling that $\, \cV (\F) = \rfraa{\L \F}{\Fp}$ [{\it cf.} Eq. (\ref{J-pot})], 
with $\, \Fp = \k^{-2}$.

It is easy to verify that for
\be \la{J-solcond}
n \,= \le(1 + \fw\ri)^{-1} \,\,, 
\ee
the equations (\ref{J-eq1a}) and (\ref{J-eq2a}) are satisfied\footnote{There is
also a solution $n = - 2$. However, that would render $H^2 < 0$, unless it is 
stipulated that $2 \fw + 3 < 0$, i.e. the non-minimal parameter $\b < 0$ ---
a possibility which we discard for the reason mentioned earlier.}, whence the 
Friedmann equation (\ref{J-eq1a}) reduces to
\be \la{J-Hub}
H^2 (a) \,=\, \fr{2 \, \k^2}{\le(n + 2\ri) \le(n + 3\ri)} 
\le[\fr \rmp {a^{n + 3}} \,+\, \L\ri] \,.
\ee
The validity condition for this, viz. $\, n > - 2$, restricts $\, \fw > \rfraa {-3} 2$. 
Further stringent bounds on $\fw$ can be obtained of course, as we shall demonstrate in 
our subsequent analysis.

Defining for convenience the Jordan frame effective DE density $\rx$ and pressure $\px$ as
\bea 
&& \rx (a) \,:=\, \fr 6 {\le(n + 2\ri) \le(n + 3\ri)} \le[\le\{a^{-n} \,-\, 
1 \,-\, \fr{n \le(n + 5\ri)} 6\ri\} \fr \rmp {a^3} \,+\, \L\ri] \,, 
\la{J-de-dens} \\
&& \px (a) \,:=\, \fr 6 {\le(n + 2\ri) \le(n + 3\ri)} \le[\fr{n \, \rmp}
{3 \, a^{n + 3}} \,-\, \L\ri] \,, \la{J-effpres}
\eea
we have the critical density of the universe  given by
\be \la{J-crit}
\r (a) \,:=\, \fr{3 H^2 (a)}{\k^2} \,=\, \rmt (a) \,+\, \rx (a) \,\,,
\ee
which satisfies the conservation relation
\be \la{J-critconsv}
\r' (a) \,+\, \fr 3 a \le[\r (a) \,+\, \px (a) \ri] =\, 0 \,\,.
\ee
It is important to note here that the above definition (\ref{J-crit}) of the critical 
density $\r$ is {\em not} the same as the one commonly used in the literature, viz.
$\rJ = 3 \F H^2$, for scalar-tensor cosmologies in the Jordan frame \cite{frni,fujii,tsuj}.
However, from a practical point of view such a definition leads to difficulties, 
especially when it comes to comparing the parametric estimations of a given scalar-tensor
DE model with those of $\L$CDM or any other minimally coupled quintessence or k-essence 
model. This is evident from the fact that for minimal coupling one uses the standard 
definition of critical density $\r$ given by Eq. (\ref{J-crit}) or (\ref{E-crit}), 
whereas with the definition $\rJ$ one has the direct dealing with the running 
gravitational coupling parameter given in terms of the field $\F$ (compare Eq. 
(\ref{J-eq1}) with the Einstein frame Friedmann equation (\ref{E-eq1}) and see that 
the coupling factor $\k^2$ is replaced by $\F^{-1}$). For e.g. the matter density 
parameter, if defined as $\Om_J = \rfraa{\rmt}{\rJ}$ for a scalar-tensor DE model in
the Jordan frame, inevitably has no correspondance with the definition $\Om = \rfraa{\rmt}
{\r}$ for say the $\L$CDM or quintessence model. Any comparison of the marginalizations 
over the matter density parameter (using observational data) in the scalar-tensor DE model 
and the $\L$CDM or quintessence model is therefore redundant. Besides, $\Om_J$ is not truely 
the matter density parameter, since it depends explicitly on the field $\F$ (by virtue of the
explicit $\F$-dependence of $\rJ$). So, the very reason for marginalizing over $\Om_J$ is
questionable, lest the technicality that needs to be sorted out carefully. On the other
hand, $\Om$ by definition has no fallacy whatsoever. Therefore, from the practical point 
of view as well as for transparency in realizing the results physically, it is imperative
to stick to the definition of $\Om$ as the matter density parameter, and hence to $\r$ as
the critical density of the universe, while studying cosmology even for the theoretical
formulations in the Jordan frame (see \cite{ssasbPP} for further elaboration).

\subsection{Effective cosmological parameters and their estimation} \la{sec:J-param}

As a function of the redshift $\, z = \le(a^{-1} - 1\ri)$, the Jordan frame matter density 
parameter is
\be \la{J-m-dp}
\Om (z) \,:=\, \fr{\rmt (z)}{\r (z)} \,=\, \Omp \, \fr{\le(1 + z\ri)^3}{\fH^2 (z)} \,,
\ee
where $\, \Omp = \Om (z=0)$, and the rationalized Hubble parameter $\fH (z)$ given by
\be \la{J-RHub} 
\fH (z) \,\equiv\, \fr{H (z)} \Hp \,= \le[1 \,+\, \fr{6 \, \Omp}{\le(n + 2\ri) 
\le(n + 3\ri)} \le\{\le(1 + z\ri)^{n+3} -\, 1\ri\}\ri]^{\rfra 1 2} \,\,,
\ee
$\Hp \equiv H (z=0)$ being the Hubble constant in the Jordan frame.

The corresponding DE EoS parameter $\wx$ is obtained as
\be \la{J-de-eos}
\wx (z) \,:=\, \fr{\px (z)}{\rx (z)} \,=\, - \, 1 \,+\, \fr{\Om (z)}
{1 \,-\, \Om (z)} \le[\fr{2 \le(1 + z\ri)^n}{n + 2} \,-\, 1\ri] \,,
\ee
and hence the EoS parameter of the system turns out to be
\be \la{J-eos}
\sw (z) \,:=\, \fr{\px (z)}{\r (z)} \,=\, - \, 1 \,+ \fr{2 \le(1 + z\ri)^n \Om (z)}
{n + 2} \,\,.
\ee
It then follows that the requirement $\, \sw (0) < \rfraa {-1} 3$ (for the universe to 
have an accelerated expansion at the present epoch) imposes the condition 
\be \la{J-accl} 
n \,>\, 3 \, \Omp \,-\, 2 \,\,.
\ee
So the relation $\, n = (1 + \fw)^{-1}$ implies {\it any one} of the following:
\ben[(a)]
\item $\fw < \le[\le(3 \Omp - 2\ri)^{-1} - 1\ri]$, for either (i) $ n > 0 \,$;  
$\Omp > \rfraa 2 3 \,$, ~ or (ii) $n < 0 \,$; $\Omp < \rfraa 2 3 \,$,
\item $\fw > - 1 \,$, for (iii) $n > 0 \,$; $\Omp < \rfraa 2 3 \,$.
\een
Moreover, as shown earlier, $\, \fw > - \rfr 3 2$. Therefore since 
$\Omp > 0$, it is easy to check that the condition (a) would hold only for $\, 
\Omp > \rfraa 2 3$, i.e. for the case (i) above. This implies an unusually 
large deviation from the concordance value ($\approx 0.3$) of the present-day 
matter density parameter. As we are not considering our model to differ such
drastically from the $\L$CDM model, by the process of elimination of the 
condition (a), we are left with $\, \fw > -1$ (the condition (b)), which 
approves $n > 0$ (the case (iii) above). Furthermore, unlike the case in the 
Einstein frame, we are having here the constraint $\, \fw > -1$, tighter than 
the previous one ($\fw > \rfraa {-3} 2$), regardless of any fiducial setting, 
such as $\, \Omp \simeq 0.3$. 

As in the Einstein frame, the $\L$CDM equations are recovered here in the limit 
the model parameter $n \rarr 0$. For instance, Eq. (\ref{J-Hub}) reduces to the 
$\L$CDM Friedmann equation
\be \la{J-Hub-lim}
\bH^2 (z) \,\equiv\, \lim_{n \rarr 0} H^2 (z) \,=\, \fr{\k^2} 3 \le[\rmp 
\le(1 + z\ri)^3 +\, \L\ri] \,.
\ee
Proceeding as in $\S$\ref{sec:E-est}, with all notations unaltered, we have the following:
\bea 
&& \hskip -0.5in \bHp = \Hp \sq{\le(1 + \fr n 2\ri) \le(1 + \fr n 3\ri)} \,\,, \la{J-Hubrel} \\
&& \hskip -0.5in \Om (z) = \fr{\le(n + 2\ri) \le(n + 3\ri) \le(1 + z\ri)^3 \Omb (z)}
{\le[6 (1 + z)^{n+3} - n (n + 5)\ri] \Omb (z) + (n + 2) (n + 3) (1 + z)^3 
[1 - \Omb (z) ]} \,,
\eea
where $\, \Omb (z) \,\equiv\, \lim_{n \rarr 0} \, \Om (z) \,$ is the same as that given 
by Eq. (\ref{E-m-dplim}).

We obtain the maximum value of the model parameter ($n$ here) to be
\be \la{J-nmax1}
\nmax \,=\, \fr{12} 5 \, \D_h \le(1 \,+\, \D_h\ri)^2 \,\,, 
\ee
by the procedure $1$ (which uses the marginalized $\L$CDM Hubble constant
$\bHp$), and
\be \la{J-nmax2}
\nmax \,=\, \fr 6 5 \, \D_m \le(1 \,+\, \D_m\ri) \,\,,
\ee
by the procedure $2$ (which uses the marginalized $\L$CDM parameter 
$\, \Omp \bh^2$).
%
\begin{table*}[!htb]
\centering
\renewcommand{\arraystretch}{1.2}
{\small
\begin{tabular}{|cc||c||c||c|c|}
\hline
\multicolumn{2}{|c||}{Observational} & $\L$CDM parameters 
& Fractional error & \multicolumn{2}{c|}{Value of $\nmax$, via} \\
\cline{5-6}
\multicolumn{2}{|c||}{datasets} & (best fit \& $68\%$ limits) & estimates 
& {\footnotesize procedure 1} & {\footnotesize procedure 2} \\
\hline\hline 
1. & \hspace{-25pt} {\footnotesize WMAP\_9y} & $\Omp = 0.2880 \pm 0.0100$ & & &  \\
& \hspace{-25pt} {\footnotesize +BAO+$\bHp$} & $\bHp = 69.33 \pm 0.88$ 
& $\D_h = 0.01269$ & $0.03123$ & \\
& & $\Omp \bh^2 = 0.1383 \pm 0.0025$ & $\D_m = 0.01808$ & & $0.02209$ \\
\hline
2. & \hspace{-15pt} {\footnotesize WMAP\_9y+SPT} & $\Omp = 0.2835 \pm 0.0094$ & & &  \\
& \hspace{-15pt} {\footnotesize +ACT+SNLS\_3y} & $\bHp = 69.55 \pm 0.78$ 
& $\D_h = 0.01121$ & $0.02751$ & \\
& \hspace{-15pt} {\footnotesize +BAO+$\bHp$} & $\Omp \bh^2 = 0.1371 \pm 0.0019$ 
& $\D_m = 0.01386$ & & $0.01686$ \\
\hline
3. & \hspace{-15pt} {\footnotesize PLANCK\_2015} & $\Omp = 0.3156 \pm 0.0091$ & & &  \\
& \hspace{-15pt} {\scriptsize TT,TE,EE}{\footnotesize +LowP} & $\bHp = 67.27 \pm 0.66$ 
& $\D_h = 0.00981$ & $0.02401$ & \\
& & $\Omp \bh^2 = 0.1427 \pm 0.0014$ & $\D_m = 0.00981$ & & $0.01189$ \\
\hline
4. & \hspace{-15pt} {\footnotesize PLANCK\_2015} & $\Omp = 0.3089 \pm 0.0062$ & & &  \\
& \hspace{-15pt} {\scriptsize TT,TE,EE}{\footnotesize +LowP} & $\bHp = 67.74 \pm 0.46$ 
& $\D_h = 0.00679$ & $0.01652$ & \\
& \hspace{-15pt} {\footnotesize +Lensing+Ext} & $\Omp \bh^2 = 0.1417 \pm 0.00097$ 
& $\D_m = 0.00684$ & & $0.00826$ \\
\hline
\end{tabular}
}
\caption{\footnotesize Best fit values and $68\%$ confidence limts of $\L$CDM cosmological parameters 
$\Omp, \bHp$ and $\Omp \bh^2$ for different observational datasets, alongwith the corresponding 
parametric upper bound $\nmax$ obtained by the procedures 1 and 2 in the Jordan frame. The 
estimated fractional errors $\D_h$ and $\D_m$, from the $\bHp$ and $\Omp \bh^2$ limits 
respectively, used in the procedures 1 and 2, are also shown for each dataset.}
\vskip -0.1in
\la{J-tab1}
\end{table*} 
%

For the datasets $1$-$4$ (see $\S$\ref{sec:E-est}), 
the calculated values of $\nmax$, via both the procedures, are shown in Table 
\ref{J-tab1}. With these $\nmax$ values, we work out the effective maximal changes 
over the $\L$CDM matter density parameter $\Omb (z)$ and the DE EoS value 
($= -1$), respectively as
\be \la{J-changes}
\d \Om (z) \,:=\, \Om (z, \nmax) \,-\, \Omb (z) \,\,, \qquad \mbox{and} \qquad 
\d \wx (z) \,:=\, 1 \,+\, \wx (z, \nmax) \,\,.
\ee
Following are the features of the Jordan frame DE evolution:

\vskip 0.1in
\no 
{\bf In the near past:}
Let us refer to Figs. \ref{J-fig1} (a) and (b), which show the variation of $\d \Om$ with $z$ in 
the range $\le[0,2\ri]$ for all the datasets, and for $\nmax$ obtained via the procedures $1$ 
and $2$ respectively. 
%
\begin{figure}[!htb]
\centering
   \begin{subfigure}{0.495\linewidth} \centering
     \includegraphics[scale=0.475]{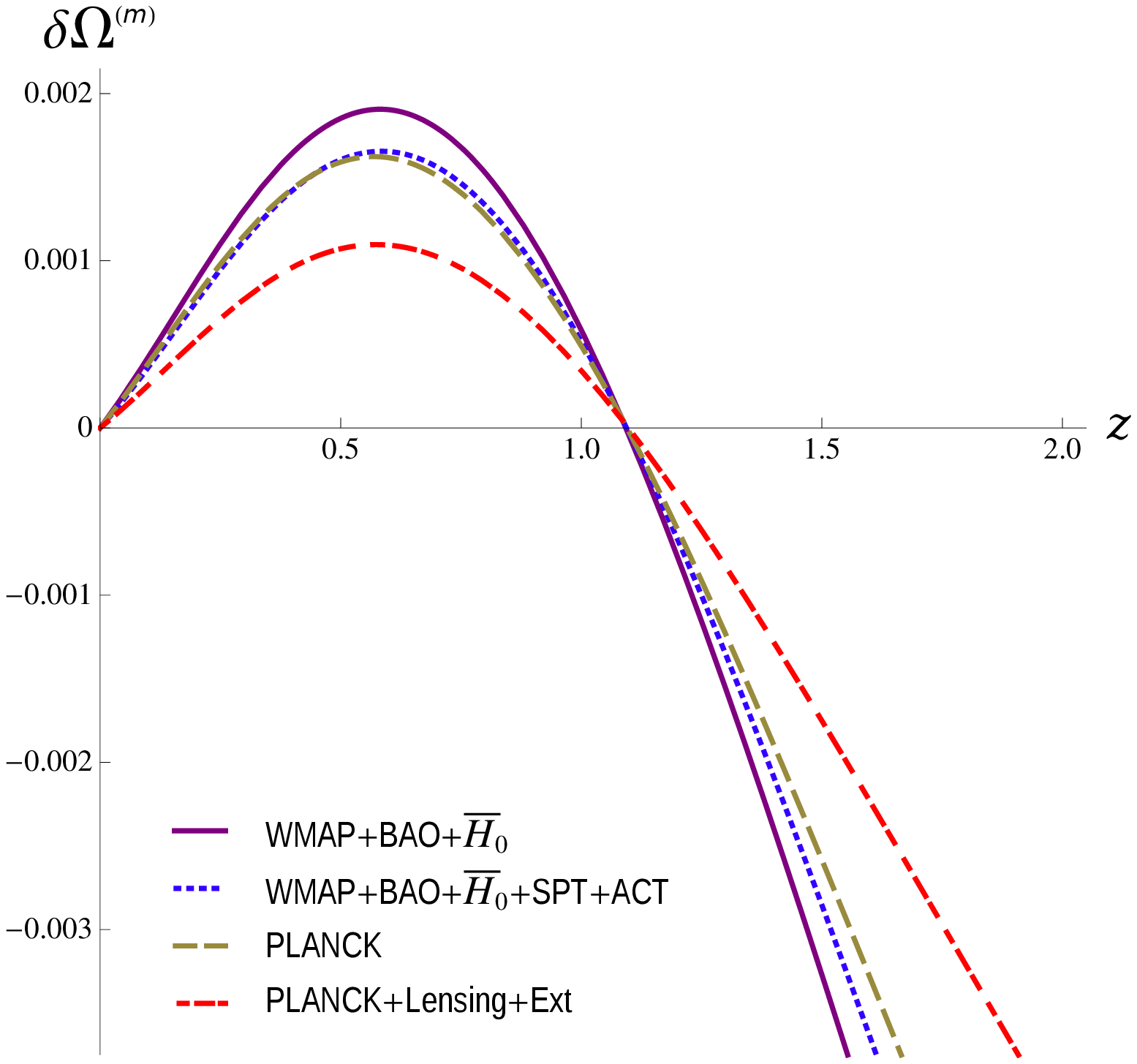}
     \caption{{\footnotesize $\d \Om (z)$ for $\nmax$ determined by procedure 1}}\label{J-fig1a}
   \end{subfigure}
   \begin{subfigure}{0.495\linewidth} \centering
     \includegraphics[scale=0.475]{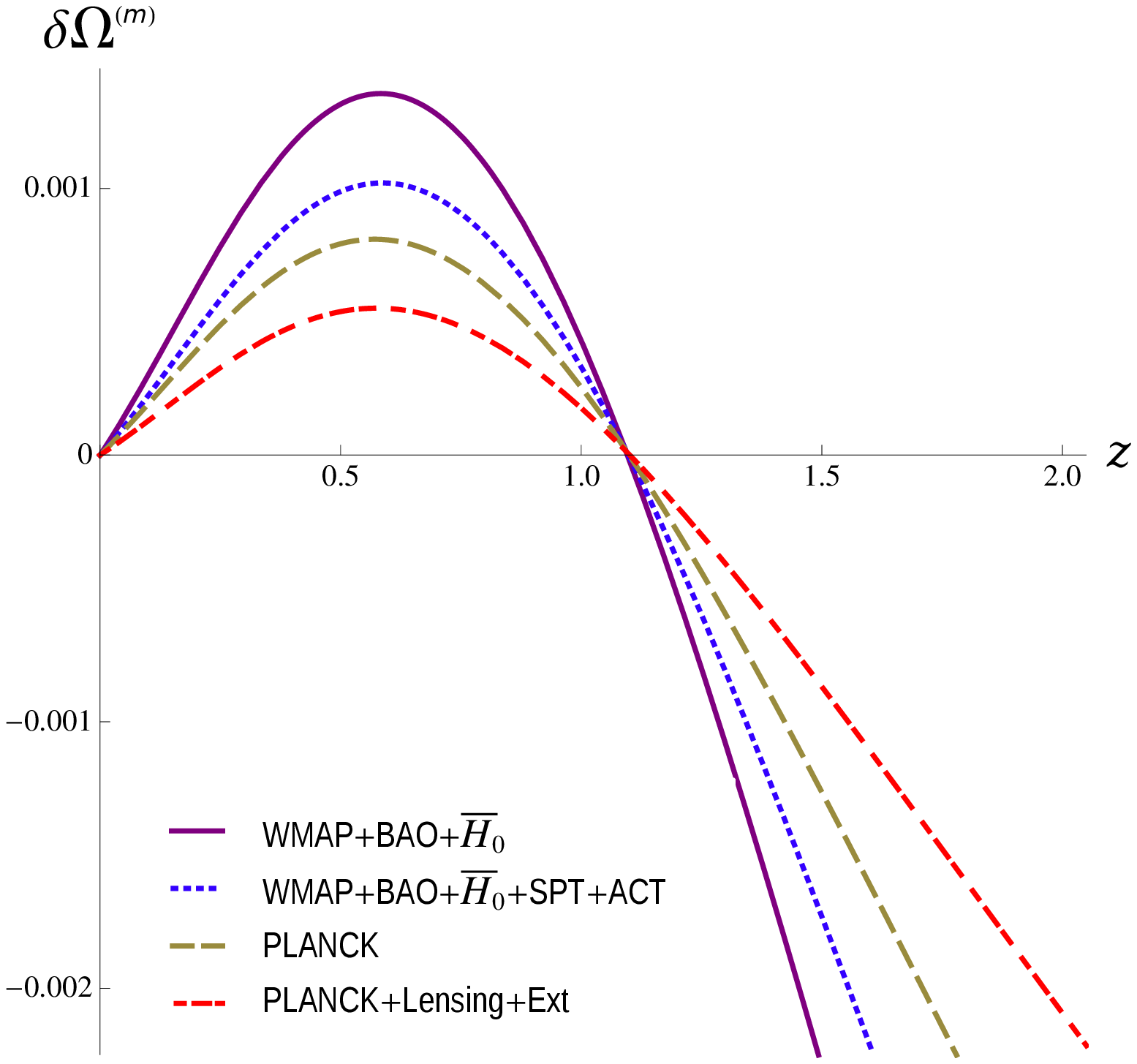}
     \caption{{\footnotesize $\d \Om (z)$ for $\nmax$ determined by procedure 2}}\label{J-fig1b}
   \end{subfigure}
\caption{\footnotesize Variation (with redshift $z$) of $\d \Om$, the Jordan frame maximal 
deviation from the $\L$CDM matter density, corresponding to $\nmax$ values obtained via the
procedures $1$ and $2$, for the different datasets.}
\vskip -0.1in
\la{J-fig1}
\end{figure}
%
\begin{figure}[!htb]
\centering
   \begin{subfigure}{0.495\linewidth} \centering
     \includegraphics[scale=0.475]{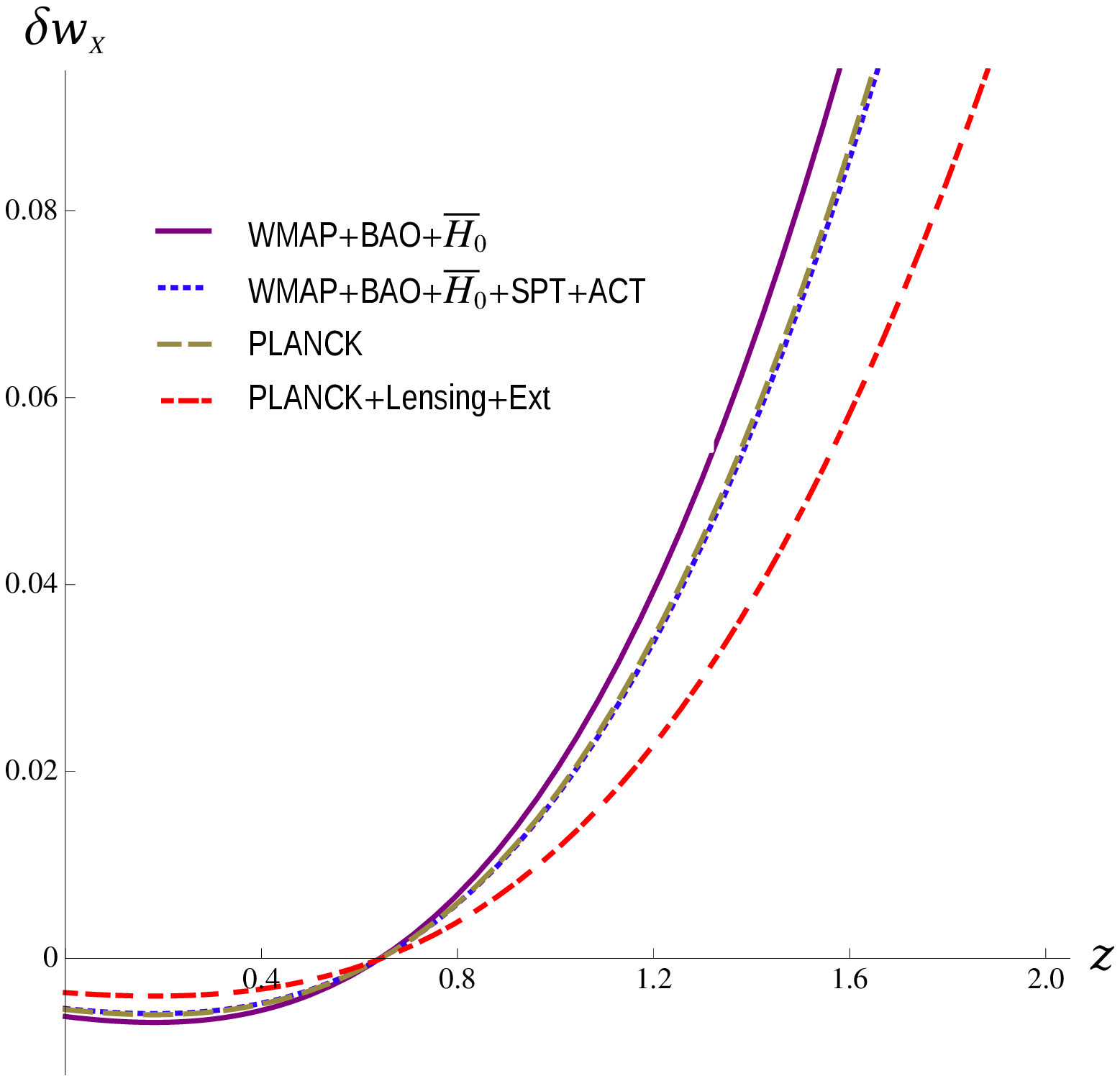}
     \caption{{\footnotesize $\d \wx (z)$ for $\nmax$ determined by procedure 1}}\label{J-fig2a}
   \end{subfigure}
   \begin{subfigure}{0.495\linewidth} \centering
     \includegraphics[scale=0.475]{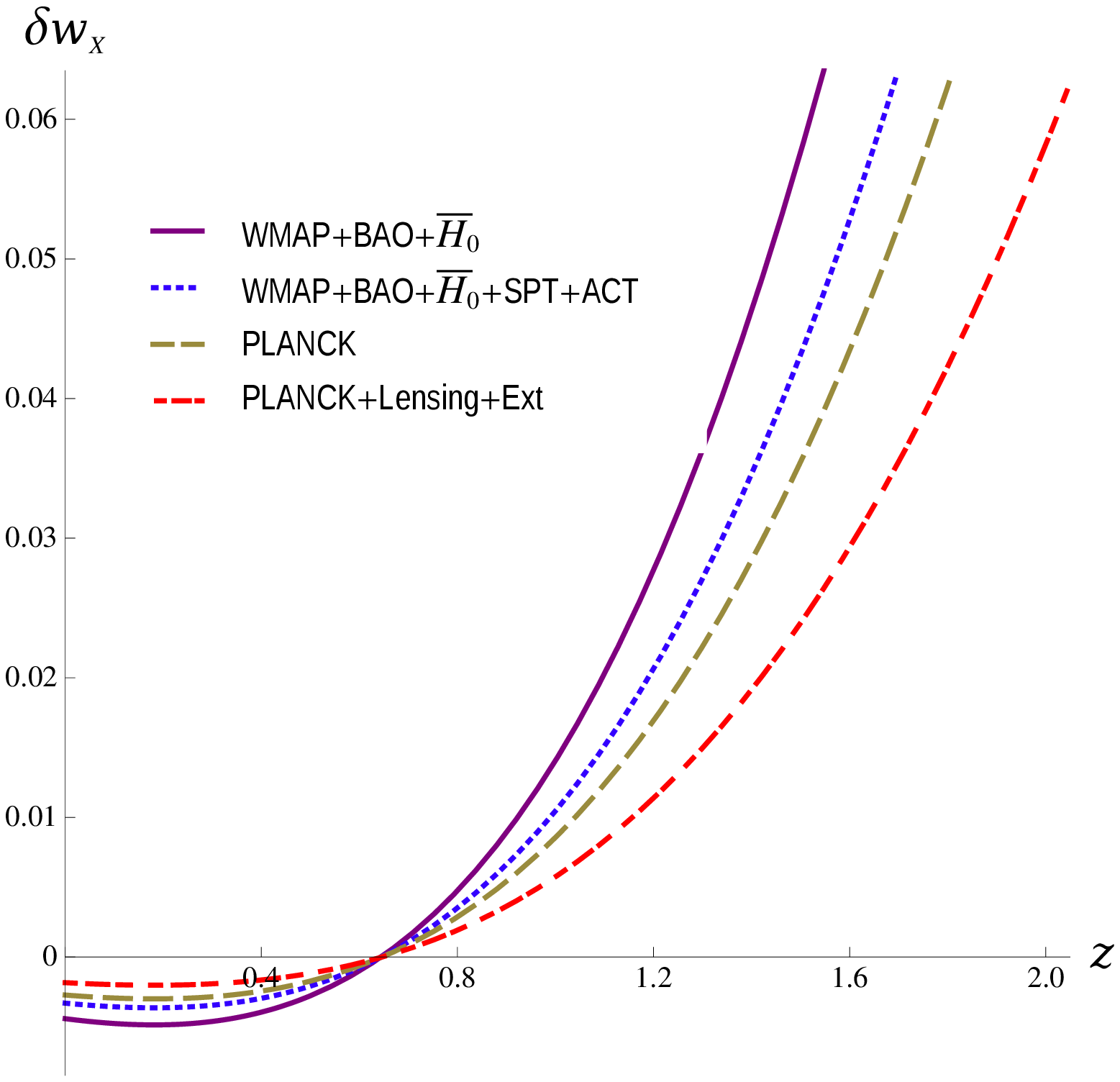}
     \caption{{\footnotesize $\d \wx (z)$ for $\nmax$ determined by procedure 2}}\label{J-fig2b}
   \end{subfigure}
\caption{\footnotesize Variation (with redshift $z$) of $\d \wx$, the Jordan frame maximal 
change over the $\L$CDM EoS ($= -1$) for the DE, corresponding to $\nmax$ values obtained 
via the procedures $1$ and $2$, for the different datasets.}
\vskip -0.15in
\la{J-fig2}
\end{figure}
%
Unlike the characteristics of the analogous quantity $\d \Omf$ in the Einstein frame, we 
see that $\d \Om > 0$ initially, and grows with $z$ as we go back in the past, reaches a 
maximum and then drops back to zero, and finally becomes negative with rapidly increasing 
magnitude as $z$ increases further. Although the $\d \Om$ curves are quite different for 
the different $\nmax$ values (determined by the two procedures), the redshifts of the 
maximum point and the zero point change very little ($z \in \le[0.5693,0.5865\ri]$ and 
$z \in \le[1.0927,1.1010\ri]$ respectively) for those $\nmax$ values. Figs. \ref{J-fig2} 
(a) and (b) show the variations of $\d \wx$ in the redshift range $\, z \in \le[0,2\ri]$ 
for all the datasets, and for $\nmax$ determined by the procedures $1$ and $2$ respectively. 
The striking feature noticed here, in comparison to the (Einstein frame) plots in Figs. 
\ref{E-fig2} (a) and (b), is that $\d \wx < 0$ (i.e. $\wx < - 1$) at the present epoch 
($z = 0$). That is, presently the universe is undergoing an effective {\em super-accelerating} 
or {\em phantom} phase of expansion. In the past, as $z$ decreases (from say, a value equal 
to $2$), $\d \wx$ changes from positive to negative (i.e. crosses over to the phantom phase), 
and attains a minimum (negative) value at a redshift close to the present epoch, and increases 
slowly thereafter (but remains negative). The redshifts of both the crossing point ($z_c 
\simeq 0.64$) and the minimum point ($z \simeq 0.18$) are nearly irrespective of the value 
of $\nmax$. However, the minimum value of $\d \wx$ is larger in magnitude for larger $\nmax$. 
There is an unevenness in the nature of the $\d \wx (z)$ curves overall for the different 
$\nmax$ values, similar to what we have seen in the Einstein frame (for different $\smax$ 
therein). This can again be attributed to the dependence of $\d \wx$ on the best fit value 
of $\Omp$.

\vskip 0.1in
\no 
{\bf Up to a moderately distant past:} 
%
\begin{figure}[!htb]
\centering
   \begin{subfigure}{0.495\linewidth} \centering
     \includegraphics[scale=0.48]{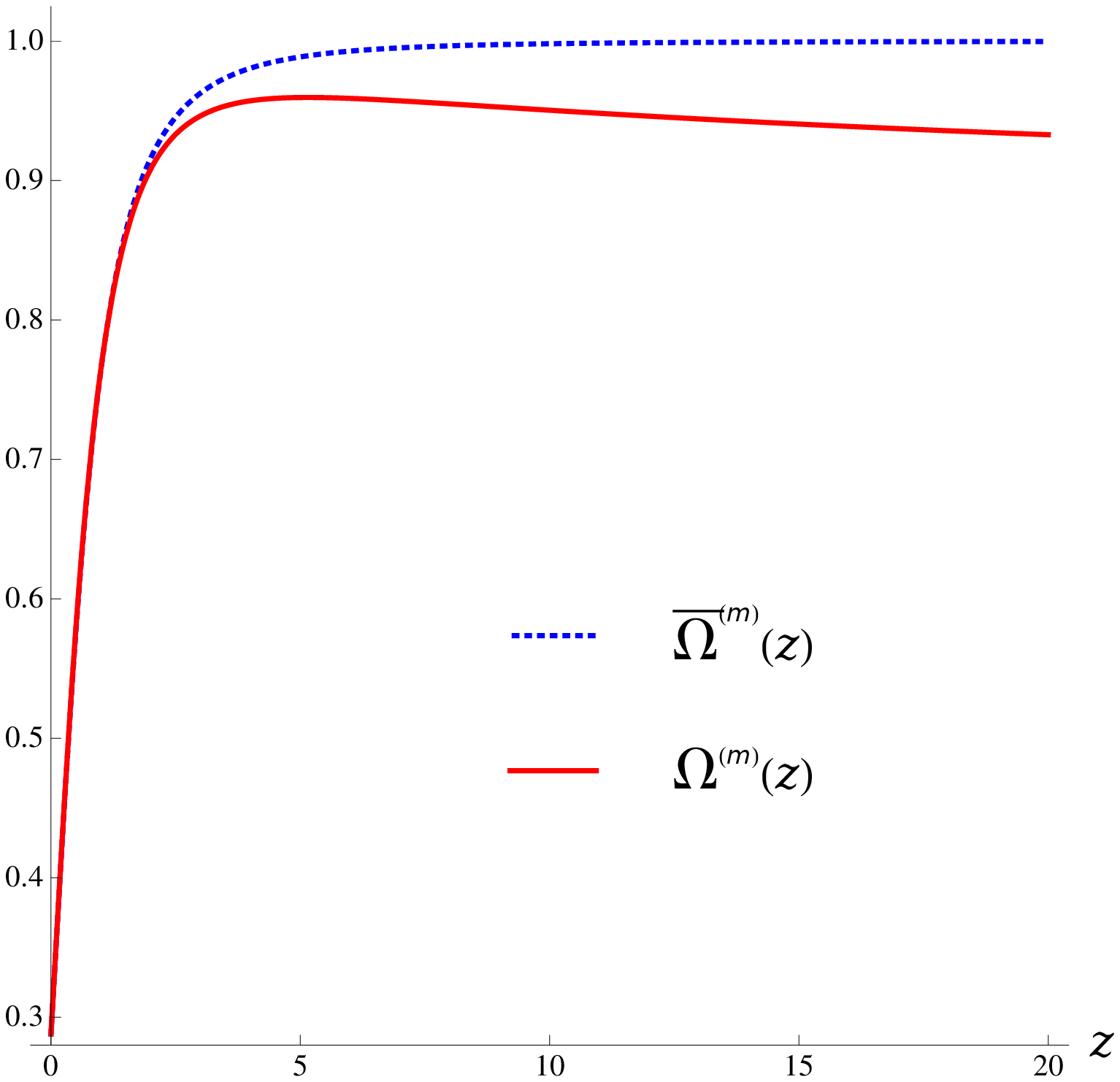}
     \caption{{\footnotesize Matter density parameters for $z \in [0, 20]$}}\label{J-fig3a}
   \end{subfigure}
   \begin{subfigure}{0.495\linewidth} \centering
   \vskip 0.1in
     \includegraphics[scale=0.49]{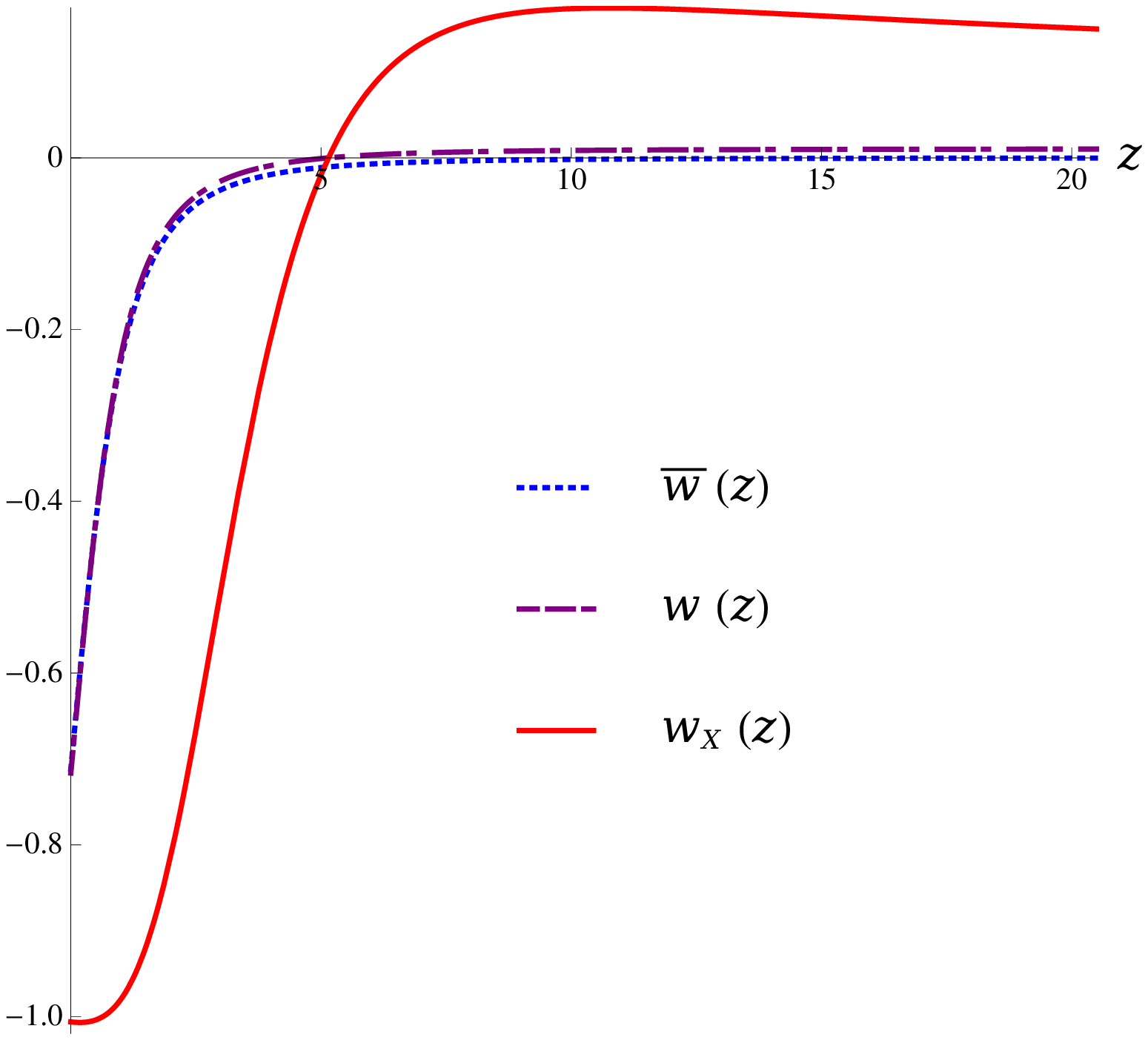}
     \caption{{\footnotesize EoS parameters for $z \in [0, 20]$}}\label{J-fig3b}
   \end{subfigure}
\caption{\footnotesize Jordan frame evolution, from the present ($z = 0$) upto a redshift $z = 20$
in the past, of (a) the matter density parameters $\Om$ and $\Omb$, and (b) the EoS parameters 
$\wx$, $\sw$ and $\bw$. The value of $\nmax$ used is the largest estimated one ($= 0.03123$ in 
Table \ref{J-tab1}).}
\vskip -0.15in
\la{J-fig3}
\end{figure}
%
For a fairly large redshift range $\, z \in 
\le[0, 20\ri]$, we have the evolution of the matter density parameters $\Om$ and $\Omb$ 
as shown in Fig. \ref{J-fig3} (a), and that of the EoS parameters $\wx, \sw$ and $\bw$ 
as shown in Fig. \ref{J-fig3} (b). 
As before, we take as the reference value of the model parameter, the largest estimated one, 
viz. $\nmax = 0.03123$, obtained via the proceduce $1$ applied to the dataset $1$ (see Table 
\ref{J-tab1}). The plots of $\Om, \wx$ and $\sw$ in Fig. \ref{J-fig3} (a) correspond to this 
value of $\nmax$. We see that up to $z \simeq 2$, $\, \Om$ changes very little over $\Omb$, 
and then deviates more and more from the latter as we go back in the past. Whereas $\Omb$ 
gets saturated to unity at redshifts $z \gtrsim 5$, $\, \Om$ attains a maximum value $0.9596$ 
at $z = 5.1503$, and then falls off slowly with increasing $z$. As to the EoS parameters in
Fig. \ref{J-fig3} (b), we see that $\sw$ evolves almost identically as $\bw$ over the entire 
range $\, z \in \le[0,20\ri]$, whereas $\wx$ shows some interesting characteristics. At the 
present epoch ($z = 0$) we have $\wx = - 1.0062$, and as we go back in the past $\wx$ drops 
to a minimum value $- 1.0069$ at $z = 0.1815$, turns back and crosses the $\L$CDM value ($= 
-1$) at $z_c = 0.6424$, continues increasing rapidly with $z$ until slowing down at $z \simeq 
4$, crossing the zero value at $z = 5.1503$ and increasing further till attaining a maximum 
value $0.1745$ at $z = 10.7731$, turning back once again finally and decreasing very slowly 
with further increase in $z$. All these are quite in contrast to what happens in the Einstein 
frame, viz. $\wx$ always stays above $- 1$ (see Fig. \ref{E-fig3} (b)).  

\vskip 0.1in
\no 
{\bf Near past to the extreme future:} 
%
\begin{figure}[!htb]
\centering
   \begin{subfigure}{0.495\linewidth} \centering
     \includegraphics[scale=0.525]{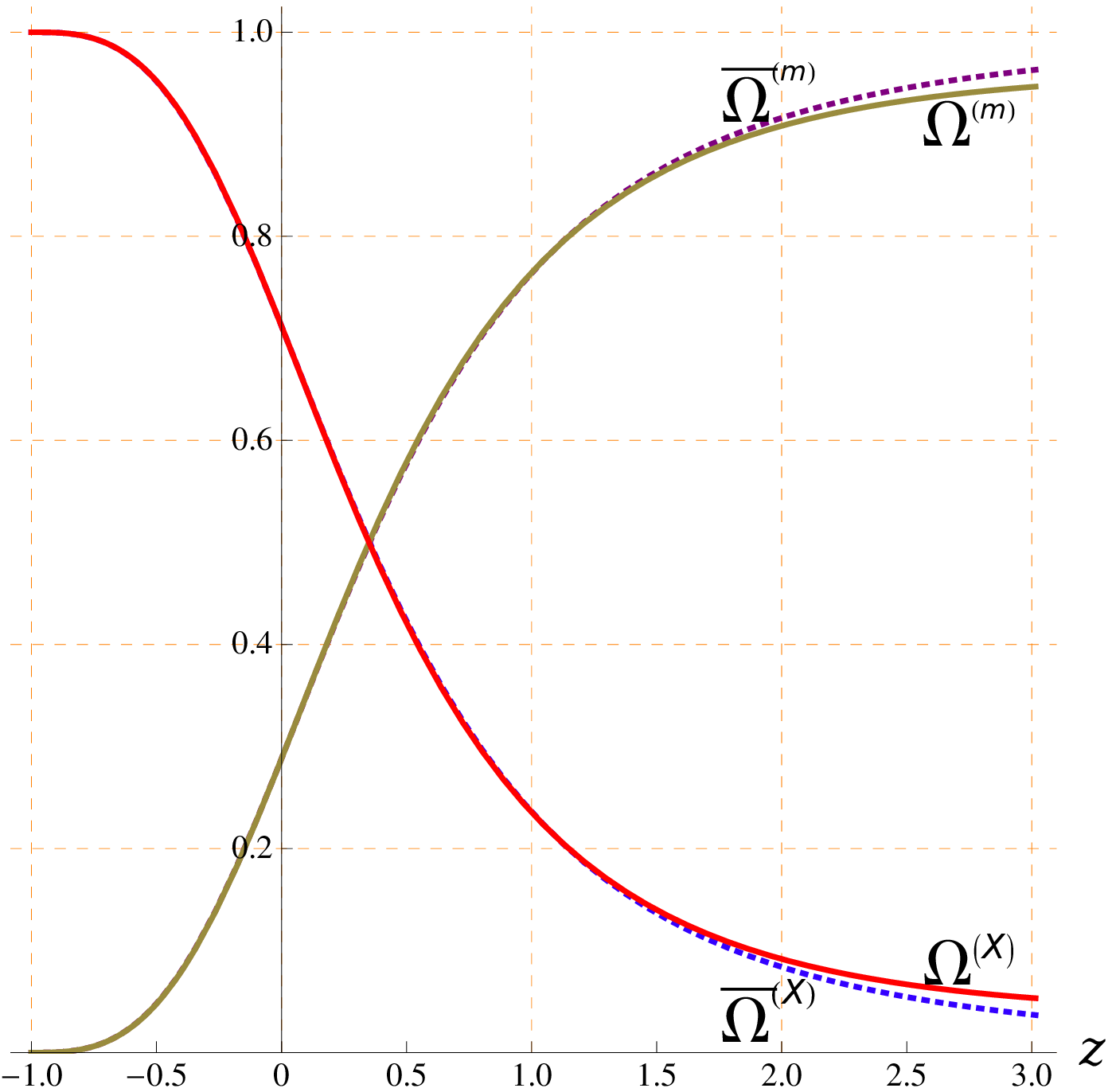}
     \caption{{\footnotesize Matter and DE density parameters for $z \in (-1, 3]$}}\label{J-fig4a}
   \end{subfigure}
   \begin{subfigure}{0.495\linewidth} \centering
     \includegraphics[scale=0.5]{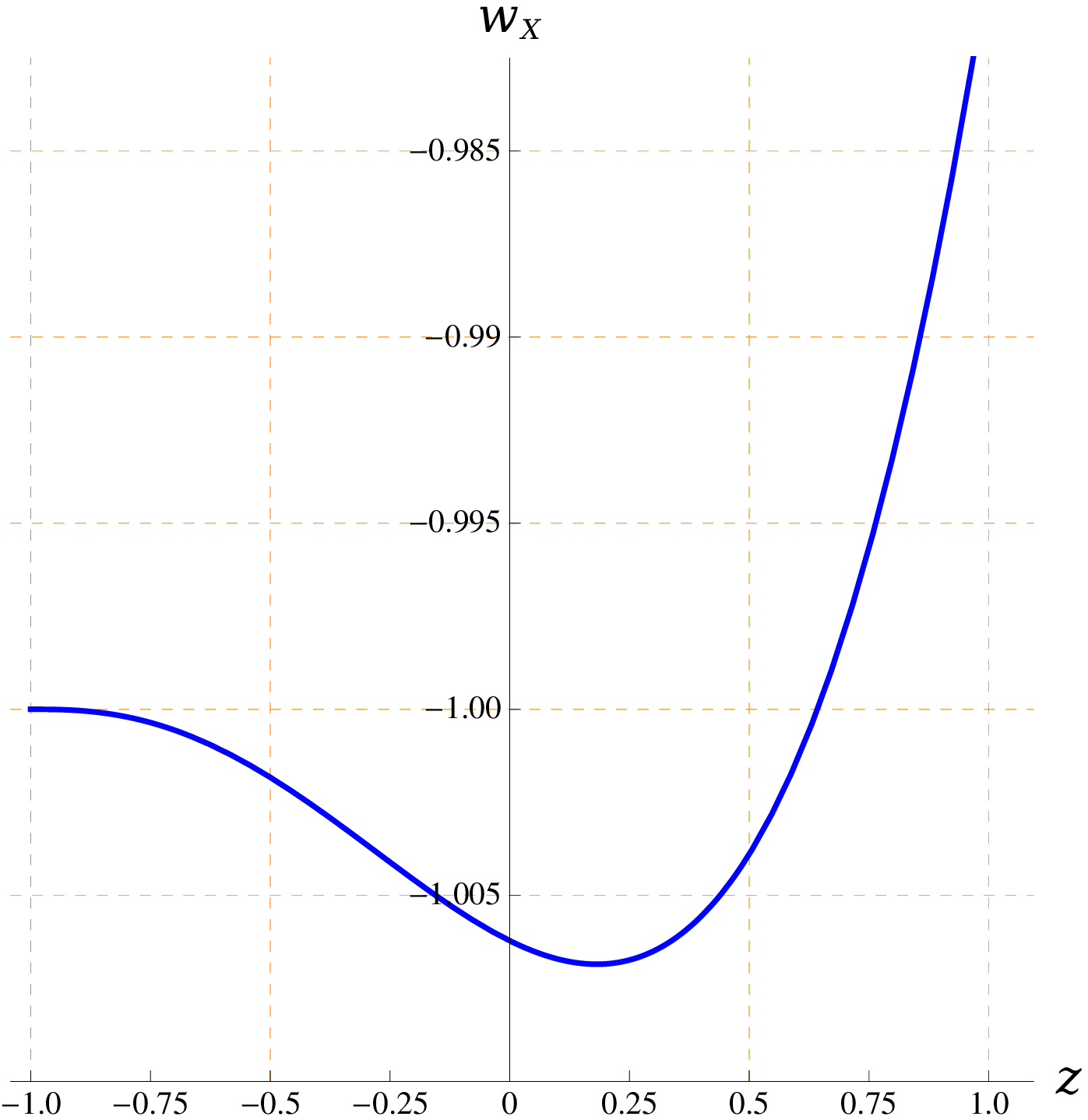}
     \caption{{\footnotesize Dark energy EoS parameter for $z \in (-1, 1]$}}\label{J-fig4b}
   \end{subfigure}
\caption{\footnotesize Jordan frame evolution up to the extreme future ($z = -1$), of (a) the 
density parameters $\Om$ and $\OX$, and their $\L$CDM analogues $\Omb$ and $\OXb$, from a redshift 
$z = 3$, and (b) DE EoS parameter $\wx$ from the near past ($z = 1$ say). The $\nmax$ value used 
is the largest estimated one ($= 0.03123$ in Table \ref{J-tab1}).}
\vskip -0.1in
\la{J-fig4}
\end{figure}
%
Finally, refer to the plots in Figs. \ref{J-fig4} (a) 
and (b), showing evolutions all the way up to the extreme future ($z = - 1)$, for the reference 
setting $\nmax = 0.03123$. 
While Fig. \ref{J-fig4} (a) shows the evolution, say from a redshift 
$z = 3$, of the matter density and DE density parameters, $\Om$ and $\OX = 1 - \Om$ respectively, 
and their $\L$CDM analogues $\Omb$ and $\OXb$ respectively, Fig. \ref{J-fig4} (b) shows the 
evolution of the DE EoS paramter $\wx$ from an epoch in the near past epoch (at $z = 1$ say), 
till $z = - 1$. 
No noticiable deviation of $\Om$ from $\Omb$ (or of $\OX$ from $\OXb$) is seen, over the entire
future regime ($z < 0$) as well as in the past till the redshift is quite high ($z \gtrsim 2$), 
much larger than that at the matter-DE equality, viz. $z_{eq} = 0.3495$ (or $0.3522$ for
$\L$CDM). So almost no improvement is there, as to the resolution of the coincidence problem 
encountered in the $\L$CDM model\footnote{It is easy to verify that this is true not only for 
the dataset 1, which yields $\nmax = 0.03123$ by the procedure 1, but also for the other datasets 
and the corresponding $\nmax$ values.}. The evolution of $\wx$ in Fig. \ref{J-fig4} (b) is in 
many respects similar to that in the Einstein frame (see Fig. \ref{E-fig4} (b)). Nevertheless, 
the striking difference to be noted is the effective phantom regime. This is not {\it transient}
either. With decreasing $z$, $\, \wx$ decreases and crosses the phantom barrier at $z_c = 0.6424$, 
continues to decrease till reaching the minimum value $\, \wxmin = - 1.0069$ at $z = 0.1815$, 
turns back and increases, but always remains below the $\L$CDM value $-1$. Only asymptotically, 
$\wx \rightarrow -1$ (from below) as $z \rightarrow -1$.

\subsection{Bounds on torsion parameters and the effective Brans-Dicke parameter} \la{sec:J-tbounds}

The Jordan frame solution ansatz (\ref{J-ans}) and the Friedmann equation (\ref{J-Hub}) 
enable us to express the norms of the torsion mode vectors $\cT_\m$ and $\cA^\m$, given 
by Eqs. (\ref{J-T-norm1}) and (\ref{A-norm}), in terms of the Hubble rate $H$ as
\bea
&& \le|\cT\ri| (z) \,=\, \fr{3 \, \dot{\F} (z)}{2 \, \F (z)} \,=\, \fr{3 n} 2 \, H (z) 
\,\,, \la{J-Tnormz} \\
&& \le|\cA\ri| =\, 4 \, \k \, \sq{6 \, \L} \,=\, 2 \sq{12} \, H (z) \, \sq{\le(n + 2\ri) 
\le(n + 3\ri) -\, 6 \le(1 + z\ri)^n \Om (z)} \,\,,
\la{J-Anormz}
\eea
in accordance with our prior assumption that $\cA^\m$ is responsible entirely for the scalar 
field mass (following e.g. the proposition 2 in $\S$\ref{sec:A-mass}). 
%
\begin{table*}[!htb]
\centering
\renewcommand{\arraystretch}{1.4}
\begin{tabular}{|cc||c c c c|}
\hline 
\multicolumn{2}{|c||}{Observational dataset} & Estimations:
& {$\rfraa{\le|\cT\ri|_{_0}\,}{\bHp} \Big|_{\nmaxb}$} 
& {$\rfraa{\le|\cT\ri|_{_0}\,}{\le|\cA\ri|} \Big|_{\nmaxb}$}
& {$\fwmin$} \\ [3pt]
\hline\hline 
1. & \hspace{-15pt} {\small WMAP\_9y+BAO+$\bHp$} & {\footnotesize (by procedure 1)} & $0.0462$ 
& $0.0032$ & $31.0205$ \\
\cline{3-6}
& & {\footnotesize (by procedure 2)} & $0.0328$ & $0.0023$ & $44.2694$ \\
\hline
2. & \hspace{-15pt} {\small WMAP\_9y+SPT+ACT} & {\footnotesize (by procedure 1)} & $0.0408$ 
& $0.0028$ & $35.3504$ \\
\cline{3-6}
& \hspace{-15pt} {\small SNLS\_3y+BAO+$\bHp$} & {\footnotesize (by procedure 2)} & $0.0251$ 
& $0.0018$ & $58.3120$ \\
\hline
3. & \hspace{-25pt} {\small PLANCK} {\scriptsize TT,TE,EE} & {\footnotesize (by procedure 1)} 
& $0.0357$ & $0.0025$ & $40.6493$ \\
\cline{3-6}
& \hspace{-25pt} {\small +LowP} & {\footnotesize (by procedure 2)} & $0.0177$ & $0.0013$ 
& $83.1043$ \\
\hline
4. & \hspace{-25pt} {\small PLANCK} {\scriptsize TT,TE,EE} & {\footnotesize (by procedure 1)} 
& $0.0246$ & $0.0018$ & $59.5327$ \\
\cline{3-6}
& \hspace{-25pt} {\small +LowP+Lensing+Ext} & {\footnotesize (by procedure 2)} & $0.0123$  
& $0.0008$ & $120.0654$ \\
\hline
\end{tabular}
\caption{\footnotesize Torsion trace parameter at the present epoch $\le|\cT\ri|_{_0}$ 
in units of $\bHp$, and the ratio of $\le|\cT\ri|_{_0}$ and the torsion pseudo-trace 
parameter $\le|\cA\ri|$, in the Jordan frame. Both the quantities are evaluated at $n = 
\nmax$. The corresponding minimum values of the effective BD parameter $\fw$ are also 
shown for the various datasets.}
\vskip -0.1in
\la{J-tab2}
\end{table*} 
%
Table \ref{J-tab2} shows the present-day trace mode norm $\le|\cT\ri|_{_0} \equiv 
\le|\cT\ri|_{z=0}$, in units of the $\L$CDM Hubble constant $\bHp$ and $\le|\cA\ri|$, 
evaluated at $n = \nmax$:
\bea 
&& \le[\fr{\le|\cT\ri|_{_0}}{\bHp}\ri]_{\nmax} =\, \fr {3 \sq{3} \, \nmax}
{\sq{2 \le(\nmax + 2\ri) \le(\nmax + 3\ri)}} \,, \qquad \mbox{and} \la{J-Tmax} \\
&& \le[\fr{\le|\cT\ri|_{_0}}{\le|\cA\ri|}\ri]_{\nmax} =\, \fr{\sq{3} \, \nmax} 
{4 \, \sq{2 \le[\le(\nmax + 2\ri) \le(\nmax + 3\ri) - \, 6 \, \Omp\ri]}} \,, 
\la{J-TAmax}
\eea
using the best fit $\bHp, \, \Omp$ and the $\nmax$ estimates (by the procedures 
$1$ and $2$), for the different datasets. Also shown in Table \ref{J-tab2}, the 
corresponding range of the lower bound on the effective BD parameter $\, \fw = 
\le(n^{-1} - 1\ri)$, viz. $\fwmin \in \le[31, 120\ri]$ approximately. So, compared 
to what have been obtained in the Einstein frame (see Table \ref{E-tab2} in 
$\S$\ref{sec:E-tbounds}), we have the values of $\fwmin$ here to be more 
in agreement with those found in independent studies \cite{acq-BD,avi-BD,chen-BD,als-BD}.

\section{Transformations relating parameters in the two frames} \la{sec:transf}

So far we have dealt with the observational aspects of the DE model that emerges out
of the MST equivalent scalar-tensor theory, assuming {\it any one} of the Einstein 
and the Jordan frames to be of physical relevance. We have obtained the field 
equations in the two frames and considered them to be describing the dynamical 
evolution of the universe in respective circumstances. Explicit solutions of such 
equations enabled us to estimate the model parameters which quantify the state of 
cosmology at any given epoch. However, the question that remains is that if the 
solutions in one frame, say the Jordan frame, are known, can we deduce the 
corresponding solutions in the other (Einstein) frame without having to go through 
the tedious process of explicitly solving the field equations in that frame. We can 
indeed do that, but only with the knowledge of how the cosmic time, the scale factor 
and the scalar field and torsion parameters transform from one frame to the other. 
Such transformations could be fixed by the relations (\ref{conf}) and (\ref{E-phi}). 
The metric redefinition (in terms of the conformal factor) yields two independent 
equations which, in view of preserving the FRW metric structure (\ref{FRW}), defines 
the cosmic time and the scale factor in a transformed frame. For the ansatz\'e chosen 
in the two frames, viz. Eqs. (\ref{E-ans}) and (\ref{J-ans}), the transformation 
equations are
\be  \la{all-transf}
\wht \,=\, \int dt \, a^{\rfra{\nbig} 2} (t) \,\,, \qquad \ha \le(\wht \,\ri) \,=\, a^{1 + 
\rfra{\nbig} 2} (t) \,\,. 
\ee
In addition, one may verify that 
\be \la{s_transf}
s \,=\, \fr n {n+2} \,\,.
\ee
Note again that the Einstein frame quantities are marked with a ($\wedge$) over them. 
Given an expression in one frame, the use of the above equations would lead to the 
corresponding expression in the other frame. The estimated upper bounds on the numerical 
values of the parameters $s$ and $n$, shown respectively in the Tables \ref{E-tab1} and 
\ref{J-tab1}, do not however comply with Eq. (\ref{s_transf}). This is expected, since 
such estimates have been made assuming either the Einstein frame or the Jordan frame as 
the physical frame in the respective analysis. This facilitated the use of the same 
observational data for computing the fractional errors $\D_h$ and $\D_m$ in each of the
frames. The idea behind our approach in $\S$\ref{sec:E-MST} and $\S$\ref{sec:J-MST}
has therefore been to differentiate between the physical viability of the two frames 
using the computed results. One could have taken the alternative route of appropriately 
scaling the observational data using the above transformation equations, considering
the Einstein and the Jordan frames to be physically equivalent. However, that would 
inevitably have lead back to the longstanding problem of dealing with the running 
coupling constants in a physical theory.

\section{Conclusion}  \la{sec:concl}

We have thus demonstrated that a non-minimal metric-scalar coupling with
space-time torsion may lead to a self-consistent DE model with slow dynamics.
The latter of course complies with the requirement of not having much
parametric distortion from the base $\L$CDM model. Such a requirement 
arizes naturally in view of (i) the wide observational support of $\L$CDM
(despite its theoretical limitations), and (ii) the miniscule experimental
signature of torsion in extensive researches till date. Moreover, one has
to get in an agreement with the high value of the lower bound on the 
effective Brans-Dicke parameter $\fw$ in the MST equivalent scalar-tensor
theory. Our analysis in $\S$\ref{sec:E-MST} and $\S$\ref{sec:J-MST} show
that such a bound is indeed tenable in our non-minimal MST formalism in the 
standard cosmological setup. While the non-minimal coupling removes the 
ambiguity in choosing the MST action, one has to take into account the 
constraints on the torsion mode parameters for maintaining the FRW metric 
structure. The result of course, is that the trace mode of torsion is 
sourced by the scalar field, which leaves us with the arbitrariness only 
in selecting the potential for the scalar. As it turned out, for an almost 
$\L$CDM like DE model, it suffices taking just a mass term for the scalar 
field. We have made propositions for such a scalar field mass (or, at 
least a part of it) out of torsion itself, in a way that the completely 
antisymmetric part of the latter gets eliminated via a suitably chosen 
constraint. Thus overall, the MST action gets reduced to the scalar-tensor 
form, alongwith the potential of the scalar field given by its mass term, 
and the cosmological matter as the non-relativistic dust. Now, a major 
obstacle in the scalar-tensor theory is to choose between the two frames, 
Einstein and Jordan,  the ``physical" one for interpreting the observational 
results. One may of course consider the two frames to be physically equivalent, 
provided in one of them the units of length, time and mass are explicit 
functions of the scalar field (and therefore time-varying) \cite{frni,fujii}. We 
have however discarded such a scenario (on account of its feasibility), and 
have also taken the non-minimal coupling parameter $\b$ to be positive definite 
(so as to ensure that the underlying quantum theory is bounded from below). 
Considering rigit sets of units in both the frames, our objective has been to 
carry out the analysis separately in the Einstein frame (supposing it to be 
the physical one), as well as in the Jordan frame (supposing it to be physical). 
This is of course a safeguard approach, in lieu of getting involved in the 
longstanding debate as to which frame is actually physical. 

From the technical point of view, instead of performing a likelihood analysis, we 
have stipulated that the cosmological parameters for our model to be kept within 
the $68 \%$ confidence limits of the corresponding ones for a reference $\L$CDM 
model. This enabled us a direct way of complying with the modern acceptance range 
of the present-day values of the cosmological parameters, via a reasonable order 
of magnitude estimation of the same. We have referred to two procedures of 
determining the statistical upper bound on our model parameter from the $\L$CDM
parametric marginalizations for a few recent datasets. Such procedures have been
followed in both the Einstein frame and the Jordan frame analysis in $\S$
\ref{sec:E-MST} and $\S$\ref{sec:J-MST} respectively. What we have observed is 
that, even with the small parametric changes, there are significant characteristic 
differences of our model with $\L$CDM. Most notably, the effective EoS parameter
$\wx$ for dark energy in the Jordan frame crosses over from a value $> -1$ to a 
value $< -1$ in the near past. It then reaches a minimum value, turns back but 
continues to remain below $-1$ throughout the present and future evolution of the 
universe. Such a crossing is usually not tenable in the scalar field models of DE, 
e.g. quintessence or k-essence, unless there is(are) {\em ghost} or {\em phantom} 
degree(s) of freedom involved, which bring(s) in instabilities \cite{vik}. However, 
in our case we have the effective cross-over to the {\em phantom regime} in a 
Brans-Dicke equivalent theory, which is at least stable against cosmological 
density perturbations \cite{frni,fujii,delcamp,bronn}. So unlike most of the scalar 
field DE formulations, our MST cosmological setup in the Jordan frame does not pre-assign
a theoretical limitation for the statistical marginalization of the cosmological 
parameters using the observational data. Another important aspect is that in the 
model-independent parametrizations of DE, the phantom barrier crossing is actually 
favoured (albeit mildly) by the recent observational data. For example, the Planck 
2015 (TT,TE,EE + LowP + Lensing + Ext) results \cite{ade-pln15-13,ade-pln15-15} show that 
$\wx$ given by the Chavellier-Polarsky-Linder (CPL) ansatz \cite{CP,linder}, has the 
best fit value at the present epoch $\, \wx (0) = - 1.019$. One may check that estimations 
close to this value would indeed be obtained from our analysis in $\S$\ref{sec:J-MST}, 
for the same Planck data (i.e. the dataset $4$ in Table \ref{J-tab1}), viz. $\, \wx (0) 
= - 1.0037$ (corresponding to $\, n = \nmax = 0.01652$, by procedure $1$), and $\, 
\wx (0) = - 1.0018$ (corresponding to $\, n = \nmax = 0.00826$, by procedure $2$). 

Finally, we should emphasize the high minimal estimates (corresponding to the 
different datasets) of the effective Brans-Dicke parameter $\fw$, found in both 
the Einstein frame and the Jordan frame analysis in $\S$\ref{sec:E-MST} and 
$\S$\ref{sec:J-MST} (see Tables \ref{E-tab2} and \ref{J-tab2}). Such high estimates 
are in agreement with the values obtained in many independent studies 
\cite{acq-BD,avi-BD,chen-BD,als-BD}, 
which are therefore in support of our MST-DE model. In fact, they also imply that 
the BD theory is not much deviated from GR. In our case, since we have torsion as 
one of the key ingredients in the effective BD formulation, the high estimates of 
the lower limit of $\fw$ simply mean that the direct effects of the torsion modes 
on the space-time geometry are very weak. Especially, for the late-time cosmologies, 
one always expects torsion to have a minor role. This is indeed reflected in the 
smallness of the upper limit on the present-day value of the torsion trace mode 
parameter shown in the Tables \ref{E-tab2} and \ref{J-tab2}. So, the effective DE 
evolution in our MST formalism must be attributed to the effect of the coupling of 
torsion with the scalar field, or more appropriately, that of the mass of the scalar, 
whether due to the torsion pseudo-trace or otherwise.       

Some open questions are in order: (i) what about the mathematical stability of 
the above formalism, i.e. against small fluctuations in the solution space? (ii)
what about the stability against density perturbations? (iii) what if the cosmological 
dust interacts explicitly (in a very non-trivial way) with torsion? (iv) what if 
we take some other cosmological fluid, e.g. Chaplygin gas, instead of dust? (vi) can
the scalar field source of the torsion trace mode have the interpretation of a
chameleon, and if so, for what potentials? (v) what if choose to consider the 
non-minimal coupling in some other form, or assign a dynamical source to the torsion 
pseudo-trace (such as the Kalb-Ramond field)? Attempts have been made to address some 
of these \cite{ssasbPP,ssasbPPfull}, and studies are in progress  
\cite{ssasbCham,ssetalDDE} which we hope to report soon.

\section*{Acknowledgement}

SS acknowledges the R \& D Grant DRCH/R \& D/2013-14/4155, Research Council, University of 
Delhi. The work of ASB is supported by the Council of Scientific and Industrial Research 
(CSIR), Government of India. The authors also thank the anonymous referee for useful 
comments that have prompted improving this paper by a great extent.

\end{document}